\documentclass[showpacs,superscriptaddress,floatfix,amsmath,amssymb,twocolumn,pra]{revtex4-1}
\usepackage{bm}
\usepackage{graphicx}
\usepackage{color}
\usepackage[normalem]{ulem}
\usepackage{epstopdf}
\usepackage{soul}
\usepackage{url}
\usepackage[normalem]{ulem}
\usepackage{verbatim}

\usepackage{hyperref}
\hypersetup{pdftex,colorlinks=true,allcolors=blue,bookmarksnumbered=true}
\usepackage{hypcap}

\def\Tr{{\rm Tr}}
\def\Re{{\rm Re}}
\def\Im{{\rm Im}}
\def\sgn{{\rm sgn}}

\begin{document}

\author{Yaxing Zhang}
\affiliation{Department of Physics and Applied Physics, Yale University, New Haven, Connecticut 06520, USA \\
and Yale Quantum Institute, Yale University, New Haven, Connecticut 06511, USA}
\author{Brian J. Lester}
\affiliation{Department of Physics and Applied Physics, Yale University, New Haven, Connecticut 06520, USA \\
and Yale Quantum Institute, Yale University, New Haven, Connecticut 06511, USA}
\author{Yvonne Y. Gao}
\affiliation{Department of Physics and Applied Physics, Yale University, New Haven, Connecticut 06520, USA \\
and Yale Quantum Institute, Yale University, New Haven, Connecticut 06511, USA}
\author{Liang Jiang}
\affiliation{Department of Physics and Applied Physics, Yale University, New Haven, Connecticut 06520, USA \\
and Yale Quantum Institute, Yale University, New Haven, Connecticut 06511, USA}
\author{R. J. Schoelkopf}
\affiliation{Department of Physics and Applied Physics, Yale University, New Haven, Connecticut 06520, USA \\
and Yale Quantum Institute, Yale University, New Haven, Connecticut 06511, USA}
\author{S. M. Girvin}
\affiliation{Department of Physics and Applied Physics, Yale University, New Haven, Connecticut 06520, USA \\
and Yale Quantum Institute, Yale University, New Haven, Connecticut 06511, USA}

\title{Engineering bilinear mode coupling in circuit QED: Theory and experiment}

\begin{abstract}
Photonic states of high-Q superconducting microwave cavities controlled by superconducting transmon ancillas provide a platform for encoding and manipulating quantum information. 
A key challenge in scaling up the platform towards practical quantum computation is the requirement to communicate on demand the quantum information stored in the cavities. It has been recently demonstrated that a tunable bilinear interaction between two cavity modes can be realized by coupling the modes to a bichromatically-driven superconducting transmon ancilla, which allows swapping and interfering the multi-photon states stored in the cavity modes [Gao et al., Phys. Rev. X 8, 021073(2018)]. Here, we explore both theoretically and experimentally the regime of relatively strong drives on the ancilla needed to achieve fast SWAP gates but which can also lead to undesired non-perturbative effects that lower the SWAP fidelity.
We develop a theoretical formalism based on linear response theory that allows one to calculate the rate of ancilla-induced interaction, decay and frequency shift of the cavity modes in terms of a susceptibility matrix. We go beyond the usual perturbative treatment of the drives by using Floquet theory, and find that the interference of the two drives can strongly alter the system dynamics even in the regime where the standard rotating wave approximation applies. The drive-induced AC Stark shift on the ancilla depends non-trivially on the drive and ancilla parameters which in turn modify the strength of the engineered interaction. We identify two major sources of infidelity due to ancilla decoherence. i) Ancilla dissipation and dephasing lead to incoherent hopping among Floquet states which occurs even when the ancilla is at zero temperature; this hopping results in a sudden change of the SWAP rate, thereby decohering the SWAP operation. ii) The cavity modes inherit finite decay from the relatively lossy ancilla through the inverse Purcell effect; the effect becomes particularly strong when the AC Stark shift pushes certain ancilla transition frequencies to the vicinity of the cavity mode frequencies. The theoretical predictions agree quantitatively with the experimental results, paving the way for using the developed theory for optimizing future experiments and architecture designs.
\end{abstract}

\maketitle

\section{Introduction}

The use of multi-photon states of superconducting microwave cavities to encode quantum information with control provided by transmon ancillas offers a promising route towards robust quantum computation~\cite{Ofek2016,Rosenblum2018,gao2018a,gao2018b,hu2018}.   
A key challenge in scaling up the architecture towards practical quantum computing is the requirement to communicate (entangle, SWAP, etc.) on demand the quantum information stored in the cavities. One promising solution towards fulfilling this requirement is to generate a set of controllable interactions between cavities that are bilinear in cavity lowering or raising operators, which includes both a beam-splitter type and two-mode squeezing interaction,
\begin{equation}
\label{eq:bilinear}
V=g_{\rm BS}(t) a^\dagger  b + g_{\rm BS}^*(t) a b^\dagger  + g_{\rm TMS}(t)a^\dagger b^\dagger +g^*_{\rm TMS}(t)ab, 
\end{equation}
where $a,b$ and $a^\dagger,b^\dagger$ are the annihilation and creation operators of the two cavity modes, and $g_{\rm BS}$ and $g_{\rm TMS}$ are the strengths of the engineered beam-splitter and two-mode squeezing interaction. 
The bilinear nature of the interactions has the crucial advantage that it does not introduce any additional nonlinearities in the system which thus remains analytically and numerically tractable for moderate-size systems.  The beam-splitter interaction is the basis for SWAP operations which can be used to route photonic signals between modules and is also a key element in important entangling operations such as deterministic controlled SWAP (Fredkin gate) and exponential SWAP (coherent superposition of SWAP and Identity) gates that have been recently experimentally realized~\cite{gao2018b}. These operations can empower novel schemes for universal bosonic quantum computation~\cite{lau2016}. The two-mode squeezing interaction, along with single-mode squeezing and beam-splitter interaction enables an essential set of operations needed for Gaussian quantum information processing~\cite{weedbrook2012} and quantum simulations of molecular spectra~\cite{huh2015a,shen2018,sparrow2018}.

Cavities have the advantage of having long lifetimes ($\sim$1 ms), but being harmonic oscillators, they require nonlinear ancillas (e.g. transmons) for universal control. The frequency mixing capability of the nonlinear transmon ancilla provides a natural way to engineer the bilinear interactions in Eq.~(\ref{eq:bilinear}) between cavities. Much like in nonlinear optics, modulating the ancillas (which play the role of nonlinear medium) with periodic drives induces effective ancilla-mediated interactions between the otherwise uncoupled cavities. The ancillas are only virtually excited, so the effects of their decoherence are partially mitigated. 
To name a few, bilinear mode interactions have been previously realized based on this method between two propagating microwave modes~\cite{abdo2013,flurin2012}, between one long-lived and one propagating mode~\cite{Pfaff2017}, and most recently between two long-lived 3D microwave cavities~\cite{gao2018a}. An alternative way to generate tunable cavity-cavity interactions is to use microwave resonators whose frequencies can be tuned into resonance via external flux drive~\cite{zakka-bajjani2011,pierre2018}, but that requires a careful analysis of the flux noise which usually limits the coherence time of the resonators. A similar method has also been applied to induce resonant couplings between two transmons~\cite{caldwell2018} and between one transmon and many cavity modes~\cite{Naik2017}.

Fast entangling and SWAP operation between the cavities requires the engineered interaction to be relatively strong, which in turn requires strong drives on the ancilla. In the presence of strong drives, properties (spectrum, decoherence rate, etc.) of the nonlinear ancilla can be strongly modified. For a two-level ancilla, well-known examples include drive-induced AC Stark shift and power broadening of the linewidth~\cite{loudon2000,schuster2005}. For a multi-level ancilla such as a weakly anharmonic transmon~\cite{koch2007}, the situation can be more complicated. On one hand, these modifications directly affect the gate rate and fidelity. On the other hand, properties of the cavities to which the ancilla is coupled to are also modified. One example is the inverse Purcell effect where the cavities inherit finite decay rate from the transmon ancilla due to the hybridization between them~\cite{reagor2016}. These non-perturbative effects due to the drives could potentially reduce the SWAP fidelity even when the rate of SWAP is enhanced. 

In this paper, we study both theoretically and experimentally a system that consists of two microwave cavities both coupled to a nonlinear transmon ancilla. It has been recently demonstrated that by driving the ancilla with two RF tones, an ancilla-mediated beam-splitter interaction arises between the two cavities which allows swapping and entangling the muti-photon states stored in the cavity modes~\cite{gao2018a}. Here, we investigate the regime of relatively strong drives needed to achieve fast SWAP gates.  In particular, we study in detail the drive dependence of the strength of the ancilla-mediated interaction and the mechanisms that affect the SWAP fidelity in this regime.

It has been shown that the dynamics of a driven multimode cQED system can be conveniently analysed based on the so-called ``black-box quantization"~\cite{Nigg2012} and perturbation theory for relatively weak drives or large drive detunings~\cite{Leghtas2015,gao2018a}. Here, we treat the drives non-perturbatively  by working in the basis of Floquet eigenstates of the driven ancilla, and show that the theory accurately captures the dynamics of the ancilla beyond the perturbative regime. Importantly, even in the regime where the standard rotating wave approximation (RWA) is applicable and thus the transmon ancilla can be treated as a weakly nonlinear oscillator~\cite{girvin2014}, the slow dynamics of the ancilla in the rotating frame of the drive can still be strongly nonlinear. The situation becomes more complicated when there are two drives where the frequency difference of the drives sets a new slow time scale. As we will show, interference between the drives can strongly alter the system dynamics and leads to effects such as non-monotonic AC Stark shift which typically does not occur when there is only one drive. Floquet theory has also been recently applied to a coupled cavity-transmon system subject to an extremely strong drive where the RWA breaks down~\cite{verney2018}; we are not considering that regime.

\begin{figure}[ht]
\centering
\includegraphics[width=8cm,height=3cm]{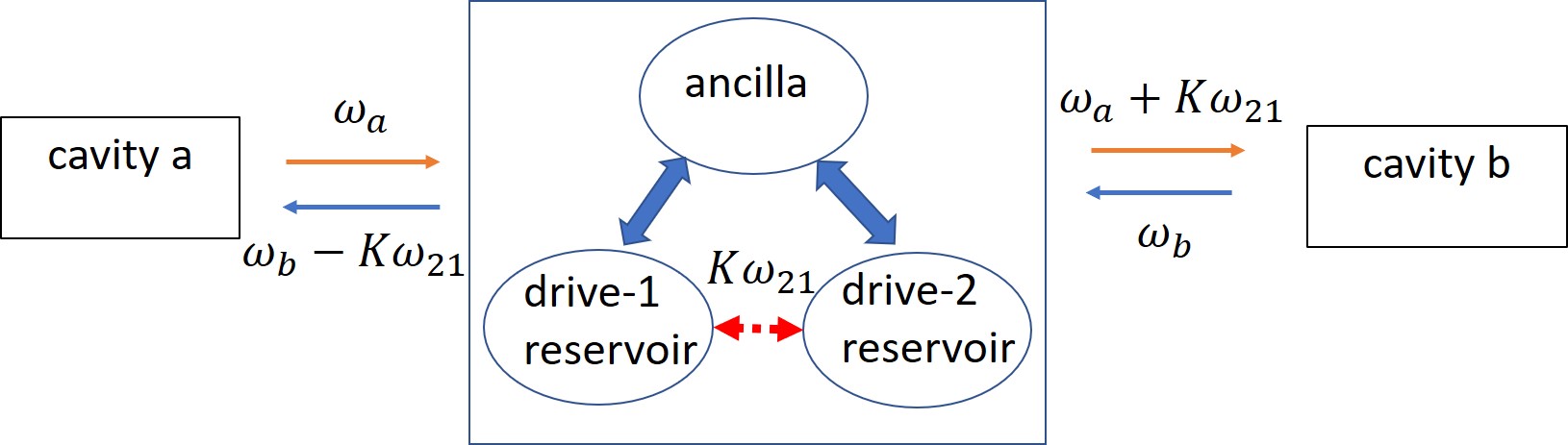} 
\caption{ A schematic showing how a beam-splitter interaction between two cavity modes at different frequencies arises due to their couplings to a driven ancilla. 
The frequency of cavity mode $a$ at the ancilla input (as a probe) is up- or down-converted by integer multiples $K$ of drive frequency difference $\omega_{21}$ at the output (as a response) which then interacts with mode $b$, whereas the frequency of cavity mode $b$ at the ancilla input (as a probe) is down- or up-converted by $K\omega_{21}$ at the output (as a response) which then interacts with mode $a$. When the condition $\omega_b -\omega_a = K \omega_{21}$ is satisfied, there arises a resonant beam-splitter interaction between the cavities $a$ and $b$. The energy needed for this process to become resonant is provided by a an indirect exchange of $K$ excitations between the two drive reservoirs (indicated by the dashed red arrow), which is a result of their individual interaction with the ancilla (indicated by the blue arrows).} 
\label{fig:schematic_linear_response}
\end{figure}

The ancilla-mediated bilinear interactions between the cavity modes are related to the linear response of the driven ancilla to the couplings to the cavities. Due to the interference between the drives and the nonlinearity, the linear response of the two-tone driven nonlinear oscillator (the transmon ancilla) has a much richer structure than a one-tone driven nonlinear oscillator (cf. \cite{dykman2012} and references therein) and is characterized by a susceptibility matrix which relates the probe (cavity $a$) and response (cavity $b$) at different frequencies; see Fig.~\ref{fig:schematic_linear_response}. As we will show, the spectra of these susceptibilities depend non-trivially on the drive and ancilla parameters.

A coherent quantum operation utilizing the ancilla-mediated interaction between the two cavities requires the driven ancilla to remain in a pure state during the operation. Finite coherence time of the driven ancilla due to dissipation and dephasing reduces the fidelity of the quantum operation mainly in the following two ways. Firstly, because of the noise that accompanies ancilla dissipation and dephasing, the ancilla can randomly hop between different Floquet eigenstates. This hopping leads to a sudden change in the strength of the ancilla-mediated interaction, thereby decohering the quantum operation.

Unlike a static system, even at zero temperature, there is a finite rate of both hopping ``up" and ``down"  in the ladder of ancilla Floquet states leading to a finite-width distribution among them, a phenomenon termed ``quantum heating"~\cite{Dykman2011,ong2013}. In our system which is effectively at zero temperature, we observed a finite steady-state population in the Floquet ``excited states" as large as 10\% for a relatively strong drive. A comparison with the theory shows that this heating is partly due to the quantum noise that accompanies dissipation and partly due to the noise that leads to ancilla dephasing. As we will show, the rate of the heating due to dephasing sensitively depends on the drive detuning from the ancilla frequency due to the strong frequency dependence of the noise spectrum. 

Secondly, even when the ancilla does not hop to another state during the operation, the cavity inherits finite decay rate from the typically lossier ancilla as a result of coupling-induced hybridization of the cavities and the driven ancilla. 
In the presence of drives, as we will show, such inherited decay can be significantly enhanced as a result of drive-assisted multi-photon resonances and care must be taken in choosing the frequencies of the drives to avoid these resonances. By the same hybridization mechanism, ancilla dephasing leads to incoherent hopping of excitations between the dressed cavities and the ancilla. This hopping effectively causes cavity photon loss if the rate of hopping back from the ancilla to the cavity is smaller than the relaxation rate of the ancilla.

The paper is organized as follows. In Sec.~\ref{sec:model}, we describe the system Hamiltonian followed by a general formulation that establishes the relations between the linear susceptibilities of the driven ancilla and the ancilla-induced bilinear interaction as well as decay and frequency shift of the cavities. We review in Appendix~\ref{sec:Cooper_pair_box} and \ref{sec:four_wave_mixing} the expansion of the Cooper-pair box Hamiltonian for the transmon ancilla and how the bilinear interaction between the cavities arises based on the four-wave mixing picture in the perturbative regime. We briefly describe the quantum noise that accompanies the ancilla-induced cavity decay in Appendix~\ref{sec:full_eom}. General relations between nonlinear susceptibilities of the ancilla and ancilla-induced Kerr of the cavities are shown in Appendix~\ref{sec:nonlinear_susceptibilities}.

In Sec.~\ref{sec:unitary}, we study the unitary dynamics of the two-tone driven ancilla. 
We start by describing the Floquet formulation; within this formulation, we study the drive-induced AC Stark shift of the ancilla transition frequencies beyond the perturbative regime as well as the process of multi-photon resonance. Finally, we derive explicit expressions for the ancilla susceptibilities in the basis of Floquet states. A detailed comparison between the theory and experiment is presented. We discuss in Appendix~\ref{sec:tight-binding} a formulation equivalent to Floquet theory by mapping to a time-independent tight-binding Hamiltonian. Ancilla dynamics in the semiclassical regime is discussed in Appendix~\ref{sec:semiclassical}.

In Sec.~\ref{sec:dissipation}, we study the Floquet dynamics of the driven ancilla in the presence of both dissipation and dephasing. Two major factors that limit the SWAP fidelity including the dissipation- and dephasing-induced hopping among Floquet states and the inverse Purcell effect are discussed in detail. In Appendix~\ref{sec:transient_susceptibilities}, we discuss the effects of ancilla decoherence on its susceptibilities in the transient regime. In Appendix~\ref{sec:two_state_model}, we discuss in detail the incoherent hopping between the cavities and the ancilla induced by ancilla dephasing. In Sec.~\ref{sec:conclusion}, we present concluding remarks.

\section{The system Hamiltonian and general formulation}
\label{sec:model}
Our goal is to engineer the tunable bilinear interactions shown in Eq.~(\ref{eq:bilinear}) between two initially uncoupled and far-detuned linear modes via a nonlinear ancilla. In this paper, we consider the linear modes to be modes of two microwave cavities and the nonlinear ancilla to be a transmon. The theoretical formulation applies also to other systems such as high frequency phononic modes controlled by a transmon ancilla~\cite{chu2017,satzinger2018}, eigenmodes of coupled cavity arrays~\cite{Naik2017}, or higher order modes of a single microwave cavity~\cite{zakka-bajjani2011,Sundaresan2015}. The Hamiltonian of the full system reads,
\begin{align}
\label{eq:Hamiltonian}
H &= H_0 + H_c + H_{\rm I}, \quad H_0/\hbar = \omega_a a^\dagger a + \omega_b b^\dagger b, \nonumber\\
H_c/\hbar &=  \omega_c c^\dagger  c -\frac{\alpha}{2} c^{\dagger2}c^2 \nonumber +(\Omega_1 e^{-i \omega_1 t}+\Omega_2 e^{-i \omega_2 t})c^\dagger+\rm{H.c.}  ,\nonumber\\
H_{\rm I}/\hbar &=  (g_a a  + g_b b ) c^\dagger + \rm{H.c.}.
\end{align}
Here, $H_0$ represents the Hamiltonian of the cavity system that consists of two modes with frequency $\omega_a$ and $\omega_b$, respectively. $H_c$ represents the Hamiltonian of the transmon ancilla whose creation and annihilation operators are denoted as $c^\dagger$ and $c$. $H_I$ represents the interaction between the two systems;  see below for a detailed explanation of the Hamiltonian $H_c$ and $H_I$ and the considered parameter regime. 


We model the ancilla as a weakly nonlinear oscillator with frequency $\omega_c$ and  Kerr nonlinearity whose strength is proportional to $\alpha$. The nonlinearity is weak in the sense that the oscillator frequency shift due to nonlinearity is much smaller than the oscillator eigenfrequency: $\alpha\langle c^\dagger c\rangle\ll \omega_c$. For a transmon, this Kerr nonlinearity comes from the expansion of a cosine potential; see Appendix~\ref{sec:Cooper_pair_box}. Without loss of generality, we will consider $\alpha>0$ as is the case for transmon. 

We consider two periodic drives on the ancilla with frequencies $\omega_{1,2}$ and amplitudes $\Omega_{1,2}$. We consider the drives to be off-resonant in the sense that the drive detunings $|\omega_{1,2}-\omega_c|$ from the ancilla frequency are much larger than the linewidth of the ancilla transitions so that the ancilla is only virtually excited by the drives.
In the mean time, we consider the drive detunings and the drive amplitudes to be much smaller than the ancilla frequency itself so that one can neglect the counter-rotating terms of the drives $\Omega_{1,2} e^{-i\omega_{1,2} t} c + \rm {h.c.}$ using the RWA; these terms are already disregarded in Eq.~(\ref{eq:Hamiltonian}). In this work, we will focus on the case where both drives are blue-detuned from $\omega_c$ so that each drive individually does not lead to bistability of the nonlinear oscillator when they become relatively strong \cite{landau1976}. Our formulation, however, applies to the general case of both red and blue detunings. 

We consider a bilinear interaction between the two cavity modes and the ancilla with a strength $g_a$ and $g_b$ as represented by $H_I$ in Eq.~(\ref{eq:Hamiltonian}). This interaction arises as a result of the coupling between the cavity electric fields and the charges on the islands of Josephson junction that supports the transmon mode. We have neglected the counter-rotating terms of this coupling such as $g_a ac+g_a^*a^\dagger c^\dagger$. This is valid when coupling strength and cavity detunings from the ancilla is smaller than the ancilla frequency, $|g_{a,b}|, |\omega_{a,b}-\omega_c|\ll \omega_c$. For the purpose of engineering unitary bilinear interactions between the cavities via the virtually-excited ancilla, we are interested in the regime where cavities $a$ and $b$ are far detuned from the ancilla so that $|g_{a,b}|\ll |\omega_{a,b}-\omega_c|$.


As a shorthand notation, we will define the detunings of the modes and the drives from the ancilla frequency as $\delta_{a,b,1,2}=\omega_{a,b,1,2}-\omega_c$ and the drive frequency difference as $\omega_{21}= \omega_2-\omega_1.$ Without loss of generality, we assume that $\omega_2>\omega_1.$ To be consistent with previously used notation~\cite{gao2018a}, we will also use a notation for the scaled drive amplitude $\xi_{1,2} = \Omega_{1,2}/\delta_{1,2}$ which can be understood as the classical response to the drives if the ancilla were linear. 

\subsection{Linear response of the two-tone driven nonlinear ancilla}
\label{sec:linear_response}
A typical approach to treat the driven multi-mode system described by Eq.~(\ref{eq:Hamiltonian}) is based on the black-box quantization~\cite{Nigg2012, Leghtas2015}. This method provides an elegant picture of multiwave mixing between the drives and cavity modes, and a straightforward way of calculating the strength of the ancilla-mediated interaction; see Appendix~\ref{sec:four_wave_mixing}. However, the approximation typically made in applying the method holds in the regime of weak ancilla anharmonicity ($\alpha \ll |\delta_{1,2}|,|\delta_{a,b}|$) and weak drive strengths. 

Our approach here is to treat the coupling between the ancilla and cavities as a perturbation, and calculate the linear response of the driven ancilla to the coupling. The linear response treatment is justified for the following two reasons: i) we are interested in the ancilla-mediated bilinear interaction between the cavities; ii) the ancilla-cavity coupling is effectively weak due to their large detuning so that higher-order response of ancilla to the cavities can be neglected. In the mean time, we are not treating the drives in linear response but instead using the Floquet theory to capture the non-perturbative effects of the drives.

Because of the nonlinearity and the drives on the ancilla, its linear response to a probe (the cavity modes) can be at different frequencies from the probe frequency. It is this frequency conversion capability that allows coupling two cavity modes at different frequencies as illustrated in Fig.~\ref{fig:schematic_linear_response}. When the probes are sufficiently weak such that linear response theory is valid and the frequency difference (or sum) of the two cavity modes matches the frequency (or sum) of the probe and response of the ancilla, there arises an effective beam-splitter (or two-mode squeezing) interaction between the two cavity modes; see next section for the derivation.

To study the linear response of the driven ancilla, we consider one additional drive (the probe) on the ancilla with a drive Hamiltonian $H_f/\hbar = -f_\omega e^{-i\omega t}c^\dagger  + \rm{H.c.}$. 
The role of $f_\omega$ is played by the fields of the cavity modes. To find the linear response, we solve the quantum Liouville equation for the ancilla plus the bath
\begin{equation}
\label{eq:Liouville}
\dot \rho = -i[H_{c+\rm{bath}}(t)+H_f(t),\rho]/\hbar,
\end{equation}
where $H_{c+\rm{bath}}$ is the total Hamiltonian of the driven ancilla and the bath it couples to that leads to ancilla decoherence. We will consider a specific model for the bath in Sec.~\ref{sec:dissipation}. Here, we proceed with a general formulation without specifying the details of the bath. $\rho$ is the total density matrix of the ancilla plus the bath. 
We now solve the density matrix to leading order in the probe field $f_\omega$: $\rho(t) \approx \rho^{(0)}(t)+\rho^{(1)}(t)$. Then we find that the linear response of the expectation value of the ancilla lowering operator $c$ to the probe can be characterized by two sets of susceptibilities (or two susceptibility matrices) and has the following form
\begin{align}
\label{eq:linear_response}
\langle c^{(1)}(t) \rangle\equiv &\,\Tr(c\rho^{(1)}(t)) \nonumber \\
= &\sum_{K=-\infty}^\infty \left[f_\omega \chi (\omega,\omega+K\omega_{21})e^{-i(\omega+K\omega_{21})t} \right.\nonumber\\
&\left.+f_\omega^* X (-\omega,2\omega_1 +K\omega_{21}-\omega)e^{-i(2\omega_1 +K\omega_{21}-\omega)t}\right],
\end{align}
where the susceptibilities are given by
\begin{align}
\label{eq:chi1_formal}
&\chi (\omega,\omega+K\omega_{21}) \nonumber\\
&=\frac{i}{\hbar} \int_0^t dt'\Tr\left( [c^{(0)}(t),c^{(0)\dagger}(t')] \rho(0)\right) e^{-i\omega(t'-t)+iK\omega_{21}t}
\end{align}
\begin{align}
\label{eq:X1_formal}
&X (-\omega,2\omega_1 +K\omega_{21}-\omega) \nonumber\\ 
&= \frac{i}{\hbar}\int_0^t dt' \Tr\left([c^{(0)}(t),c^{(0)}(t')]\rho(0)\right)e^{i\omega(t'-t)+i (2\omega_1 + K\omega_{21})t}
\end{align}
The susceptibilities $\chi $ and $X $ both have two arguments: the first is the probe frequency and the second is the response frequency. 
Following the convention used in nonlinear optics~\cite{boyd}, we use a positive frequency $\omega$ to indicate the response to a field with complex amplitude $f_\omega$ and a negative frequency $-\omega$ to indicate the response to a field with complex amplitude $f_\omega^*$. The Heisenberg operator $c^{(0)}(t)$ in the commutator evolves under the unitary evolution governed by $H_{c+\rm{bath}}(t)$. 
We note that when only drive-1 is present, all susceptibilities with $K\neq 0$ vanish.

The physical meanings of the two classes of susceptibilities $\chi $ and $X $ are as follows. Susceptibility $\chi (\omega,\omega+K\omega_{21})$ characterizes the frequency conversion process in which the probe frequency is up- or down-converted by integer multiples of $\omega_{21}$. Susceptibility $X (-\omega,2\omega_1+K\omega_{21}-\omega)$ characterizes the process where $(2-K)$ excitations in drive-1(with frequency $\omega_1$) and K excitations in drive-2 (with frequency $\omega_2$) are converted into one excitation at the probe frequency and one at the response frequency. Importantly, the total number of excitations is always conserved in the framework of the RWA. In the next section, we will show that the strength of the susceptibilities at the frequency of the cavity modes quantifies the strength of the ancilla-induced bilinear interaction between the modes.

To get some intuition about the drive-dependence of the susceptibilities, we qualitatively discuss here the situations of no drive, one drive and two drives on the ancilla. A more detailed discussion is given in Sec.~\ref{sec:susceptibilities_Floquet} and Sec.~\ref{sec:susceptibilities_dissipation}. In the absence of external drives, only the diagonal part of the susceptibility matrix $\chi $ is non-zero, i.e. $\chi (\omega,\omega)\neq 0$. The absorption spectrum $\Im \chi (\omega,\omega)$ has peaks at frequencies corresponding to transitions between neighboring levels of the ancilla and the spectrum $\Re \chi (\omega,\omega)$ has characteristic dispersive line shapes at the same frequencies. The two spectra are related via Kramers-Kronig relations. In the presence of one drive with frequency $\omega_1$, extra peaks emerge in the spectrum $\Im \chi {(\omega,\omega)}$ at frequencies corresponding to transitions between non-neighboring levels of the ancilla assisted by the drive. Also, the susceptibility $X (-\omega,2\omega_1-\omega)$ becomes non-zero. In the presence of two drives, all the susceptibilities in Eqs.~(\ref{eq:chi1_formal},\ref{eq:X1_formal}) with $K\neq 0$ becomes generally non-zero and their spectrum can have a much richer structure due to the interference between the two drives.

\subsection{Effective equations of motion for the cavity modes}
\label{sec:equations_of_motion}
The fields of the cavity modes perturb the ancilla; the back action from the ancilla induces frequency shift and decay of the cavity modes as well as interactions between them when a certain frequency matching condition is satisfied. We now make the connection of the linear susceptibilities to the ancilla-induced back action. 

The Heisenberg equations of motion for the cavity mode lowering operators $a,b$ in the interaction picture read,
\begin{align}
\label{eq:Heisenberg}
\dot {\tilde a} =  -ig_a^* e^{i\omega_a t} c , \nonumber\\
\dot {\tilde b} =  -ig_b^* e^{i\omega_b t} c,
\end{align}
where $\tilde a = ae^{i\omega_a t},\tilde b = be^{i\omega_b t}$. 

In the spirit of linear response theory, we will make the substitution in Eq.~(\ref{eq:Heisenberg}): \[c(t) \approx c^{(0)}(t) + \langle  c^{(1)}(t) \rangle\] where $\langle c^{(1)}\rangle$ is given by Eq.~(\ref{eq:linear_response}) with $f_\omega$ replaced by $-g_a\tilde a$ and $-g_b \tilde b$, and $\omega$ replaced by $\omega_a$ and $\omega_b$, respectively. This procedure is equivalent to the standard Born-Markov approximation applied in tracing over the ancilla degree of freedom. The approximation relates to the fact that the dynamics of the cavity modes in the interaction picture is much slower than the relaxation dynamics of the ancilla or the rate determined by the detuning between the cavity modes and the ancilla. Under this approximation, we should only keep slowly varying terms after substituting the expression for $\langle c^{(1)}\rangle$.

\subsubsection{Ancilla-induced beam-splitter interaction between the cavity modes}
We now consider separately the two cases of engineering beam-splitter and two-mode squeezing interaction between the two cavity modes. In the first case, an excitation of one cavity mode is converted into an excitation of the other cavity mode at a different frequency. The energy offset is  compensated by exchanging (indirectly) excitations between the reservoirs of the two drives. Therefore,  frequencies of the cavity modes must satisfy the condition
\begin{align}
\label{eq:frequency_matching_BS}
\omega_b - \omega_a \approx K\omega_{21},
\end{align}
for any integer $K$. When the above condition is satisfied, we obtain approximate equations of motion for the cavity modes after disregarding rapidly oscillating terms, 
\begin{align}
\label{eq:eom_BS}
\dot {\tilde a} &= -\frac{\kappa^{(0)}_a}{2}\tilde a - (i \delta\omega_a+ \frac{\delta\kappa_a}{2}) \tilde a  -( i g_{\rm BS}+ \kappa_{\rm BS})\tilde b e^{-i\delta_{\rm BS} t} , \nonumber\\ 
\dot {\tilde b} &= -\frac{\kappa^{(0)}_b}{2}\tilde b - (i \delta\omega_b + \frac{\delta\kappa_b}{2}) \tilde b - (i g^*_{\rm BS}+\kappa^*_{\rm BS})\tilde a e^{i\delta_{\rm BS} t},
\end{align} 
where we denote the detuning from the frequency matching condition as $\delta_{\rm BS} = \omega_b - \omega_a -K\omega_{21}$. We have also included the intrinsic decay of the cavity modes with a rate $\kappa^{(0)}_{a,b}$. For simplicity, we have not written explicitly the noise that accompanies $\kappa^{(0)}_{a,b}$ and that accompanies $\delta\kappa_{a,b}$ and $\kappa_{\rm BS}$ which come from the terms proportional to $c^{(0)}(t)$. A detailed discussion of the quantum noise is presented in Appendix~\ref{sec:full_eom}.

Substitution of ancilla operator $c$ with its linear response $\langle c^{(1)}\rangle$ results in a shift in the frequency of the cavity modes 
\begin{align}
\label{eq:frequency_shift}
\delta\omega_{a,b} =  -|g_{a,b}|^2 \Re \chi (\omega_{a,b},\omega_{a,b})
\end{align}
and a modification to the cavity decay rate
\begin{align}
\label{eq:delta_kappa_formal}
\delta\kappa_{a,b} =2 |g_{a,b}|^2 \Im \chi (\omega_{a,b},\omega_{a,b}).
\end{align}
The above frequency shift encodes the drive-induced AC Stark shift on the cavities and the decay is related to the inverse Purcell effect we mentioned in the introduction. We will discuss these in more detail in Sec.~\ref{sec:susceptibilities_Floquet} and Sec.~\ref{sec:susceptibilities_dissipation}, respectively. 

In addition, because of the frequency matching condition in Eq.~(\ref{eq:frequency_matching_BS}), there arises an effective beam-splitter interaction between the cavity modes whose unitary and non-unitary part are related to the susceptibility $\chi $ via: 
\begin{align}
\label{eq:g_BS}
g_{\rm BS} = -g_a^* g_b [\chi (\omega_b,\omega_a)+(\chi (\omega_a,\omega_b))^*]/2
\end{align}
\begin{align}
\kappa_{\rm BS} = g_a^* g_b [\chi (\omega_b,\omega_a)-(\chi (\omega_a,\omega_b))^*]/2i. 
\label{eq:kappa_BS}
\end{align}
Here we have neglected the finite detuning $\delta_{BS}$ and made the approximations $ \chi (\omega_b,\omega_b-K\omega_{21})\approx \chi (\omega_b,\omega_a)$ and $\chi (\omega_a,\omega_a+K\omega_{21})\approx \chi (\omega_a,\omega_b)$. This is consistent with the approximation made in substituting operator $c(t)$ with its linear response which requires the linear susceptibility $\chi (\omega,\omega+K\omega_{21})$ to be sufficiently smooth over the scale of $\delta_{\rm BS}$, in other words, $\delta_{\rm BS}$ needs to be much smaller than the ancilla relaxation rate or the detuning of the cavity modes from the ancilla frequency. 

Of primary interest in this paper is to engineer a relatively strong unitary beam-splitter interaction characterized by $g_{\rm BS}$ in Eq.~(\ref{eq:eom_BS}). For this purpose, as we will show, it is important to design the cavity frequencies $\omega_{a,b}$ to be far away from any resonant structures of the susceptibility $\chi (\omega_a,\omega_b)$, so that $\kappa_{\rm BS}$ is largely suppressed and $g_{\rm BS}$ is relatively strong. 
In the case where the unitary beam-splitter interaction dominates, the solution to Eq.~(\ref{eq:eom_BS}) reads,
\begin{align}
\label{eq:sol_BS}
\tilde a(t) &= \tilde a(0)\cos(|g_{\rm BS}| t) - i e^{i \phi (g_{\rm BS})} \tilde b(0)\sin(|g_{\rm BS}|t) \nonumber \\
\tilde b(t) &= \tilde b(0)\cos(|g_{\rm BS}| t) - i e^{-i\phi(g_{\rm BS})} \tilde a(0)\sin(|g_{\rm BS}|t).
\end{align}
where the phase $\phi_{\rm BS}$ is the argument of the complex beam-splitter rate $\phi_{\rm BS} \equiv \rm {arg}(g_{\rm BS})$, i.e. $\exp(i\phi_{\rm BS})=g_{\rm BS}/|g_{\rm BS}|$. It depends on the phases of the couplings $g_a,g_b$ and the relative phases of the two drives: $\phi_{\rm BS} = \rm {arg}( g_a^* g_b (\Omega_1 \Omega_2^*)^K )$ or $\rm {arg}( g_a^* g_b (\Omega_1 \Omega_2^*)^K )+\pi$. $g_a,g_b$ can always be made real by choosing a gauge for the modes $a$ and $b$, then $\phi_{\rm BS}$ only depends on the relative phases of the two drives. At $t = \pi/4|g_{\rm BS}|$, Eq.~(\ref{eq:eom_BS}) corresponds to a 50:50 beam splitter; at $t = \pi/2 |g_{\rm BS}|$, it corresponds to a SWAP of the states between the two modes. 

\subsubsection{Ancilla-induced two-mode squeezing interaction between the cavity modes}
In the case of engineering two-mode squeezing interaction, excitations of the two cavity modes are simultaneously converted into or from the excitations of the drives where the total number of excitations remains the same. A most general condition for this process to become resonant is
\begin{align}
\label{eq:frequency_matching_TMS}
\omega_a + \omega_b &\approx (2-K)\omega_1+K\omega_2 \nonumber \\
& = 2\omega_1 + K\omega_{21},
\end{align}
for any integer $K$.  When this condition is satisfied, one obtains a similar set of equations of motion for the cavity modes as in Eq.~(\ref{eq:eom_BS}) with the beam-splitter interaction replaced by the two-mode squeezing interaction,
\begin{align}
\label{eq:eom_TMS}
&\dot {\tilde a} = -\frac{\kappa^{(0)}_a}{2}\tilde a - (i \delta\omega_a+ \delta\kappa_a) \tilde a  -( i g_{\rm TMS}+ \kappa_{\rm TMS})\tilde b^\dagger  e^{i\delta_{\rm TMS} t}, \nonumber\\ 
&\dot {\tilde b}^\dagger = -\frac{\kappa^{(0)}_b}{2}\tilde b^\dagger  + (i \delta\omega_b - \delta\kappa_b) \tilde b^\dagger  + (i g^*_{\rm TMS}+\kappa^*_{\rm TMS})\tilde a e^{-i\delta_{\rm TMS} t},
\end{align}
where we denote the detuning from the frequency matching condition in Eq.~(\ref{eq:frequency_matching_TMS}) as $\delta_{\rm TMS} = \omega_a + \omega_b -2\omega_1 - K\omega_{21}$. 

The unitary and non-unitary parts of the two-mode squeezing interaction are related to the susceptibility $X $ via: 
\begin{align}
\label{eq:g_TMS}
g_{\rm TMS} = -g_a^* g_b^* [X (\omega_b,\omega_a)+X (\omega_a,\omega_b)]/2, 
\end{align}
\begin{align}
\kappa_{\rm TMS} = g_a^* g_b^* [X (\omega_b,\omega_a)-X (\omega_a,\omega_b)]/2i. 
\label{eq:kappa_TMS}
\end{align}
Similar to Eqs.~(\ref{eq:g_BS},\ref{eq:kappa_BS}), we have made the approximations $X (\omega_b,2\omega_1+K\omega_{21}-\omega_b)\approx X (\omega_b,\omega_a)$ and $X (\omega_a,2\omega_1+K\omega_{21}-\omega_a)\approx X (\omega_a,\omega_b).$ Here we emphasize that two-mode squeezing interaction can arise in the case of only one drive when the condition $\omega_a+\omega_b\approx 2\omega_1$  is satisfied. 

When the unitary two-mode squeezing interaction dominates, we obtain the solution to the equations of motion in Eq.~(\ref{eq:eom_TMS}) to be,
\begin{align}
\label{eq:sol_TMS}
\tilde a(t) &= \tilde a(0)\cosh(|g_{\rm TMS}| t) - i e^{i \phi (g_{\rm TMS})} \tilde b^\dagger (0)\sinh(|g_{\rm TMS}|t), \nonumber \\
\tilde b^\dagger (t) &= \tilde b^\dagger (0)\cosh(|g_{\rm TMS}| t) + i e^{-i\phi(g_{\rm TMS})} \tilde a(0)\sinh(|g_{\rm TMS}|t),
\end{align}
where $\phi_{\rm TMS} \equiv \rm {arg}(g_{\rm TMS}) = \rm {arg}(g_a^* g_b^* (\Omega_1 \Omega_2)^K )$ or $\rm {arg}(g_a^* g_b^* (\Omega_1 \Omega_2)^K )+\pi$.

We note that when the condition $2\omega_a = 2\omega_1 + K\omega_{21}$ is satisfied, there arises a single-mode squeezing term $\tilde a^2 + a^{\dagger 2} $ in the Hamiltonian. The above results for two-mode squeezing [Eqs.~(\ref{eq:eom_TMS},\ref{eq:g_TMS},\ref{eq:sol_TMS})] apply to single-mode squeezing as well with $b$ replaced by $a$ everywhere. 

Formally, Eqs.~(\ref{eq:frequency_shift},\ref{eq:delta_kappa_formal},\ref{eq:g_BS},\ref{eq:kappa_BS},\ref{eq:g_TMS},\ref{eq:g_TMS},\ref{eq:kappa_TMS}) comprise a set of key results of this paper. They allow us to calculate the strengths of the ancilla-induced interactions between the cavity modes as well as ancilla-induced frequency shifts and decay rates of the cavity modes in the presence of ancilla drives. One can also establish relations between the nonlinear susceptibilities of the ancilla and the ancilla-induced conservative and dissipative nonlinearity of the cavity modes. These ancilla- induced cavity nonlinearities can be useful for engineering nonlinear interactions between cavities and self interaction for a single cavity mode; see Appendix~\ref{sec:nonlinear_susceptibilities}.

\section{Floquet theory of the two-tone driven nonlinear ancilla}
\label{sec:unitary}
In this section, we neglect the coupling between the ancilla and its environment and study the unitary dynamics of the driven ancilla. 
In Sec.~\ref{sec:dissipation}, we will discuss the effects of ancilla decoherence. 

As a qualitative picture, the two off-resonant drives on the ancilla have two major effects. Firstly, they both lead to the AC Stark shift of the ancilla energy levels. This AC Stark shift results from the drive-induced mixing between unperturbed ancilla eigenstates. In a Floquet language, the AC Stark shift is embedded in the drive-dependence of the quasienergies of the driven ancilla. Due to the interference between the two drives, as we will show, the dependence of the AC Stark shift on the drive amplitudes displays interesting behaviors that are absent in the case of one drive. 

Secondly, interference of the two drives leads to a nontrivial periodic-modulation of the ancilla Floquet states at the difference frequency $\omega_{21}\ll \omega_{1,2}$. We emphasize that this modulation occurs at a frequency much smaller than the usual periodic modulation of Floquet states at the drive frequency (sometimes termed ``micro motion") when there is only one drive. Thus it can have a significant effect on the ancilla dynamics even in the regime where the RWA applies. 

Because of this periodic modulation of ancilla Floquet states at frequency $\omega_{21}$, linear response of the ancilla initialized in a given Floquet state can oscillate at a frequency different from the probe frequency by integer multiples of $\omega_{21}$. This lies behind the frequency conversion capability of the driven ancilla as described in Sec.~\ref{sec:model}.

In this section, we will first describe the Floquet formulation of the two-tone driven ancilla. Then we will explore, within this formulation, the drive-induced AC Stark shift of ancilla levels and the linear susceptibilities of the driven ancilla. We will also present a comparison between the theory and experiment on both the AC Stark shift and the ancilla-induced beam-splitter rate between two off-resonant cavity modes within a cQED setup.

\subsection{The Floquet formulation}
\label{sec:Floquet_formulation}
At first sight, since the ancilla Hamiltonian $H_c(t)$ in Eq.~(\ref{eq:rotated_Hc}) is modulated with two drives whose frequencies are generally incommensurate, one may need a generalization of the standard Floquet theory that only applies to a periodic Hamiltonian to the case of quasi-periodic Hamiltonian as a result of two or more incommensurate modulations. Such generalization leads to extra dimensions in the Floquet space which is analogous to the Bloch theory in solids of higher than one dimension  \cite{Ho1983, Casati1989, Martin2017}. However, in our case, because we have neglected non-RWA terms for both drives, our Hamiltonian can be treated in fact by the standard one-frequency Floquet theory. This can be seen by going to the rotating frame at one of the drive frequencies, for instance, $\omega_1$. The resulting Hamiltonian reads 
\begin{align}
\label{eq:rotated_Hc}
\tilde H_c(t)/\hbar =  -\delta_1 c^\dagger  c -\frac{\alpha}{2} c^{\dagger 2}c^2+ (\Omega_1^* + \Omega_2^* e^{i\omega_{21}t})c+\rm {h.c.}, 
\end{align}
We emphasize that here we can not apply the rotating wave approximation a second time to eliminate the time-dependence in $\tilde H_c$ because $\omega_{21}$ can be of the same order of magnitude as $\delta_1$ and $\alpha$.

Hamiltonian $\tilde H_c(t)$ is periodic in time with periodicity $\tau = 2\pi/\omega_{21}$. According to the standard Floquet theory, the eigenstates of $\tilde H_c(t)$ are given by Floquet states~\cite{Shirley1965,zeldovich,ritus1967,sambe1973}
\begin{align}
\label{eq:Floquet_states}
\Psi_m(t) = e^{-i\epsilon_m t/\hbar}u_m(t), u_m(t+\tau) = u_m(t).
\end{align}
where $\epsilon_m$ is called the quasienergy. $u_m(t)$ is a periodic function of time with the same period as the Hamiltonian $\tilde H_c(t)$ and satisfies the Schr\"odinger equation,
\begin{align}
\label{eq:Schrodinger_Floquet}
(\tilde H_c(t) - i\hbar^{-1}\partial_t) u_m(t) = \epsilon_m u_m(t).
\end{align}
By writing the function $u_m(t)$ in terms of its Fourier components, one can map the time-dependent Schr\"odinger equation in Eq.~(\ref{eq:Schrodinger_Floquet}) to a time-independent tight-binding Hamiltonian~\cite{Shirley1965}; see Appendix~\ref{sec:tight-binding}. This mapping is particularly useful in the absence of ancilla decoherence in which case one can calculate the ancilla susceptibilities from the tight-binding Hamiltonian based on simple time-independent perturbation theory. 

The Floquet states are analogous to Bloch states in crystals. Importantly, for a driven system with Hilbert space of dimension $N$, there are $N$ independent Floquet states $\Psi_m(t)$, just like there are $N$ independent stationary states in the absence of driving. Analogous to the crystal momentum, the quasienergy $\epsilon_m$ is defined modulo $\hbar\omega_{21}$ in the ``reduced Brillouin zone" scheme. This definition introduces a discontinuity in the quasienergies as they cross the Brillouin zone boundary. 

For the purpose of analytical analysis, we will instead use the ``extended Brillouin zone" scheme in which $\epsilon_m$ ranges from $-\infty$ to $\infty$. This scheme is particularly useful when the width of Brillouin zone $\hbar\omega_{21}$ is small compared to other characteristic energy scales of system such as $\hbar\delta_1$ and $\hbar\alpha$. In this scheme, for each state $u_m$ with quasienergy $\epsilon_m$ that satisfies the Schr\"odinger equation (\ref{eq:Schrodinger_Floquet}), there is a set of states $u_m' = u_m \exp(iK\omega_{21}t)$ for any integer $K$ with quasienergy $\epsilon_m' = \epsilon_m + K\hbar\omega_{21}$ that also satisfies Eq.~(\ref{eq:Schrodinger_Floquet}). One can show that $u_m'$ and $u_m$ correspond to the same Floquet state $\Psi_m = \exp(-i\epsilon_mt/\hbar)u_m$ and thus are physically equivalent~\cite{sambe1973}. In the analysis, we are free to choose any set of states $|u_m(t)\rangle$ and associated quasienergies $\epsilon_m$ as long as they yield a set of independent Floquet states $\Psi_m$; the value of any physical quantity will be independent of the choice. 

Numerically, Floquet states and quasienergies can be found by diagonalizing the unitary operator 
\begin{align}
\label{eq:Uc}
U_c(0,\tau) = \hat T \exp[-i\int_0^\tau dt \tilde H_c(t)].
\end{align} 
Eigenvalues $z_m$ of $U_c(0,\tau)$ are related to the quasienergies $\epsilon_m$ through the relation: $\epsilon_m = i\hbar\omega_{21} (\ln z_m)/2\pi$. The corresponding Floquet states $\Psi_m$ can be found from the eigenstates $\phi_m$ of $U_c(0,\tau)$ through the relation: $\Psi_m(t) = U_c(0,t)\phi_m.$ We use the numerical software Quantum Toolbox in Python (QuTiP) \cite{johansson2012} to find Floquet states and quasienergies of the Hamiltonian $\tilde H_c(t)$ to implement the above procedure.

\subsection{The AC Stark shift}
\label{sec:Stark_shift}
Periodic drives shift the energy levels of a quantum system, an effect known as the "AC Stark shift." For a driven weakly nonlinear oscillator (the ancilla), the AC Stark shift has two major contributions. Firstly, periodic drives ``dispersively" shift energy levels by non-resonantly coupling neighboring levels. In a quantum language, this process does not involve absorption or emission of drive photons by the oscillator. Sometimes this effect by itself is called AC Stark shift. Secondly, periodic drives can induce resonant transitions between the oscillator levels when the drive frequencies or integer multiples of drive frequencies match the transition frequencies. In a frame that rotates with the drive where the resonating states become degenerate, the periodic drives induce a gap between them which is often termed the ``Rabi splitting."  We will discuss both of these effects in this section.

The AC Stark shift in the energy levels is manifested in the shift of quasienergies of the driven ancilla as the drive parameters are changed. In order to map from the quasienergies to the energy levels of the ancilla, we choose a set of states $u_m$ with quasienergies $\epsilon_m$ that connect to the ancilla Fock states $|m\rangle$ at zero drive amplitudes ($\Omega_1 = \Omega_2 = 0$). For this choice, one can express the quasienergies $\epsilon_m$ as
\begin{align}
\label{eq:quasienergies}
\epsilon_m = -m\hbar\omega_1 + E_m+\delta E_m(\Omega_1,\Omega_2) 
\end{align}
where $E_m$ is the $m$-th bare energy level of the ancilla $E_m/\hbar =m \omega_c - \alpha m(m-1)/2$ and $\delta E_m$ is the AC Stark shift to this level. At zero drive amplitudes, $\delta E_m(0,0) = 0$; state $|u_m\rangle$ becomes the Fock states $|m\rangle$ of the ancilla and $\epsilon_m$ becomes the energy of the ancilla in the rotating frame of drive-1. 
One can interpret Eq.~(\ref{eq:quasienergies}) as saying that the drives have shifted the bare energy levels of the undriven ancilla by $\delta E_m$.
We note that for positive detunings $\delta_{1,2}>0$, the order of quasienergy level is trivially flipped compared to that of the Fock states, that is, $\epsilon_n > \epsilon_{n+1}$. This is simply a consequence of being in the rotating frame. In writing down Eq.~(\ref{eq:quasienergies}), we are using the extended Brillouin zone scheme where $\epsilon_m$ ranges from $-\infty$ to $\infty$. 

Throughout this paper, we will use a shorthand notation to denote the quasienergy difference of the driven ancilla and the energy differences (transition frequencies) of the undriven ancilla:
\begin{align*}
\epsilon_{mn} \equiv \epsilon_m - \epsilon_n,\, E_{mn} \equiv E_m - E_n.
\end{align*}
The ancilla frequency $\omega_c$ in Eq.~(\ref{eq:rotated_Hc}) is equivalent to $E_{10}/\hbar$.

\subsubsection{Multiphoton resonance}
\label{sec:multi-photon_resonance}
For the considered case of $\delta_{1,2}>0$, each drive is off-resonant with all the transition frequencies of the ancilla between neighboring levels. The situation is more complicated when both drives are present. Being off-resonant individually, the two drives can however ``cooperatively" resonate with one of the ancilla transitions. For the case where $\omega_c<\omega_1<\omega_2$, one can have a process where the ancilla resonantly gets excited from the n-th to the m-th level by absorbing $m-n+K$ drive-1 photons and emitting $K$ drive-2 photons. The resonance condition for this process is $E_{mn}/\hbar + K\omega_2 = (K+m-n)\omega_1$. 
In terms of quasienergies in Eq.~(\ref{eq:quasienergies}), the resonance condition becomes 
\begin{align}
\label{eq:resonance_condition}
\epsilon_{mn}/\hbar = -K\omega_{21}, 
\end{align}
meaning that a resonance occurs when there are two levels whose quasienergies differ by integer multiples of $\omega_{21}$. Note that the above resonance condition also takes into account the drive-induced dispersive shift of the ancilla energy levels. 

We emphasize that the resonance process discussed above conserves the total excitation number and thus is allowed within the RWA. This is in contrast to the multi-photon resonance that occurs in atomic gas experiments which often requires very intense laser light and going beyond the RWA; cf.~\cite{story1994}. This is also different from a recent study of Floquet resonances of a two-level system modulated at a frequency much lower than the instantaneous transition frequency~\cite{russomanno2017}.

\begin{figure}[ht]
\centering
\includegraphics[width=9.cm]{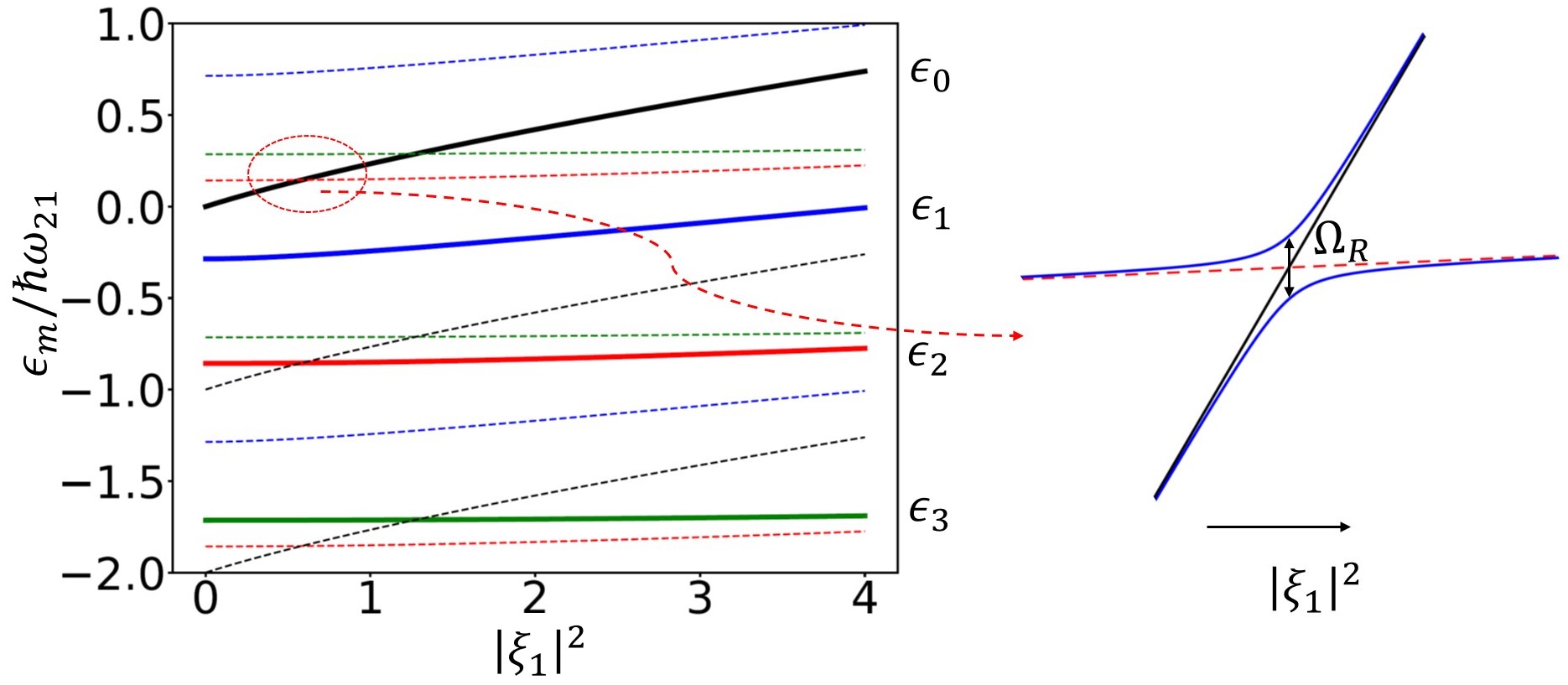} 
\caption{Illustration of quasienergy level anti-crossings. Shown on the left is an example of quasienergy spectrum for the case of one blue-detuned drive (drive-1) as a function of the scaled drive power $|\xi_1|^2$. The detuning of the drive is fixed at $\delta_1 = \alpha.$ From top to down, the thick solid lines refer to quasienergy levels $\epsilon_m$ in Eq.~(\ref{eq:quasienergies}) with $m =$ 0 (black), 1(blue), 2 (red), 3 (green). Turning on the second drive (drive-2) leads to anti-crossings of quasienergy levels when projected into the same Brillouin zone; equivalently, there occurs anti-crossing between replicas of quasienergy levels $\epsilon_m$ which are shifted from $\epsilon_m$ by integer multiples of $\hbar\omega_{21}$ as shown by the thin dashed lines. As an illustration, we choose the detuning of drive-2 to be $\delta_2 = 4.5\alpha$. On the right is a schematic for the anti-crossing between quasienergy levels at $\epsilon_0$ and $\epsilon_2 + \hbar\omega_{21}$. The gap $\Omega_R$ of the anti-crossing for a weak drive-2 is given by Eq.~(\ref{eq:Rabi_splitting}).}
\label{fig:level_anti-crossing}
\end{figure}

If the drive parameters (detuning and amplitude) are such that the oscillator is close to the above resonance condition, further tuning the drive parameters results in an anti-crossing of the quasienergy levels, when projected into the same Brillouin zone. Another way to think about it is that, in the extended Brillouin zone scheme, there are actually infinitely many replicas of each quasienergy level $\epsilon_m$ separated by a distance $\hbar\omega_{21}$ as illustrated in Fig.~\ref{fig:level_anti-crossing}. Even though there is no direct anti-crossing between $\epsilon_m$ and $\epsilon_n$, there can be anti-crossing between $\epsilon_m$ and one of the replicas of $\epsilon_n$ at $\epsilon_n - K\hbar\omega_{21}$. 

The gap of the anti-crossing between the two quasienergy levels determines the frequency of Rabi oscillation in the two-level manifold if the oscillator is initially in a superposition of them. Near the anti-crossing, one can describe the two-level manifold by a Hamiltonian 
\begin{equation}
\label{eq:LZ}
H_R = \frac{\hbar}{2} \left (
\begin{matrix}
\Delta & \Omega_R\\
\Omega_R^* & -\Delta 
\end{matrix} \right ),
\end{equation}
where $\Delta$ is the detuning between two levels (in the absence of Rabi splitting), and $\Omega_R$ is the Rabi splitting. Importantly, both the detuning $\Delta$ and the Rabi splitting $\Omega_R$ depend on the drive strengths. The detuning $\Delta$ only depends on the drive powers through the drive-induced dispersive energy shift whereas the Rabi splitting $\Omega_R$ depends on the drive amplitudes and therefore carries the phases of the drives.

For a relatively weak drive-2 (which is farther detuned from $\omega_c$) but arbitrary drive-1, the Rabi splitting can be calculated based on the degenerate perturbation theory in the eigenbasis of the Hamiltonian $\tilde H_c$ at $\Omega_2 = 0$. For a pair of quasienergy levels whose quasienergy difference $\epsilon_{mn}/\hbar \approx -K\omega_{21}$, we find the corresponding Rabi splitting to be \cite{larsen1976}
\begin{align}
\label{eq:Rabi_splitting}
  \Omega_R = 2 \sum_{n_1 n_2...n_{K-1}}\frac{\prod_{j=1}^K \Omega_2^*\langle \phi_{n_j}| c |\phi_{n_{j-1}}\rangle }{\prod_{j=1}^{K-1} (\epsilon_{n_0}- \epsilon_{n_j} - j\omega_{21})}
\end{align}
where $n_0 = n, n_K = m$, and states $\phi_n$ and quasienergies $\epsilon_n$ are given by the stationary eigenstates and eigenenergies of the Hamiltonian $\tilde H_c(t)$ at $\Omega_2 = 0$, respectively. 

Equation~(\ref{eq:Rabi_splitting}) shows that the strength of the Rabi splitting for the case $K\geq 2$ is suppressed when the frequency difference $\omega_{21}$ of the two drives is large. This is because for large $\omega_{21}$, satisfying the resonance condition ~(\ref{eq:resonance_condition}) requires states that are far from each other ($m-n$ is large)  and therefore involves a relatively large number of drive photons. For not extremely strong drives, the Rabi splitting is typically weak. When both drives are weak, one can show that the Rabi splitting is proportional to the drive amplitudes raised to a power given by the number of drive photons involved in the resonance process: $\Omega_R \propto \Omega_1^{m-n-K}(\Omega_2^*)^K$. Importantly, $\Omega_R = 0$ if $\alpha = 0$ because the oscillator is linear.

Because of the dispersive shift of the quasienergy levels as the drives are turned on, the oscillator initially in the ground state inevitably goes through several of these level anti-crossings, as shown also in Fig.~\ref{fig:level_anti-crossing}. Near each anti-crossing, the corresponding Hamiltonian Eq.~(\ref{eq:LZ}) can be approximated as in the Landau-Zener problem: $\Delta$ is approximated as a function linear in time \[\Delta(t)\approx  st, s>0,\] where $t=0$ is the time when the anti-crossing occurs; $\Omega_R$ is approximated to be  a constant. We note that this approximation relies on that the Rabi splitting being sufficiently small so that the region of anti-crossing is narrow and one can neglect the time-dependence in $\Omega_R$. This approximation typically applies when one or both of the drives are relatively weak. Where the approximation applies, the probability for the oscillator to make a diabatic transition is given by the Landau-Zener formula~\cite{zener1932a},
\begin{align}
P_{\rm diab} = \exp(-\pi |\Omega_R|^2/2s).
\end{align}
As we will show, for a broad range of drive parameters used in the experiment, the oscillator will make a diabatic transition when it goes through an anti-crossing, except for some special situations; see below.

We now discuss the possible situations where the approximation that leads to the Landau-Zener formula breaks down. The first one is that the drive frequencies are such that the oscillator is very close to some lower-order multi-photon resonance before the drives are turned on. Then as the drives are turned on, it is possible that the Rabi splitting $\Omega_R$ changes faster in time than the detuning between the two resonating levels. In this case, the standard Landau Zener analysis does not apply.  A situation of this sort was studied for a parametrically driven oscillator in Ref.~\cite{zhang2017c}. Another possibility is that the oscillator comes close to a level anti-crossing near the peak of the drive pulse where the drive amplitude changes much slower in time than at the pulse edge. In this case, the transition region may not be narrow in time and the full time-dependence in $\Delta$ and $\Omega_R$ needs to be taken into account~\cite{rubbmark1981}. 


Finally, we comment that for the case where one drive is red-detuned from $\omega_c$, because of the non-equidistance of the oscillator levels, this drive can become resonant with one or several of the oscillator transitions depending on the ratio of detuning and anharmonicity~\cite{dykman2005}. As the drive strength increases, there can occur systematic level crossings between the oscillator quasienergy levels even when there is only one drive. We will not discuss this situation. 

\subsubsection{Dispersive AC Stark shift}
In this section, we will discuss the drive-induced dispersive shift of the oscillator levels. In view of the typical parameters used in the experiment (see below), we will focus on the regime where the drive farther-detuned from the ancilla (drive-2) is relatively weak compared to the frequency difference of two drives ($|\Omega_2| \lesssim \omega_{21}$) so that the drive-induced Rabi splitting (Eq.~\ref{eq:Rabi_splitting}) is much weaker than the dispersive AC Stark shift. 

To conveniently present the dispersive AC Stark shift in the considered parameter regime and compare with experiments, we define the Floquet state $\Psi_m$ with quasienergy $\epsilon_m$ in Eq.~(\ref{eq:quasienergies}) in the following dynamical way: away from any level anti-crossings, $\Psi_m$ is the \textit{adiabatic} Floquet state of the oscillator that smoothly connects to the Fock state $|m\rangle$ at zero drive amplitudes and $\delta E_m(\Omega_1,\Omega_2)$ refers to the energy (or quasienergy) shift of this state with respect to the zero drive amplitudes limit; across the level anti-crossing, we consider $\Psi_m$ to be the \textit{diabatic} state given that the avoided crossing is rather weak. We note that this definition inevitably introduces a discontinuity in $\delta E_{m}$ across the level anti-crossing; the size of discontinuity depends on the size of the gap at the anti-crossing. However, this definition ensures that across weak avoided crossings, the wavefunction of the state $\Psi_m$ does not change dramatically. In the rest of the paper, we will simply refer to the the state $\Psi_m$ defined this way as the state that adiabatically connects to the vacuum state $|0\rangle$ as the drives are turned on. This definition of the state $\Psi_m$ is also illustrated in Fig.~\ref{fig:level_anti-crossing}. A similar construction of adiabatic Floquet states is studied in Ref.~\cite{weinberg2017}. 

In order to observe the dispersive AC Stark shift $\delta E$ as defined above, it is important to carefully choose the rate of turning on the drives. Generally speaking, the rate of ramping up the drive amplitudes needs to be smaller than the typical quasienergy spacings which is set by the drive detunings $\delta_{1,2}$ and ancilla anharmonicity $\alpha$; at the same time, the rate of the ramps needs to be larger than the typical gap $\Omega_R$ of the anti-crossings. For the considered parameter regime where the gaps are small, the allowed range for the rate of the ramps can be quite broad. 

The shifts $\delta E_m$ in the energy levels leads to shifts in the transition frequencies of the ancilla. In the limit of weak drives, the Stark shift of transition frequencies between neighboring levels of ancilla can be found by solving Eq.~(\ref{eq:Schrodinger_Floquet}) in the Fourier domain perturbatively in the drive amplitudes. To second order in the drive amplitudes, we find that
\begin{align}
\label{eq:Stark_shift_weakdrive}
&\delta E_{n(n-1)}/\hbar \approx -2\alpha \sum_{j=1,2}|\Omega_j|^2 \nonumber\\
&\times \frac{\delta_j-\alpha}{(\delta_j+n\alpha)[\delta_j+(n-1)\alpha][\delta_j+(n-2)\alpha]}
\end{align}
where $\delta E_{n(n-1)}\equiv \delta E_n - \delta E_{n-1}.$ Importantly, the shift in the transition frequency vanishes if the ancilla is linear ($\alpha = 0$). The expression above holds for any $n\geq 1$.

For positive drive detunings ($\delta_{1,2}>0$), the magnitude of the shift $\delta E_{n(n-1)}$ decreases as $n$ increases. For weak anharmonicity, $\delta E_{n(n-1)}$ for different $n$ become close to each other and the expression reduces to that obtained in the four-wave mixing picture; see Eq.~(\ref{eq:Stark_shift_small_alpha}) in Appendix~\ref{sec:four_wave_mixing}. Interestingly, when the detuning and anharmonicity are of the same size, the shift $\delta E_{n(n-1)}$ for $n>2$ may have opposite sign from $\delta E_{10}$ depending on the magnitude of $\delta_{1,2} $ and $\alpha$. 

The situation is more complicated for the case of a negative driving detuning ($\delta_1$ or $\delta_2 < 0$). The AC Stark shift of certain transition frequencies can be greatly enhanced when $-\delta_{1,2}$ is close to integer multiples of ancilla anharmonicity $\alpha$ as can be seen from Eq.~(\ref{eq:Stark_shift_weakdrive}). Such an enhancement of AC Stark shift is a sign of the drive being resonant with one of the ancilla transition frequencies between neighboring levels. In the following, we will focus on the simpler case of positive detunings.

\begin{figure}[ht]
\centering
\includegraphics[width=6 cm]{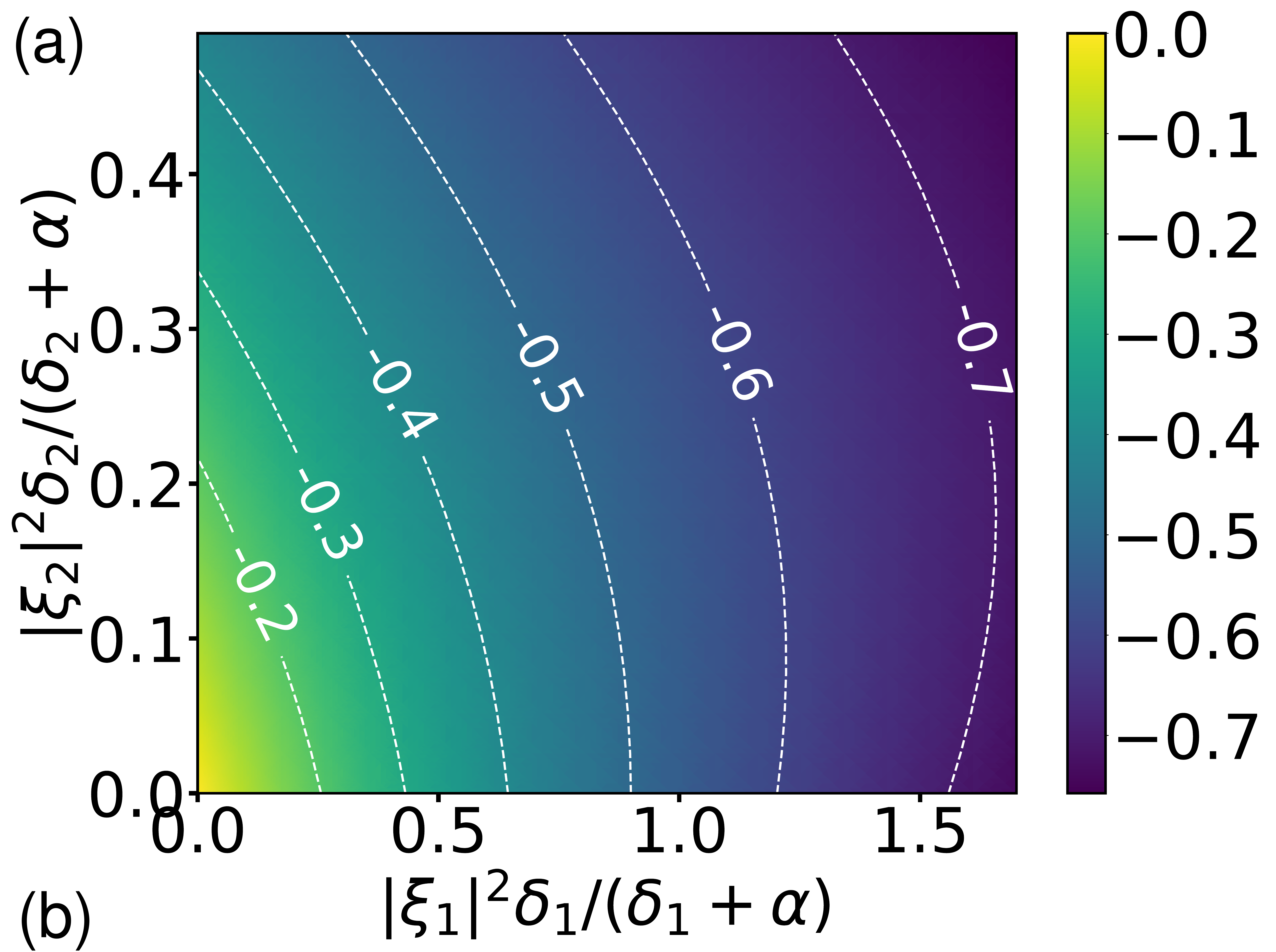} \\
\includegraphics[width=5.5 cm]{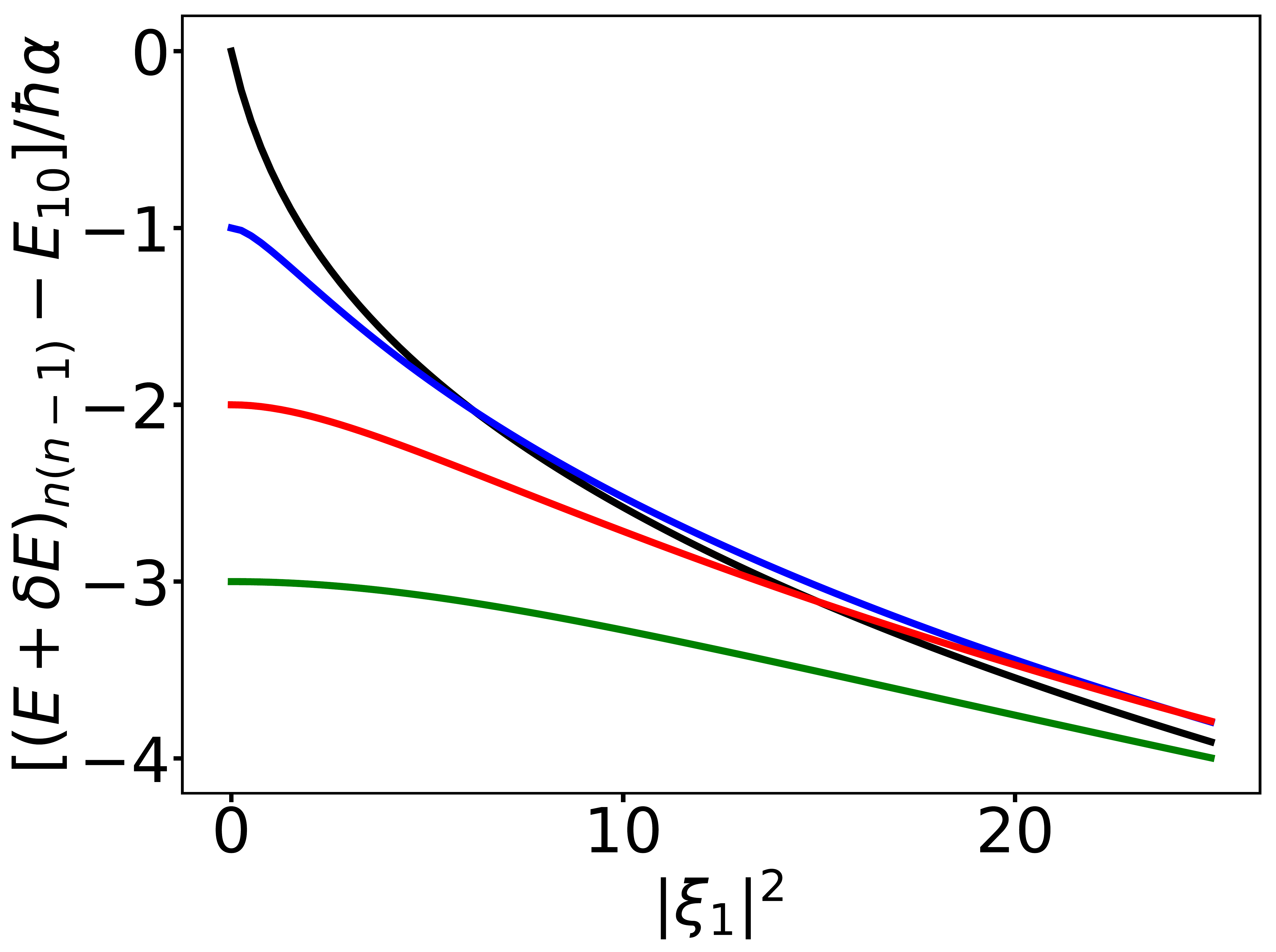} 
\caption{(a) the scaled AC Stark shift $\delta E_{10}/2\hbar\alpha$ of the transition frequency from the first excited to the ground state of the ancilla as a function of scaled power of the two drives. The solid lines are contours of constant $\delta E_{10}$. For weak drives, the contours are straight lines well described by Eq.~(\ref{eq:Stark_shift_weakdrive}). For strong drives, the contours become curved and the Stark shift becomes nonlinear in the drive power. For the purpose of comparing with the perturbative result of the Stark shift at weak drives $\delta E_{10}/\hbar\approx -2\alpha \sum_{j=1,2}|\xi_j|^2 \delta_j /(\delta_j+\alpha)$, we have chosen a proper scaling for the $x$ and $y$ axes so that for small values of $x$ and $y$, the scaled Stark shift is simply equal to $-(x+y)$. The detunings of the two drives are $\delta_1 = \alpha,\delta_2 = 4.5\alpha$. (b) the AC-Stark-shifted transition frequency $(E+\delta E)_{n(n-1)}/\hbar$ of the ancilla in the presence of one drive (drive-1). The transition frequencies are counted from the bare ancilla frequency $E_{10}/\hbar \equiv \omega_c$ and scaled by $\alpha$. The black, blue, red and green lines (from top to down at low drive power) refer to $n =  1,2,3,4$, respectively. The detuning of the drive $\delta_1 = \alpha$. In panel (a), due to level anti-crossings, the quantity $\delta E_{10}$ is generally discontinuous at particular values of drive amplitudes. We make the contour plot by choosing a discrete set of drive amplitudes and interpolate among them; the discontinuities appear to be smeared out by such interpolation.} 
\label{fig:Stark_shift}
\end{figure}

For stronger drives, the AC Stark shifts of the transition frequencies become nonlinear in the drive powers as shown in Fig.~\ref{fig:Stark_shift}. This nonlinear dependence can be understood as the drive-induced shift in the transition frequencies modifying the drive detunings which in turn modify the effective strength of drives on the ancilla. Therefore, roughly speaking, the AC Stark shift becomes nonlinear in the drive powers when the drive-induced frequency shift becomes comparable to the drive detuning from the frequency $\omega_c$ of the undriven ancilla.  

Because of its nonlinear dependence on the drive powers, the Stark shift when both drives are present is not a simple sum of Stark shifts due to each individual drive at the same amplitude. A somewhat striking effect is that when one drive is relatively strong, the AC Stark shift $\delta E_{10}$ due to a second drive can become a non-monotonic function of its drive power as can be seen in Fig.~\ref{fig:Stark_shift}(a). We attribute such a behavior to the modification of the ancilla anharmonicity due to the first strong drive; see below. 

Another interesting effect that occurs at relatively strong drive is that, due to the differences in the AC Stark shifts $\delta E_{n(n-1)}$ for different $n$, the effective anharmonicity of ancilla (the non-equidistance of levels) can be modified. This occurs even when there is only one drive. As shown in Fig.~\ref{fig:Stark_shift}(b), for the case of positive detuning ($\delta_1>0$),  as the power of drive-1 increases, the transition frequencies between lower levels can even become smaller than those between higher levels. Eventually at stronger drive, one can show that the ancilla levels become close to being equidistant but with a negative anharmonicity  (compared to the sign of $\alpha$); see Appendix~\ref{sec:auxiliary_nonlinearity}.

\begin{figure}[ht]
\centering
\includegraphics[width =6cm]{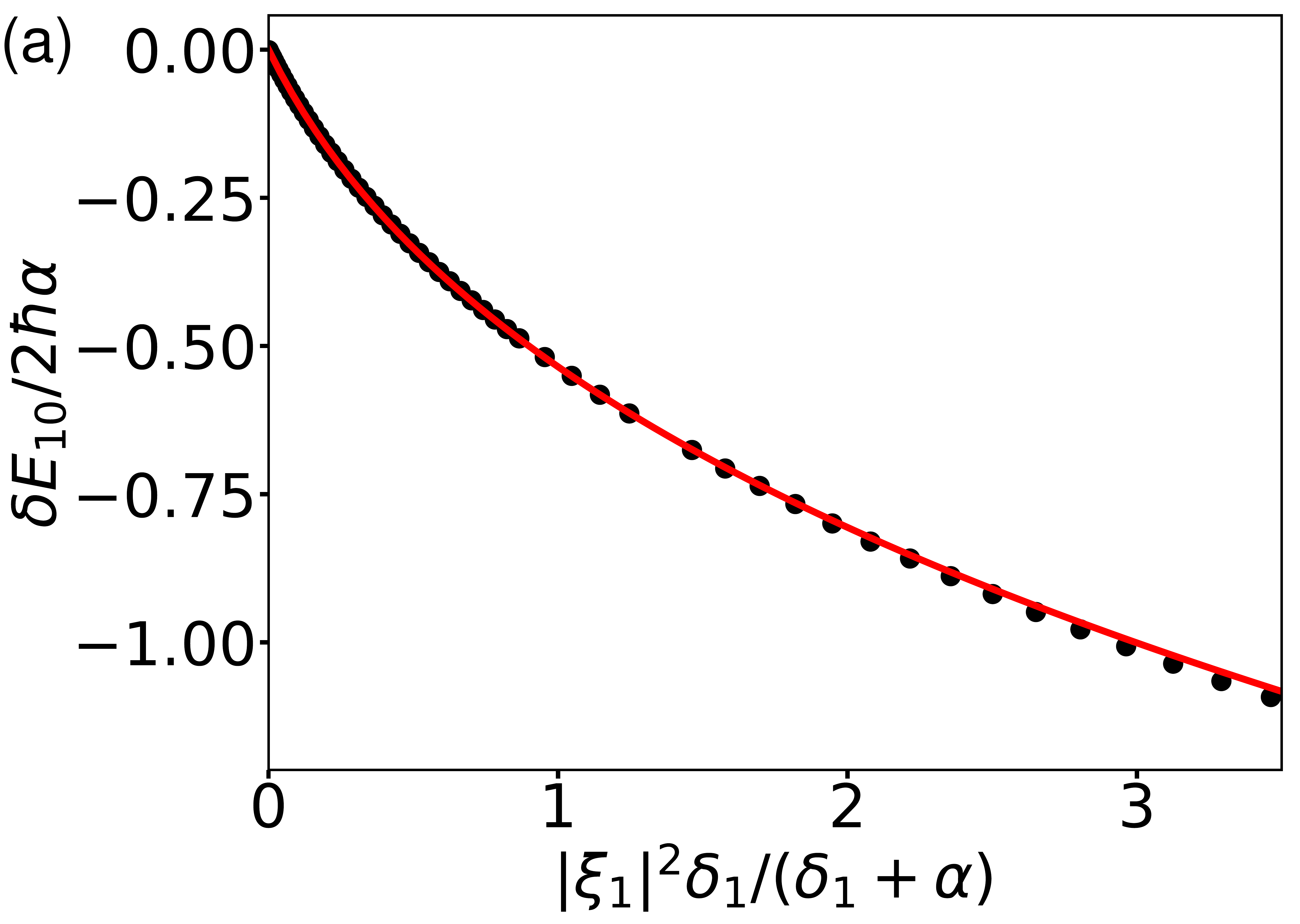}  \\
\includegraphics[width = 5.8cm]{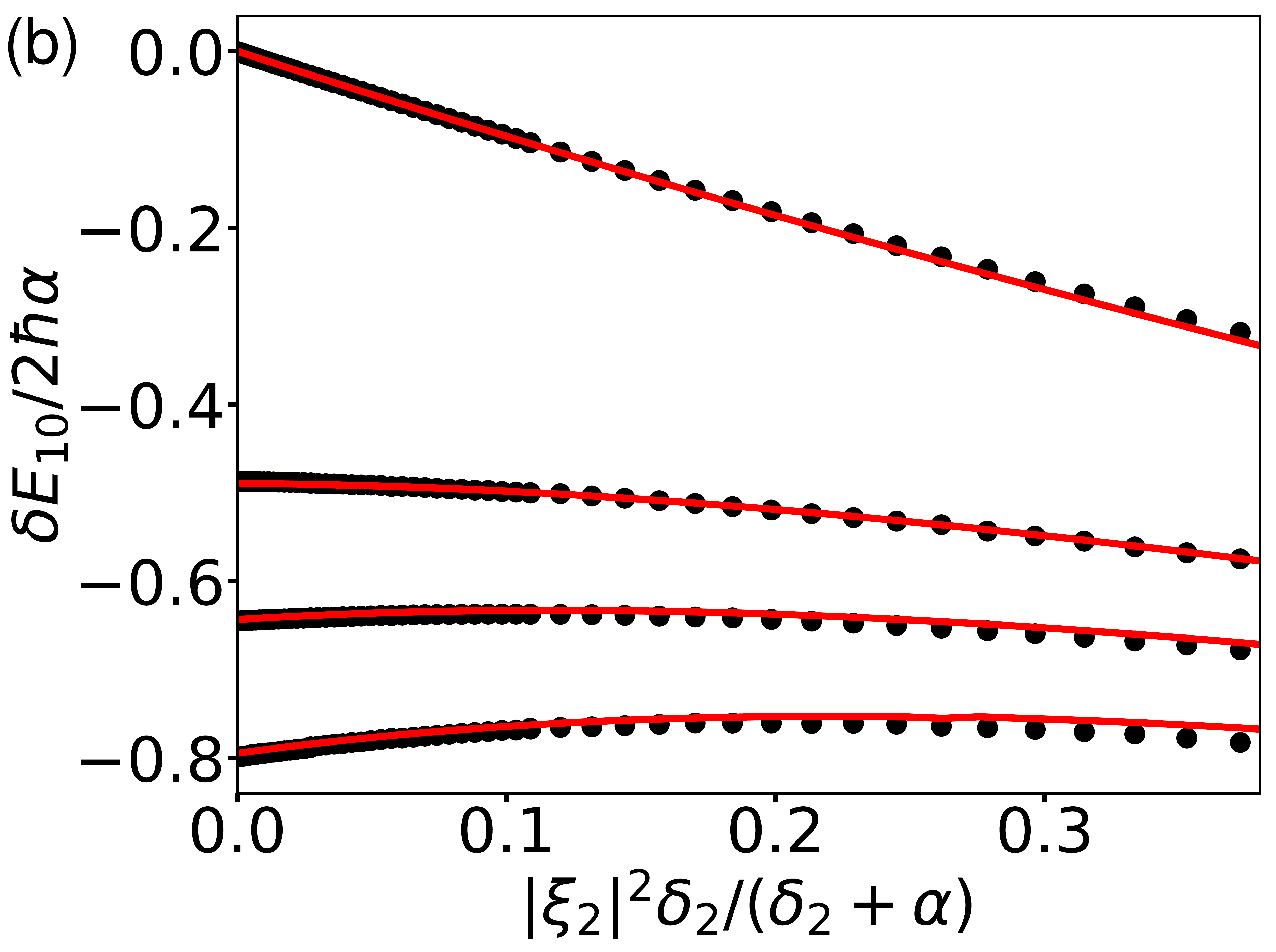} 
\caption{Comparison of the scaled AC Stark shift $\delta E_{10}/2\hbar\alpha$ between the theory (solid lines) and the experiment (black dots). (a) AC Stark shift $\delta E_{10}$ as a function of the scaled drive-1 power while drive-2 is turned off. (b) AC Stark shift $\delta E_{10}$ as a function of the scaled drive-2 power for various drive-1 strengths. From top to bottom, the scaled drive-1 power $|\xi_1|^2 \delta_1/(\delta_1+\alpha) = 0,0.87,1.35,1.95.$ For a relatively large drive-1 power, the AC Stark shift due to drive-2 becomes non-monotonic in its power. The detunings of the two drives are the same as in Fig.~\ref{fig:Stark_shift}: $\delta_1 = \alpha,\delta_2 = 4.5\alpha$. The theoretical curve in the top panel and the top curve in the bottom panel are used to calibrate the amplitudes of the two drives, namely, to find out the conversion factor between the drive amplitudes $\Omega_{1,2}$ in the Hamiltonian and the readout of the pulse generator in the experiment.}
\label{fig:Stark_shift_exp}
\end{figure}

\subsubsection{Comparison with experiment}
\label{sec:Stark_shift_exp}
In this section, we present the comparison between theory and experiment for the dispersive AC Stark shift. In the experiment, the ancilla is a Y-shaped transmon superconducting qubit which is coupled to two microwave cavities. The transmon has an anharmonicity \[\alpha/2\pi = 71.68~\rm{MHz}\]  and frequency \[\omega_c/2\pi = 5.963~\rm{GHz}.\] More details of the experiment setup can be found in Ref.~\cite{gao2018a}. 

The procedure to measure the AC Stark shift of the driven ancilla is as follows. The ancilla is first initialized in the vacuum state $|0\rangle$. Then the RF drives on the ancilla are turned on with a cosine-shaped envelope. The time for the drive amplitudes to reach the peak value from zero is kept a constant (200 ns). During the time the drives are present, we perform spectroscopy on the ancilla by sending in a $\pi$ pulse whose length is close to the duration of drives on the ancilla $\sim 1 \rm{ \mu s} $. We sweep the frequency of the spectroscopy tone and when it matches the Stark-shifted ancilla transition frequency $E_{10}/\hbar$, the ancilla will be excited from the ground state.  We then measure the transmon population in the ground state using a dispersive readout after we have turned off the drives with a symmetric ramp down. This allows us to locate the transition frequency $E_{10}/\hbar$ of the ancilla in the presence of the RF drives. 

The two RF drive amplitudes are calibrated independently by fitting the measured transition frequency $E_{10}/\hbar$ as a function of experimental drive powers to the result of the Floquet theory. Then using the obtained calibration, we compare the theory and experiment on the Stark shift  $\delta E_{10}/\hbar$ when both drives are present; see Fig.~\ref{fig:Stark_shift_exp}. We obtain excellent agreement between theory and experiment. 

We note that in the process of ramping up the drives, the ancilla passes through several level anti-crossings as illustrated in Fig.~\ref{fig:level_anti-crossing}. On the one hand, as indicated by the agreement between theory and experiment, for the chosen rate of ramping up and down the drives and a broad range of drive parameters, the ancilla indeed remains in the state $|\Psi_0\rangle$  that adiabatically connects to the vacuum state $|0\rangle$ while away from level-anti-crossing and makes a diabatic transition while passing through the level anti-crossing. The same process occurs during the ramping down of the drives, and the ancilla returns back to the vacuum state $|0\rangle$. On the other hand, we also find that for some particular combinations of the drive amplitudes, the ancilla does not end up in the vacuum state after ramping up and down the drives. This situation occurs because the ancilla comes close to a level anti-crossing near the peak of the drive envelope where the drive amplitudes change rather slowly or stay constant, and the probability of diabatic versus adiabatic transition become comparable. We present an analysis of this situation in Appendix~\ref{sec:level_anti-crossing}.

\subsection{Linear susceptibilities of the driven ancilla in the Floquet picture}
\label{sec:susceptibilities_Floquet}
In Sec.~\ref{sec:model}, we established the general relation between the linear susceptibilities of the driven ancilla and the ancilla-induced bilinear interaction between the cavity modes. In this section, we will derive general expressions for the susceptibilities in the basis of Floquet states and discuss different asymptotic limits, in particular, how they relate to the formula we obtained based on the four-wave mixing picture (Appendix~\ref{sec:four_wave_mixing}).  We will also present a comparison between the theory and experiment on the rate of the ancilla-induced beam-splitter interaction between the two cavity modes.
\subsubsection{General expressions} 
In the absence of ancilla decoherence, we can calculate the linear susceptibilities from Eqs.~(\ref{eq:chi1_formal}, \ref{eq:X1_formal}) where the Heisenberg operators $c^{(0)}(t),c^{(0)\dagger}(t)$ evolve under the unitary operation
\begin{align*}
c^{(0)}(t) = U_c^\dagger (0,t) cU_c(0,t) e^{-i\omega_1 t},
\end{align*} 
where we have transformed into the rotating frame of drive-1 and $U_c(0,t)$ is given in Eq.~(\ref{eq:Uc}). 

Then assuming that the ancilla is initially in a given Floquet state $\Psi_m$ and after disregarding rapidly oscillating terms, we find the linear susceptibilities to be
\begin{align}
\label{eq:chi1}
&\chi _m(\omega,\omega+K\omega_{21}) \nonumber\\
&= -\sum_{n\neq m,K'}  \left[\frac{c_{mn,K'-K}(c^\dagger )_{nm,-K'}}{(\omega-\omega_1) + K'\omega_{21}+(\epsilon_{mn}/\hbar) }\right. \nonumber\\
&+\left.\frac{(c^\dagger )_{mn,-K'}c_{nm,K'-K}}{-(\omega-\omega_1+K\omega_{21})+(K-K')\omega_{21}+(\epsilon_{mn}/\hbar)} \right] 
\end{align}
\begin{align}
\label{eq:X1}
&X _m(-\omega,2\omega_1+K\omega_{21}-\omega) \nonumber\\
&= -\sum_{n\neq m,K'}\left[\frac{ c_{mn,K'-K} c_{nm,-K'}}{-(\omega-\omega_1)+K'\omega_{21}+(\epsilon_{mn}/\hbar)}\right. \nonumber\\
&\left.+\frac{c_{mn,-K'}c_{nm,K'-K} }{-[K\omega_{21}-(\omega-\omega_1)]+(K-K')\omega_{21}+(\epsilon_{mn}/\hbar) } \right]. 
\end{align}
The subscript $m$ in the susceptibilities indicates that the initial state of the ancilla at $t=0$ is $\Psi_m$.
If the ancilla is in a mixed state, then an ensemble average over them is needed. $c_{mn,K}$ is the $K$-th Fourier component of matrix element $\langle u_m(t) | c | u_n(t)\rangle$ of operator $c$ between state $u_m$ and $u_n$ :\[c_{mn,K} \equiv \frac{\omega_{21}}{2\pi}\int _0^{2\pi/\omega_{21}}\langle u_m(t)| c | u_n(t)\rangle e^{-iK\omega_{21}t}dt.\]
A useful property of $c_{mn,K}$ is that $(c^\dagger )_{mn,K} = (c_{nm,-K})^*.$

The strength of ancilla-induced bilinear interaction and linear frequency shift can be calculated from Eqs.~(\ref{eq:chi1},\ref{eq:X1}) above using the general relations Eqs.~(\ref{eq:frequency_shift},\ref{eq:g_BS},\ref{eq:g_TMS}) found in Sec.~\ref{sec:model}. In the absence of ancilla decoherence, one can show that the susceptibilities have a symmetry: 
\begin{align*}
\chi _m(\omega,\omega+K\omega_{21})&=[\chi _m(\omega+K\omega_{21},\omega)]^*, \\
X _m(-\omega,2\omega_1+K\omega_{21}-\omega)&=X _m(-2\omega_1-K\omega_{21}+\omega,\omega). 
\end{align*}
It follows from these symmetry relations that the rate of ancilla-induced beam-splitter and two-mode squeezing interaction simplifies to 
\begin{align}
\label{eq:gBS_simplified}
g_{\rm BS} = -g_a^* g_b [\chi_m(\omega_a,\omega_b)]^*
\end{align}
where $\omega_b = \omega_a + K\omega_{21}$, and
\begin{align}
g_{\rm TMS} = -g_a^* g_b^*X _m(\omega_a,\omega_b)
\end{align}
where $\omega_a+\omega_b = 2\omega_1 + K\omega_{21}$.

As noted in Sec.~\ref{sec:Floquet_formulation}, in evaluating the susceptibilities using Eqs.~(\ref{eq:chi1},\ref{eq:X1}), we are free to choose any set of states $|u_m(t)\rangle$ which yield a set of independent Floquet states $\Psi_m$. The values of susceptibilities are independent of the choice because of the summation over $K'$. For the purpose of the analytical calculation, we will choose, as in previous section, a set of $|u_m(t)\rangle$ that adiabatically connect to ancilla Fock states $|m\rangle$ in the absence of drives and their quasienergies $\epsilon_m$ are given by Eq.~(\ref{eq:quasienergies}).  For numerical analysis, it is often more convenient to choose the set of the states $u_m$ with quasienergies $\epsilon_m$ in a given Brillouin zone (i.e., the reduced Brillouin zone scheme) according to the procedure described below Eq.~(\ref{eq:Uc}).

Equations~(\ref{eq:chi1},\ref{eq:X1}) have a structure similar to standard second-order perturbation theory, the squared matrix element divided by the energy difference. Indeed, one can also derive them using time-independent perturbation theory by mapping the Hamiltonian $\tilde H_c$ to a time-independent tight-binding Hamiltonian; see Appendix~\ref{sec:tight-binding}. 

\begin{figure}[ht]
\centering
\includegraphics[width=8.5cm,height=3.cm]{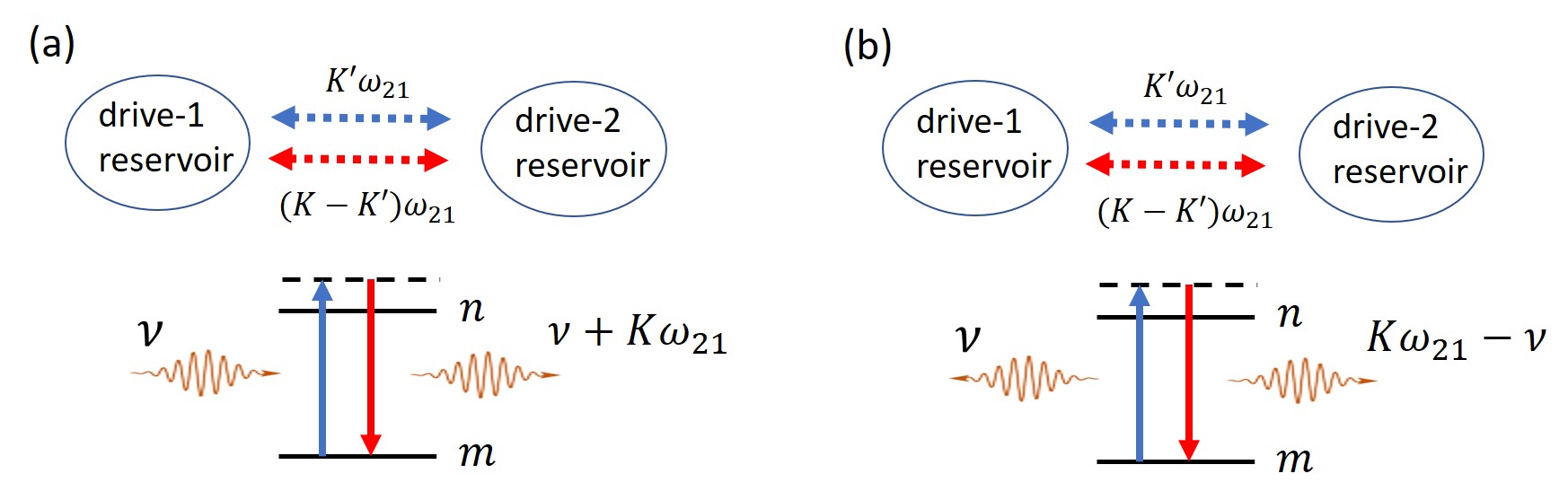} 
\caption{A schematic showing how the linear susceptibilities arise from virtual transitions between quasienergy states of driven ancillas in the rotating frame of drive-1 (a) A schematic showing the process that gives the first term in $\chi (\omega,\omega+K\omega_{21})$. $\nu$ is the frequency of probe field in the rotating frame of drive-1: $\nu = \omega-\omega_1.$ (b) A schematic showing the process that gives the first term in $X (\omega,2\omega_1+K\omega_{21}-\omega)$ }
\label{fig:schematic_transitions}
\end{figure}

By thinking of the classical drives as quantized fields, the expressions for the susceptibilities $\chi $ and $X $ can be interpreted as follows. As illustrated in Fig.~\ref{fig:schematic_transitions}(a), the first term in $\chi $ arises from a process in which the driven ancilla first makes a virtual transition from the $m$ to $n$-th quasienergy level accompanied by a virtual absorption of an incident probe photon and an exchange of $K'$ excitations between the two drive reservoirs.  Then, the driven ancilla undergoes a virtual transition back to the $m$-th quasienergy level accompanied by an emission of a probe photon at frequency $\omega+K\omega_{21}$, and an exchange of $K-K'$ excitations between the two drive reservoirs. The net result is that the incident probe photon has been up/down converted by a frequency $K\omega_{21}$ and there is an overall exchange of K excitations between the two drive reservoirs to conserve the energy.

The process illustrated in Fig.~\ref{fig:schematic_transitions}(a) for $K\neq 0$ is analogous to the Umklapp scattering process in phonon transport in which phonons with wave vectors adding up to $k$ can be scattered into phonons with wave vector adding up to $k+G$ where $G$ is a reciprocal lattice vector. The second term in $\chi $ arises from the time-reversed process of the first term.  The susceptibilities $X $ can also be understood in the same way as shown in Fig.~\ref{fig:schematic_transitions}(b). The net result is that the driven ancilla simultaneously emits two probe photons, one at frequency $\omega$ and the other at frequency $2\omega_1 + K\omega_{21}-\omega$.

\subsubsection{The limit of weak anharmonicity}
\label{sec:weak_anharmonicity}
As discussed previously, the capability of frequency conversion of the ancilla originates from its finite anharmonicity. To gain some insights on the magnitude of the linear susceptibilities, we consider in this section the limit of weak anharmonicity. In this limit, it is convenient to go to a displaced frame with a displacement given by the response of the ancilla in the absence of anharmonicity and then treat the anharmonicity as a perturbation. The displacement transformation reads
\begin{align*}
D = \exp[\xi (t) c^\dagger  - \xi^*(t)c],\, \xi(t) = \xi_1 + \xi_2 e^{-i\omega_{21}t},\\ 
\end{align*}
where $\xi_{1,2}$ is the scaled drive amplitude: $\xi_{1,2} = \Omega_{1,2}/\delta_{1,2}$.

The Hamiltonian after the transformation $H_D = D^+ \tilde H_c D - i\hbar D^+ \dot D$ reads
\begin{align}
\label{eq:H_D}
&H_D = -\delta_1 c^\dagger  c + \delta_1 |\xi(t)|^2 - \alpha \left[ \frac{1}{2} c^{\dagger 2}c^2 +  2|\xi|^2 c^\dagger c +\right.\nonumber\\
& \left.\left(\frac{1}{2} \xi^2 c^{\dagger 2} +  \xi c^{\dagger 2}c + |\xi|^2\xi c^\dagger  + \rm {h.c.}\right)\right] 
\end{align}  
In the limit $\alpha\rightarrow 0$, Hamiltonian $H_D$ is diagonalized in the Fock basis, and thus the Floquet states of $\tilde H_c$ are simply displaced Fock states $|u_m(t)\rangle = D|m\rangle$, and their quasienergies remain equidistant with a distance between neighboring levels given by $\delta_1$, regardless of the values of driving strengths. The two drives do not interfere with each other as a consequence of superposition principle that a linear oscillator obeys.

It also follows from Eq.~(\ref{eq:H_D}) that in the limit $\alpha = 0$, all matrix elements in Eq.~(\ref{eq:chi1},\ref{eq:X1}) are zero except $c_{m(m+1),0}$. Therefore, among all linear susceptibilities, the only non-zero one is $\chi _m(\omega,\omega) =  - (m+1)/(\omega-\omega_c)$. This is simply the dispersive shift to the frequency of the cavity modes due to coupling to the ancilla. 

The interplay of drives and finite anharmonicity leads to two major effects: (i) periodic modulation of the frequency of the ancilla through the term $|\xi^2(t)|c^\dagger c$ in Eq.~(\ref{eq:H_D}) (ii) squeezing of the Fock states through the term $c^{\dagger 2}\xi^2(t) + c^2\xi^{*2}(t)$. The periodic modulation in ancilla frequency leads to periodic modulation of the phase evolution of the Fock states. Neglecting other effects, the Floquet states $u_m(t) \approx \exp[2im\alpha\int^t dt' |\xi(t')|^2] |m\rangle$. Note that the time-dependence in $|\xi(t)|^2$ comes from the interference between the two drives. Such modulation leads to a finite matrix element $c_{m(m+1),K}$ for non-zero $K$, which results in a non-zero susceptibility $\chi (\omega,\omega+K\omega_{21})$. It is straightforward to show that the squeezing terms in Eq.~(\ref{eq:H_D}) lead to a finite matrix element $c_{m(m-1),K}$, which results in non-zero susceptibility $X _m(-\omega,2\omega_1+K\omega_{21}-\omega)$.
To the lowest order in the anharmonicity $\alpha$, one can show that using perturbation theory
\begin{align}
\label{eq:susceptibility_weak_alpha}
&\chi _m (\omega,\omega+K\omega_{21}) \propto |\alpha\xi_1\xi_2|^{|K|}, \nonumber \\
&X _m (-\omega,2\omega_1+K\omega_{21}-\omega) \propto |\alpha|^{\frac{|K-2|+|K|}{2}}|\xi_2|^{|K|} |\xi_1|^{|K-2|},
\end{align}
for any integer $K$. The power in the driving amplitudes of the expressions above is simply the minimum number of drive photons involved in the underlying process represented by the susceptibilities as illustrated in Fig.~\ref{fig:schematic_transitions}. 
One can show that if one goes to next-to-leading order in $\alpha$, there are terms in the susceptibilities proportional to drive amplitudes raised to higher powers than that in Eq.~(\ref{eq:susceptibility_weak_alpha}).  As a result, the perturbation theory in $\alpha$ breaks down at large drive powers. 

The terms linear and cubic in ancilla operators $c,c^\dagger $ in Eq.~(\ref{eq:H_D}) also lead to frequency modulation and squeezing of the ancilla if one goes to second order in $\alpha$, but they do not contribute to the susceptibilities to leading order in $\alpha$ as shown above. We show in Appendix~\ref{sec:one_drive_quantization} that   the terms linear in $c,c^\dagger$ can be eliminated by modifying the displacement transformation, so that $\xi$ is the full classical response of the nonlinear ancilla to the drives. This way, the non-perturbative effects of the nonlinearity can be partially captured. 

\subsubsection{Susceptibilities $\chi _0$ and $X _0$}
Of primary interest to us are the susceptibilities $\chi _0$ and $X _0$ where the ancilla is in the Floquet state $\Psi_0$. As described in Sec.~\ref{sec:Stark_shift}, state $\Psi_0$ can be prepared from the ancilla vacuum state by slowly turning on the drives (but not too slow compared to the gap of quasienergy level anti-crossing and ancilla relaxation rate; see Sec.~\ref{sec:heating}). In this section, we will study in detail the parameter dependence of the susceptibilities $\chi _0$ and $X _0$.

Explicit expressions for $\chi _0$ and $X _0$ can be obtained in the limit of weak drives by solving Eq.~(\ref{eq:Schrodinger_Floquet}) for the states $u_m$ perturbatively in the driving strengths. For the case of $K = 1$, we find that to leading order in the drive amplitudes,
\begin{align}
\label{eq:BS_weakdrive}
\chi _0 (\omega,\omega+\omega_{21}) \approx &2\alpha \frac{\xi_1^*\xi_2}{\delta(\delta+\omega_{21})} \frac{\delta+\delta_2}{\delta+\delta_2+\alpha}.
\end{align} 
\begin{align}
\label{eq:TMS_weakdrive}
X _0(-\omega,\omega_1 +\omega_2-\omega) \approx 2\alpha \frac{\xi_1\xi_2}{\delta (\delta_1+\delta_2-\delta)}\frac{\delta_1+\delta_2}{\delta_1+\delta_2+\alpha}
\end{align}
where $\delta \equiv \omega - \omega_c.$ One can show the rate of beam-splitter and two-mode squeezing interaction obtained from the above susceptibilities reduce to those obtained based on the four-wave mixing to leading order in the anharmonicity $\alpha$; see Appendix~\ref{sec:four_wave_mixing}.

Also of interest to us is the susceptibility $\chi_0(\omega,\omega)$ which relates to the ancilla-induced frequency shift of the cavity modes through Eq.~(\ref{eq:frequency_shift}). To leading order in the drive amplitudes, we find that
\begin{align}
\label{eq:chi0_weakdrive}
\chi_0(\omega,\omega) \approx - \frac{1}{\delta} + \sum_{j=1,2}\frac{2\alpha |\xi_j|^2 (\delta+\delta_j)}{\delta^2(\delta+\delta_j+\alpha)}. 
\end{align}
We note that the ancilla-induced cavity frequency shifts are generally of the same size as the ancilla-mediated interaction between the cavities. Therefore, to ensure resonant interaction between the cavities, it is important to fine-tune the drive frequencies so that the frequency matching conditions in Eqs.~(\ref{eq:frequency_matching_BS},\ref{eq:frequency_matching_TMS}) are satisfied. 

\begin{figure}[ht]
\centering
\includegraphics[width=5.5cm]{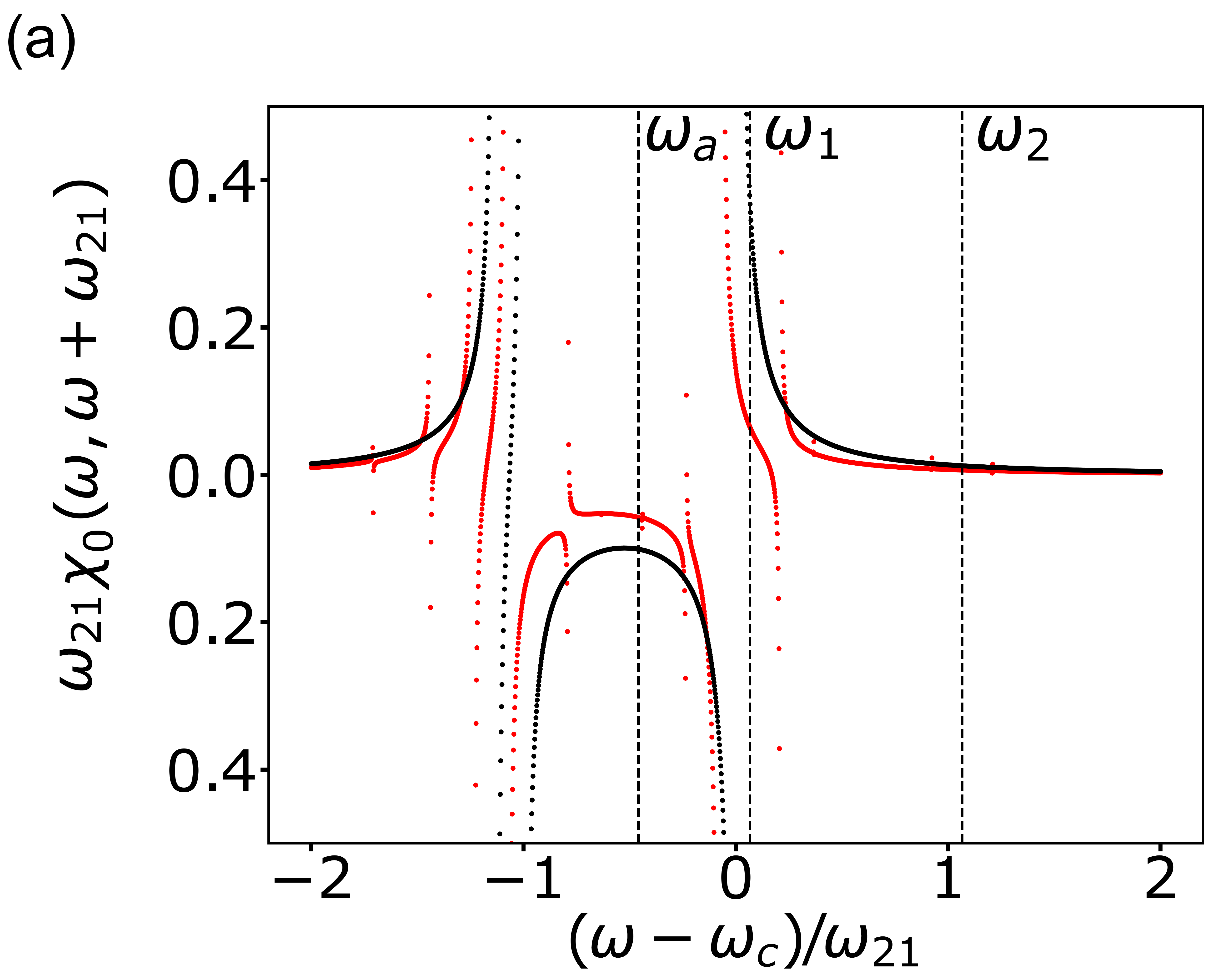} \\
\includegraphics[width=6 cm]{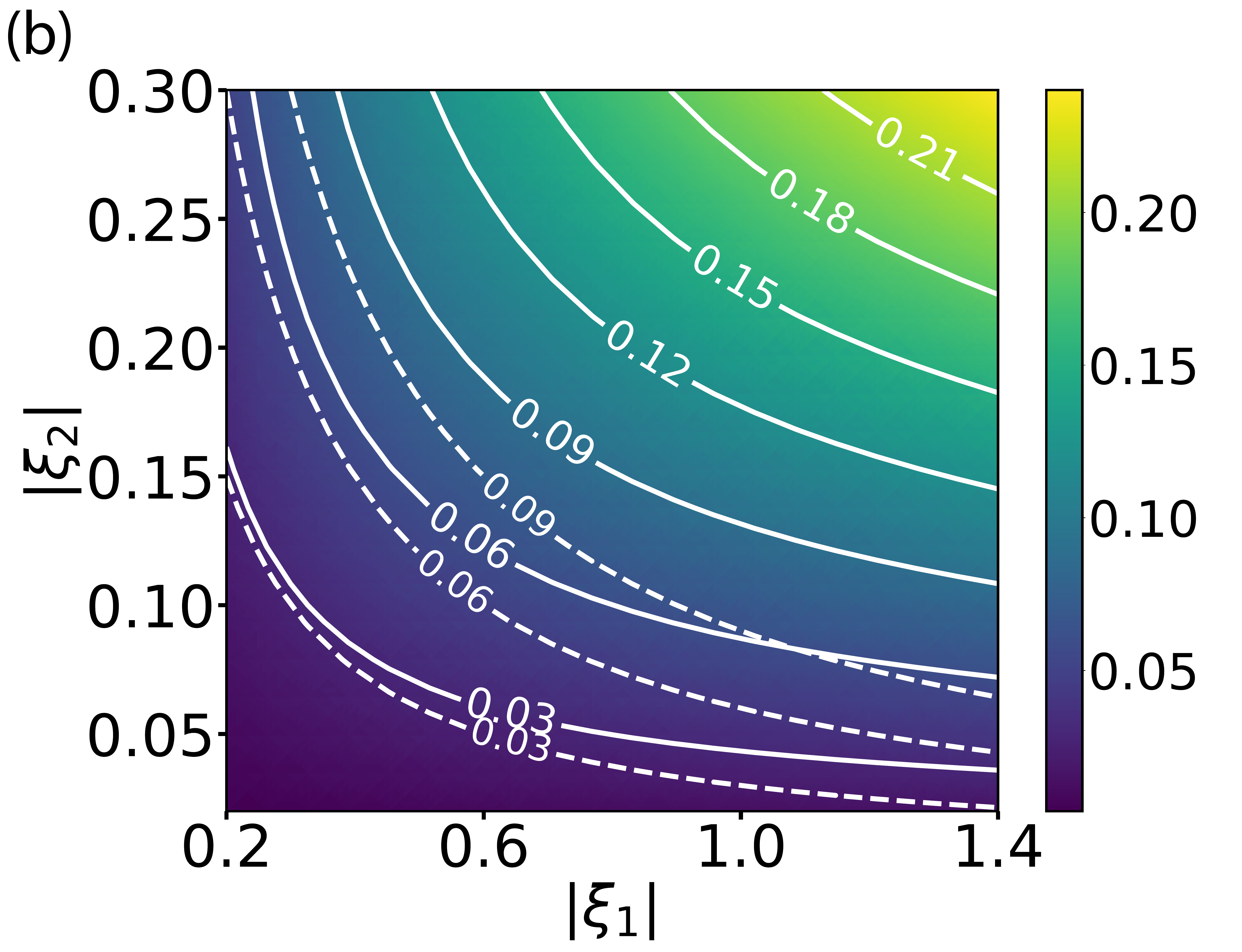} 
\caption{Comparison between perturbation theory and full Floquet theory on the susceptibility $\chi _0(\omega,\omega+\omega_{21})$ which is responsible for the ancilla-mediated beam-splitter interaction between two cavity modes with frequencies $\omega_b-\omega_a = \omega_{21}$. (a) Susceptibility $\chi _0(\omega,\omega+\omega_{21})$ as a function of probe frequency $\omega$ at fixed drive strengths: $|\xi_1| = 1.5,|\xi_2| = 0.14$. The detunings of the drives are $\delta_1/\alpha = 1,\delta_2/\alpha = 16.1$ as indicated by the vertical dashed lines. The red and black dots refer to results of the full Floquet theory and perturbation theory [Eq.~(\ref{eq:BS_weakdrive})], respectively.  We have chosen the relative phase of the two drives to be zero, so that $\chi_0(\omega,\omega+\omega_{21})$ is real in the absence of ancilla decoherence. (b) The scaled beam-splitter interaction strength $|\overline g_{BS}|$ between the two cavity modes as a function of driving amplitudes $|\xi_1|$ and $|\xi_2|$ calculated using Floquet theory. $\overline g_{BS}$ is related to the beam-splitter strength $g_{\rm BS}$ in Eq.~(\ref{eq:g_BS}) by a constant factor: $g_{\rm BS}=\zeta \overline g_{\rm BS}, \zeta = -\frac{2\alpha g_a^*g_b}{\delta_a\delta_b} \frac{\delta_a+\delta_2}{\delta_a+\delta_2+\alpha}$. The scaling factor is chosen so that at weak drives, $\overline g_{\rm BS} \approx \xi_1\xi_2^*.$ The solid lines are contours of constant $\overline g_{\rm BS}$. As a comparison, the dashed lines are contours of constant $|\xi_1\xi_2|$. The frequencies of cavity modes are detuned from the ancilla frequency by $\delta_a/\alpha = -6.9, \delta_b = \delta_a+\omega_{21}$ as also indicated as a vertical dashed line in the top panel. }
\label{fig:beam-splitter}
\end{figure}

An important feature of the spectrum $\chi_0(\omega,\omega+\omega_{21})$ and $X_0(\omega,\omega_1+\omega_2-\omega)$ is that there are multiple peaks with dispersive lineshape. We show an example of the spectrum $\chi_0(\omega,\omega+\omega_{21})$ in  Fig.~\ref{fig:beam-splitter}(a). Those peaks are related to resonant absorption or emission of the probe field; see Sec.\ref{sec:dissipation}. Such dispersive structure can already be seen from the formula ($\ref{eq:BS_weakdrive},\ref{eq:TMS_weakdrive}$). The locations of the peaks are shifted as the drive strengths increase due to the AC Stark shift of ancilla transition frequencies. We note that, in addition to capturing the effects of AC Stark shift, the Floquet calculation based on Eqs.~(\ref{eq:chi1},\ref{eq:X1}) contains more peaks than the perturbation theory Eqs.~(\ref{eq:BS_weakdrive},\ref{eq:TMS_weakdrive}) due to transitions between state $\Psi_0$ and ``far away'' states that only become strong at large drives.

At strong drive powers, the susceptibilities $\chi_0(\omega,\omega+\omega_{21})$ and $X_0(\omega,\omega_1+\omega_2-\omega)$ become nonlinear in the drive amplitudes. The nonlinear dependence on the drive amplitudes arises in two ways: first, energy denominators in Eqs.~(\ref{eq:chi1},\ref{eq:X1}) depend on the drives through the AC Stark shift in the quasienergies; second, the matrix elements generally depend nonlinearly on the drive amplitudes. The drive-dependence of the AC Stark shift has been analyzed in Sec.~\ref{sec:unitary}. In order to quantify the latter effect, we choose to probe the ancilla at a frequency [labeled as $\omega_a$ in Fig.~\ref{fig:beam-splitter}(a)] that is far from any resonance so that the nonlinear dependence of the susceptibilities on the drive amplitudes mainly comes from the matrix elements; see below.  

\begin{figure}[ht]
\centering
\includegraphics[width=6cm]{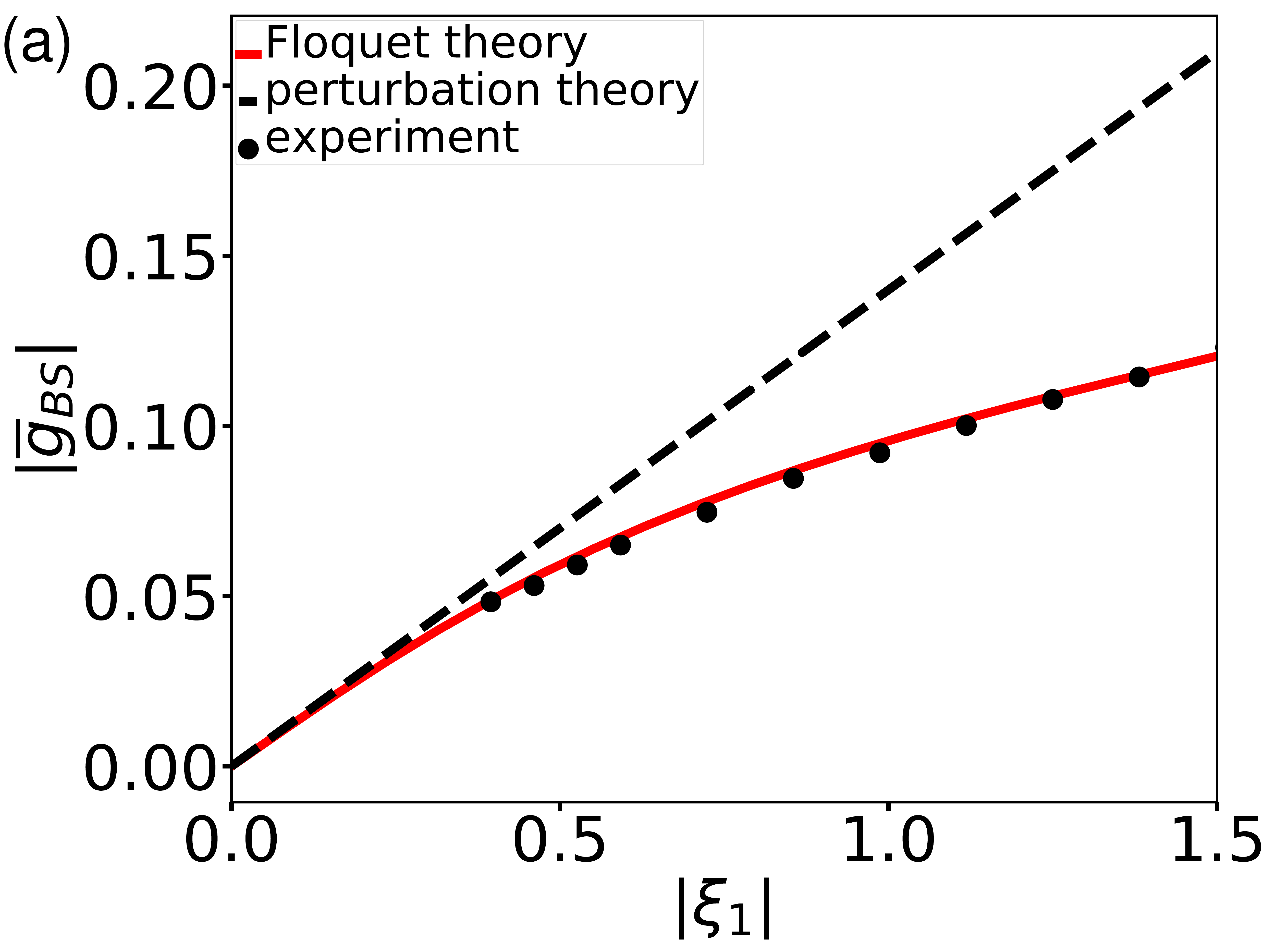} \\
\includegraphics[width =  5.8cm]{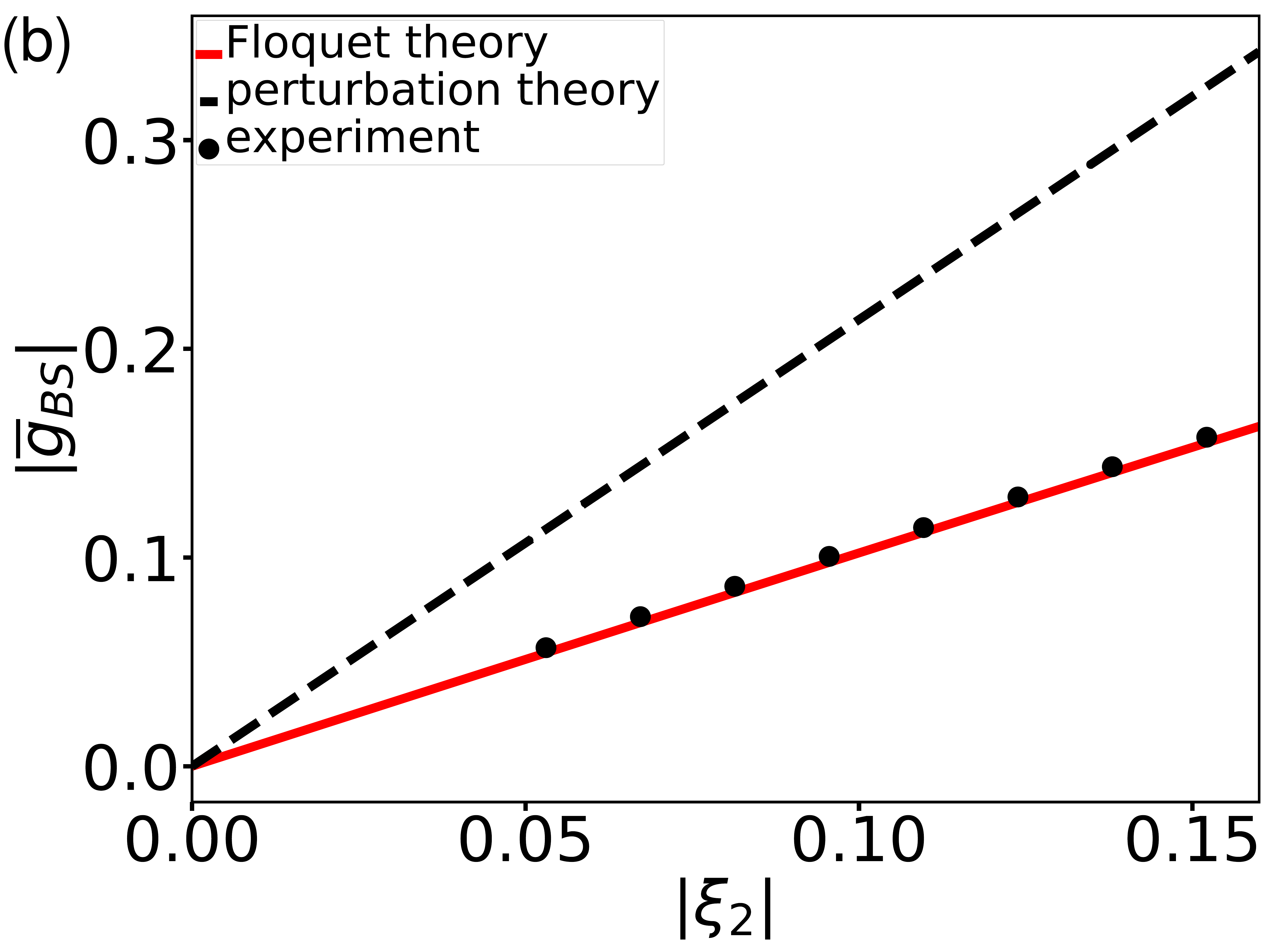} 
\caption{Comparison between the theory and experiment on the ancilla-induced beam-splitter rate $g_{\rm BS}$ between the two cavity modes whose frequencies satisfy $\omega_b-\omega_a = \omega_{21}$. The red solid and black dashed lines refer to the results of Floquet theory and perturbation theory in Eq.~(\ref{eq:BS_weakdrive}), respectively. The black dots refer to the experimental results. The detunings of the drives and cavity modes from the ancilla frequency are the same as in Fig.~\ref{fig:beam-splitter}. (a) $|\overline  g_{\rm BS}(\xi_1,\xi_2)|$ as a function of $|\xi_1|$ at fixed $|\xi_2| = 0.14.$ (b) $|\overline g_{\rm BS}(\xi_1,\xi_2)|$ as a function of $|\xi_2|$ at fixed $|\xi_1| = 2.14.$ The dots are experimental results. In the experiment, parameter $|\zeta|/2\pi = 0.33$~MHz corresponding to cavity-ancilla coupling strengths $|g_a/\delta_a|=0.047, |g_b/\delta_b|=0.054$.} 
\label{fig:beam_splitter_exp}
\end{figure}

We show in Fig.~\ref{fig:beam-splitter}(b) and \ref{fig:beam_splitter_exp} the dependence of the engineered beam-splitter rate on the drive amplitudes. As in the Stark shift analysis, we focus on the situation of two blue-detuned drives ($\delta_{1,2}>0$) where one drive is close to the ancilla frequency and relatively strong, whereas the other drive (drive-2) is far detuned and relatively weak. The deviation of the Floquet calculation from the perturbation theory in Eq.~(\ref{eq:BS_weakdrive}) is most pronounced when the near-detuned drive becomes strong. Interestingly, the beam-splitter strength becomes sublinear in the drive amplitude $\xi_1$ for large $\xi_1$. Such sublinear dependence can be well captured by replacing $\xi_1$ in Eq.~(\ref{eq:BS_weakdrive}) with the full classical reponse $\overline \xi_1$ of the ancilla to drive-1 which relates to $\xi_1$ via the relation: $\overline \xi_1 = \xi_1/(\alpha |\overline \xi_1|^2/\delta_1+1)$; see Appendix~\ref{sec:one_drive_quantization} for details. For weak drive, $\overline \xi_1 \approx \xi_1$; at strong drive, $\overline \xi_1$ becomes smaller than $\xi_1$ and scales as $\xi_1^{1/3}$ when $\alpha \overline \xi_1 /\delta_1 \gg 1$. 
We also note that although the beam-splitter strength remains linear in $\xi_2$ (see Fig.~\ref{fig:beam_splitter_exp}b), it deviates from the perturbation theory because of the non-perturbative effect of drive-1.

To confirm the theory, we performed experiments to engineer a beam-splitter interaction between two off-resonant microwave cavities based on the aforementioned cQED setup \cite{gao2018a}. By initializing one of the cavity modes in the Fock state $|1\rangle$ and then measuring the oscillations of its photon number population after the beam-splitter interaction has been turned on, we can extract the strength $g_{\rm BS}$ of beam-splitter interaction. We find excellent agreement between experiments and the theory on $g_{\rm BS}$ as a function of drive strengths; see Fig.~\ref{fig:beam_splitter_exp}.

\section{Floquet dynamics in the presence of dissipation and dephasing}
\label{sec:dissipation}
Coherent quantum operations between the cavity modes based on ancilla-mediated interactions require the ancilla to be in a pure Floquet state during the operation. However, because of the finite coherence time of the ancilla, it can undergo transitions from one Floquet state to another during this time, thereby reducing the coherence of the desired operation. In this section, we discuss the effects of ancilla dephasing and dissipation on the engineered bilinear interaction between cavity modes. Since the ancilla typically has a  much shorter coherence time than the cavity modes in a typical cQED setup, its decoherence is one of the dominant factors that limit the fidelity of the operation. 

The major effects of ancilla decoherence are two-fold. Firstly, due to the coupling between the ancilla and the cavity modes, the cavity modes inherit finite dissipation and dephasing rates from the ancilla through the ``inverse Purcell effect." This effect  becomes particularly strong when the frequency of the cavity modes is close to some resonance that excites the ancilla to higher levels with or without absorption of drive photons. 
Secondly, both dissipation and dephasing can induce transitions among the Floquet states of the driven ancilla, even when the environment that leads to ancilla dissipation and dephasing is at zero temperature. This leads to an effective ``heating" of the ancilla. We will show that the transition rates have non-trivial dependence on the drive powers and frequencies.  In the following, we first present the model we use to describe ancilla dissipation and dephasing. Then we will address the two effects separately and present a comparison between theory and experiment. Lastly, we discuss the ancilla-induced dephasing of the SWAP operation as a result of its random transitions among the Floquet states.

\subsection{The model of ancilla dissipation and dephasing}
We will assume that the ancilla is weakly coupled to a thermal bath and the coupling is linear in the dynamical variables of the ancilla. Therefore, the ancilla decays by emitting one excitation at a time to the bath. We will also consider the possibility that the ancilla is dispersively coupled to a bath that leads to dephasing. The total Hamiltonian of the ancilla plus the baths reads,
\begin{align}
\label{eq:Hamiltonian_ancilla_plus_bath}
H_{c+\rm{bath}} = \tilde H_c(t) + H_{\rm bath} + H_i,   \nonumber\\
H_i = (ce^{-i\omega_1 t} + c^\dagger e^{i\omega_1t}) h_1 + c^\dagger c h_2,
\end{align}
where we have made a unitary transformation $U=\exp(-ic^\dagger c\omega_1t)$ to go to the rotating frame of drive-1 and $\tilde H_c$ is the ancilla Hamiltonian in the rotating frame as given in Eq.~(\ref{eq:rotated_Hc}). $h_1,h_2$ are bath operators that lead to ancilla relaxation and dephasing, respectively. 

To find the time evolution of the reduced density matrix of the ancilla, we follow the standard procedure to eliminate the bath degrees of freedom based on the Markov approximation. For Floquet systems, a rather clear derivation can be found in Ref.~\cite{Hone2009} and references therein. Here we sketch the main steps involved in the derivation, tailored for a periodically driven weakly nonlinear oscillator. We first go to the interaction picture and solve iteratively the equation for the total density matrix to second order in $H_i$; after taking the trace over the bath degrees of freedom, we obtain
\begin{align}
\dot {\bar \rho}_c =  - \frac{1}{\hbar^2} \int _{-\infty}^t dt' \Tr_b \left([\bar H_i(t),[\bar H_i(t'), \bar \rho(t')]]\right),
\end{align}
where the bar over the operators indicates the interaction picture and $\Tr_b$ indicates trace over bath degrees of freedom; $\bar \rho_c  \equiv \Tr_b \bar \rho$.

Next we make two approximations: 1) the total density matrix $\bar\rho(t')$ factorizes  $\bar\rho(t')\approx \rho_{\rm bath}\otimes\bar\rho_c(t')$ where $\rho_{\rm bath}$ is the bath density matrix in equilibrium at $t = -\infty$; 2) the rate of change of the reduced density matrix $\bar\rho_c(t)$ is much smaller than the relaxation rate of the bath, so that one can make the Markov approximation that $\bar \rho_c(t')\approx \bar \rho_c(t)$. The two approximations ultimately rely on the coupling between the ancilla and the baths being weak.  After making these two approximations and going back to the Schr\"odinger picture, we obtain
\begin{align}
\label{eq:reduced_density_matrix}
&\dot\rho_c(t) = -\frac{i}{\hbar}[\tilde H_c(t),\rho_c] - \frac{1}{\hbar^2}\int_{-\infty}^t dt' \bigg\{ \langle h_1(t)h_1(t') \rangle \nonumber \\
&\times \left[(ce^{-i\omega_1t}+\rm{H.c.})(c(t,t')e^{-i\omega_1 t'}+ \rm{H.c.})\rho_c(t) \right. \nonumber\\
&\left.- (c(t,t')e^{-i\omega_1t'}+\rm{H.c.})\rho_c(t)(ce^{-i\omega_1 t}+\rm{H.c.})\right] \nonumber\\
& - \langle h_2(t)h_2(t')\rangle [ c^\dagger c (c^\dagger c)(t,t')\rho_c(t)-(c^\dagger c)(t,t')\rho_c(t)c^\dagger c ] \nonumber \\
&+\rm{H.c.}\bigg\}
\end{align}
where $c(t,t')\equiv U_c(t',t)cU_c^\dagger (t',t), U_c(t',t) = \hat T \exp[-i \int_{t'}^t  dt'' \tilde H_c(t'')]$ and similarly for $(c^\dagger c)(t,t')$. It is important to notice that because of the time ordering as denoted by the operator $\hat T$, $U_c^\dagger (t',t)\neq U_c(t,t').$ In arriving at Eq.~(\ref{eq:reduced_density_matrix}), we have assumed that there is no correlation between the bath variables $h_1$ and $h_2$. In accordance with the rotating wave approximation we have made in  treating the drives and nonlinearity of the ancilla, we can also neglect the cross term between $c$ and $c(t,t')$, $c^\dagger $ and $c^\dagger (t,t')$ in the second and third line of the equation above. 

In the limit that the spectral density of the bath (the Fourier transform of correlator $\langle h_{1,2}(t)h_{1,2}(0) \rangle$) is sufficiently smooth (or almost constant) over the scale of the ancilla anharmonicity and drive detunings, Equation~(\ref{eq:reduced_density_matrix}) reduces to the familiar Lindbladian master equation: 
\begin{align}
\label{eq:master_equation_Markovian}
&\dot \rho = -i[\tilde H_c(t),\rho]/\hbar +  (n_{\rm th}+1)\mathcal D[\sqrt{\gamma}c]\rho + n_{\rm th} \mathcal D[\sqrt{\gamma}c^\dagger ] \rho  \nonumber\\ 
&+ \mathcal D[\sqrt{2\gamma_{\rm ph}} c^\dagger c]\rho, \quad \mathcal D[c]\rho \equiv c \rho c^\dagger  - \frac{1}{2}\{c^\dagger c,\rho\}.
\end{align} 
Here $\gamma$ and $\gamma_{\rm ph}$ are the ancilla decay and dephasing rate, respectively. They are given by $\gamma = 2\hbar^{-2}\Re \int_0^\infty e^{i\omega_c t}\langle [h_1(t),h_1(0)]\rangle.$ and $\gamma_{\rm ph} = \hbar^{-2}\Re \int_0^\infty \langle h_2(t)h_2(0)\rangle dt$. We have assumed that the bath that leads to ancilla relaxation is in thermal equilibrium with the thermal population $n_{\rm th} = [\exp(\hbar\omega_c/k_BT)-1]^{-1}.$
  
For large drive detunings, however, the assumption of constant spectral density of the bath might break down, particularly for the bath that causes dephasing. In the following, we relax this assumption and consider the more general situation. To capture the frequency dependence in the spectral density of the bath, it is convenient to write the density matrix $\rho_c$ in the basis of Floquet states of the Hamiltonian $\tilde H_c(t)$: \[\rho_c(t) = \sum_{mn} \rho_{mn}(t) |u_m(t)\rangle\langle u_n(t)|. \]
In such a basis, Eq.~(\ref{eq:reduced_density_matrix}) has the form
\begin{align}
\label{eq:masterequation_Floquet}
\dot \rho_{mn} = -i \epsilon_{mn} \rho_{mn}/\hbar + \mathcal M_{m'n'}^{mn}(t)\rho_{m'n'},
\end{align}
where the rank-4 tensor $\mathcal M_{m'n'}^{mn}\propto 1/\hbar^2$ can be found straightforwardly by inserting $\rm{I} = \sum_m |u_m(t)\rangle \langle u_m(t)|$ into Eq.~(\ref{eq:reduced_density_matrix}) and using the relation $U_c^\dagger (t',t)|u_m(t)\rangle = e^{i\epsilon_m(t-t')}|u_m(t')\rangle$. Its magnitude depends on the matrix elements of operators $c,c^\dagger ,c^\dagger c$ in the Floquet basis and the spectral density of the baths at certain frequencies; see below. $\mathcal M_{m'n'}^{mn}(t)$ is periodic in time with a periodicity $\tau$ due to the periodicity in the basis states $u_m(t)$. This is in contrast to systems in equilibrium where the corresponding tensor $\mathcal M$ is time-independent. In the following, we will give explicit expressions for the tensor $\mathcal M$ in the limit of weak damping and dephasing.

\subsection{The limit of weak damping and dephasing}
\label{sec:weak_damping}
Equation~(\ref{eq:masterequation_Floquet}) greatly simplifies in the limit of weak damping and dephasing where the quasienergy spacing and their non-equidistance is much larger than the broadening of the quasienergy levels due to coupling to the bath. Due to the fast oscillation in the off-diagonal element $\rho_{mn}$ with a rate set by $\epsilon_{mn}/\hbar$, one can neglect couplings between diagonal and off-diagonal elements of $\rho_c$. Furthermore, when the level spacings $\epsilon_{mn}$ are sufficiently non-equidistant compared to their broadening,  one can as well neglect the couplings among the off-diagonal elements of $\rho_c$. In the same weak damping and dephasing limit, one can also disregard the time dependence in $\mathcal M$ by averaging over a period $\tau$ as long as $\omega_{21}\gg \gamma,\gamma_{\rm ph}$. After these approximations, Eq.~(\ref{eq:masterequation_Floquet}) reduces to
\begin{align}
\label{eq:diagonal}
 &\dot \rho_{mm} = -\sum_n W_{mn}\rho_{mm} + \sum_n W_{nm}\rho_{nn},
\end{align}
\begin{align} 
\label{eq:off_diagonal}
&\dot \rho_{mn} = -i\epsilon_{mn}\rho_{mn}/\hbar - V_{mn} \rho_{mn}, m\neq n
\end{align}
Equations~(\ref{eq:diagonal}) and (\ref{eq:off_diagonal}) capture the main effects of ancilla decoherence on the Floquet dynamics: 1) there is incoherent hopping between different Floquet states with a hopping rate given by $W_{mn}$ due to dissipation and dephasing; 2) the coherence between quasienergy states acquires a finite decay rate given by $V_{mn}$. These two effects are responsible for the aforementioned heating and the inverse Purcell effect, respectively. We will study them in detail in the next few sections. 

The parameter regime where Eqs.~(\ref{eq:diagonal},\ref{eq:off_diagonal}) hold readily applies to the current cQED experiments. The characteristic quasienergy spacing $\epsilon_{mn}$ is set by the drive detunings $\delta_{1,2}$ and the ancilla anharmonicity $\alpha$. Typical anharmonicity of transmon ancilla used in cQED ranges from tens to hundreds of MHz and is orders of magnitude larger than its dephasing and dissipation rate which is typically tens of kHz. The detunings of the drives from the ancilla frequency can be chosen to be of the same size as the ancilla anharmonicity. We emphasize that the approximations that lead to Eqs.~(\ref{eq:diagonal},\ref{eq:off_diagonal}) break down near quasienergy level anti-crossings where the level spacings become smaller than their widths. 

\subsubsection{Incoherent hopping between Floquet states}
Due to the noise that accompanies the dissipation and dephasing, there occurs incoherent hopping between different Floquet states as described by Eq.~(\ref{eq:diagonal}).  The hopping rate $W_{mn}$ from the $m$-th to the $n$-th Floquet state is found to be
\begin{align}
\label{eq:hopping_rate}
W_{mn} =& W^{\gamma}_{mn} +W^{\gamma_{\rm ph}}_{mn},\nonumber\\
W^{\gamma}_{mn}= & \sum_K\bigg[ |(c^\dagger )_{nm,K}|^2n_{\rm th} \gamma(\omega_1+\epsilon_{nm}/\hbar+K\omega_{21}) \nonumber \\
&+|c_{nm,K}|^2(n_{\rm th}+1) \gamma(\omega_1-\epsilon_{nm}/\hbar-K\omega_{21}) \bigg], \nonumber \\
W^{\gamma_{\rm ph}}_{mn}=&2 \sum_K |(c^\dagger c)_{nm,K}|^2\gamma_{\rm ph}(\epsilon_{nm}/\hbar+K\omega_{21}).
\end{align} 
Here the frequency-dependent dissipation and dephasing rates are given by
\begin{align}
\gamma(\omega) &= 2\hbar^{-2}\Re \int_0^\infty dte^{i\omega t}\langle [h_1(t),h_1(0)]\rangle,\nonumber\\\gamma_{\rm ph}(\omega) &= \hbar^{-2}\Re \int_0^\infty dt e^{-i\omega t} \langle h_2(t)h_2(0)\rangle .
\end{align}
We have neglected the frequency dependence in $n_{\rm th}$ in the considered parameter regime $|\delta_{1,2}|,\alpha\ll \omega_c$. 

The formula for the hopping rate $W_{mn}$ has the same form as the usual transition rates given by Fermi' s golden rule, the squared matrix element times the density of states at the energy the bath provides to or receives from the ancilla. The hopping induced by dissipation is accompanied by absorption and emission of an excitation near frequency $\omega_c$ into or from the bath as represented by the first and second terms in $W^{\gamma}$, respectively. In contrast to undriven oscillators, the hopping generally occurs not just between neighboring levels but also between levels separated by more than one transition frequency $\omega_c$; the extra energy needed for the transition to occur is provided by the drives, which are embedded in the Floquet states $u_m$. Because of the second drive, there is also a summation over $K$ which indicates an exchange of $K$ excitations between the two drive reservoirs. 

An important feature of dissipation-induced hopping is that even at zero temperature ($n_{\rm th}$ = 0) where the ancilla can only emit excitation to the bath, it can still ``hop up" in the ladder of Floquet states. Let us consider for instance the hopping from the state $u_m$ to $u_{m+1}$ and the simple case where only drive-1 is present [thus $K =0$ in Eq.~(\ref{eq:hopping_rate})]. For weak drive, state $u_m$ is close to ancilla Fock state $|m\rangle$. In hopping from state $u_m$ to $u_{m+1}$ at zero temperature, the ancilla absorbs two drive excitations at frequency $\omega_1$ and emits one excitation to the bath at frequency $2\omega_1-E_{(m+1)m}/\hbar$. Indeed, one can show that the relevant matrix element for this process $c_{(m+1)m,0}\propto \alpha\Omega_1^2$ for weak drive; see also Eq.~(\ref{eq:heating_rate_weakdrive}).

The frequency noise (dephasing) of the ancilla also induces hopping between the Floquet states with a hopping rate given by $W^{\gamma_{\rm ph}}_{mn}$ in Eq.~(\ref{eq:hopping_rate}). Importantly, the hopping induced by frequency noise does not involve exchange of excitation between the ancilla and the bath near frequency $\omega_c$. Instead, the hopping occurs because the ancilla makes a transition to a neighboring level by absorbing or emitting a near-resonant drive excitation ($\delta_{1,2}\ll \omega_c$) and the extra energy is absorbed by or emitted to the bath. Therefore, to leading order in the drive amplitudes, the relevant matrix elements $(c^\dagger c)_{m(m\pm1),0}\propto \Omega_1$; see Eq.~(\ref{eq:heating_rate_weakdrive}). A transition to a ``far away'' level is also possible by absorbing or emitting multiple drive excitations. 

An important complication that must be considered is that spectral density of the noise that leads to dephasing is typically strongly frequency dependent. The measured dephasing rate from Ramsey fringe and spin echo experiments is a measure of the noise spectrum at very low frequencies, whereas the inelastic transitions described above rely on the spectral density of the noise bath at much higher frequencies; see next section for a detailed discussion.

\subsubsection{Heating from the Floquet ``ground state"}
\label{sec:heating}
As described previously, even at zero temperature, the ancilla can hop from one Floquet state to another and thereby forms a finite-width distribution over the Floquet states after a relaxation time of the ancilla.  In order to perform a coherent quantum operation between cavity modes utilizing the ancilla-mediated interactions, one would like to prepare the ancilla in the Floquet state that has the smallest escape rate. 
Normally, this state is also the most populated state when the driven ancilla reaches its steady state. We will call this state the Floquet ``ground state.'' 

For a driven nonlinear oscillator where the drives are blue detuned ($\delta_{1,2}>0$), as we will show, the Floquet ground state is the state $\Psi_0$ that adiabatically connects to the ancilla vacuum state $|0\rangle$ as the drive amplitudes increase or decrease. 
The situation is more complicated when the drive is red-detuned. In this case, there occur systematic level anti-crossings depending on the ratio of drive detuning and anharmonicity \cite{dykman2005} and the oscillator may undergo a sharp transition to states with large photon number as the drive amplitude increases \cite{siddiqi2004a}. We will not discuss this situation here. 

Of primary interest to us is the ``heating rate'' from the Floquet ground state $\Psi_0$ after we have prepared the ancilla in that state by slowly turning on the drives. In the following, we will focus on the interesting case of $n_{\rm th} = 0$ where the heating is solely due to the quantum noise that accompanies dissipation and frequency noise. In the limit of weak drives, Floquet state $|\Psi_0\rangle$ is mostly Fock state $|0\rangle$ and has a small amount of coherent admixture with other Fock states. This admixture results in a finite transition rate $W_{0n}$ from $\Psi_0$ to $\Psi_n$ for any $n$. For weak drives, the transition to the neighboring state $\Psi_1$ dominates and the rate is
\begin{align}
\label{eq:heating_rate_weakdrive}
&W_{01}^{\gamma} \approx  \alpha^2 \bigg[  \Big|\frac{\xi_1^2}{2\delta_1+\alpha}\Big|^2\gamma(2\omega_1-\omega_c) +  \Big|\frac{\xi_2^2}{2\delta_2+\alpha}\Big|^2 \nonumber \\
&\times \gamma(2\omega_2-\omega_c) + \Big|\frac{\xi_1\xi_2} {\delta_1+\delta_2+\alpha}\Big|^2\gamma(\omega_1+\omega_2-\omega_c)\bigg] \nonumber \\
&W^{\gamma_{\rm ph}}_{01} \approx 2  \bigg[|\xi_1|^2\gamma_{\rm ph}(-\delta_1)+ |\xi_2|^2 \gamma_{\rm ph}(-\delta_2) \bigg]
\end{align}

An important difference between dissipation- and dephasing-induced heating is that the former requires nonlinearity while the latter does not.  One way to understand this is to  consider the limit of zero anharmonicity; see Sec.~\ref{sec:weak_anharmonicity}. In this limit, the Floquet states are simply displaced Fock states. It is not hard to show that the dissipative dynamics in the displaced frame is exactly the same as in the lab frame without drive; therefore, dissipation can only bring the ancilla down in the Floquet ladder. In contrast, since displaced Fock states (in particular, the coherent state) are superpositions of Fock states, dephasing can cause transitions among these states. Another enlightening way to see the difference is to consider the limit of constant $\gamma(\omega)$ and $\gamma_{\rm ph}(\omega)$. In this limit, the total rate $\sum_{n\neq m}W_{mn}$ of leaving the state $\Psi_m$ can be summed up to be equal to the variance of the operator $c$ and $c^\dagger c$ for the dissipation- and dephasing-induced transitions, respectively:
\begin{align}
\label{eq:heating_rate_sum}
\sum_{n\neq m} W^{\gamma}_{mn} = &\gamma \tau ^{-1}\int_0^\tau dt \Big(\langle u_m(t) | c^\dagger c|u_m(t)\rangle \nonumber \\
&- |\langle u_m(t)|c|u_m(t)\rangle|^2\Big) \nonumber \\
\sum_{n\neq m} W^{\gamma_{\rm ph}}_{mn} =&2 \gamma_{\rm ph} \tau^{-1}\int_0^\tau dt \Big[\langle u_m(t) | (c^\dagger c)^2|u_m(t)\rangle \nonumber \\
&- \langle u_m(t)|c^\dagger c|u_m(t)\rangle^2 \Big]
\end{align}
In the limit of zero anharmonicity where $\Psi_0$ is a coherent state, Eq.~(\ref{eq:heating_rate_sum}) shows that $\sum_{n\neq 0} W^{\gamma}_{0n}=0$ and $\sum_{n\neq 0} W^{\gamma_{\rm ph}}_{0n} = 2\gamma_{\rm ph} (|\xi_1|^2+|\xi_2|^2).$ The heating due to the interplay of drive and broadband dephasing noise has also been studied theoretically and observed experimentally for a linear oscillator in the classical regime~\cite{zhang2014}. 

\subsubsection{Comparison with experiment}
\label{sec:heating_exp}
\begin{figure}[ht]
\centering
\includegraphics[width=5.5cm]{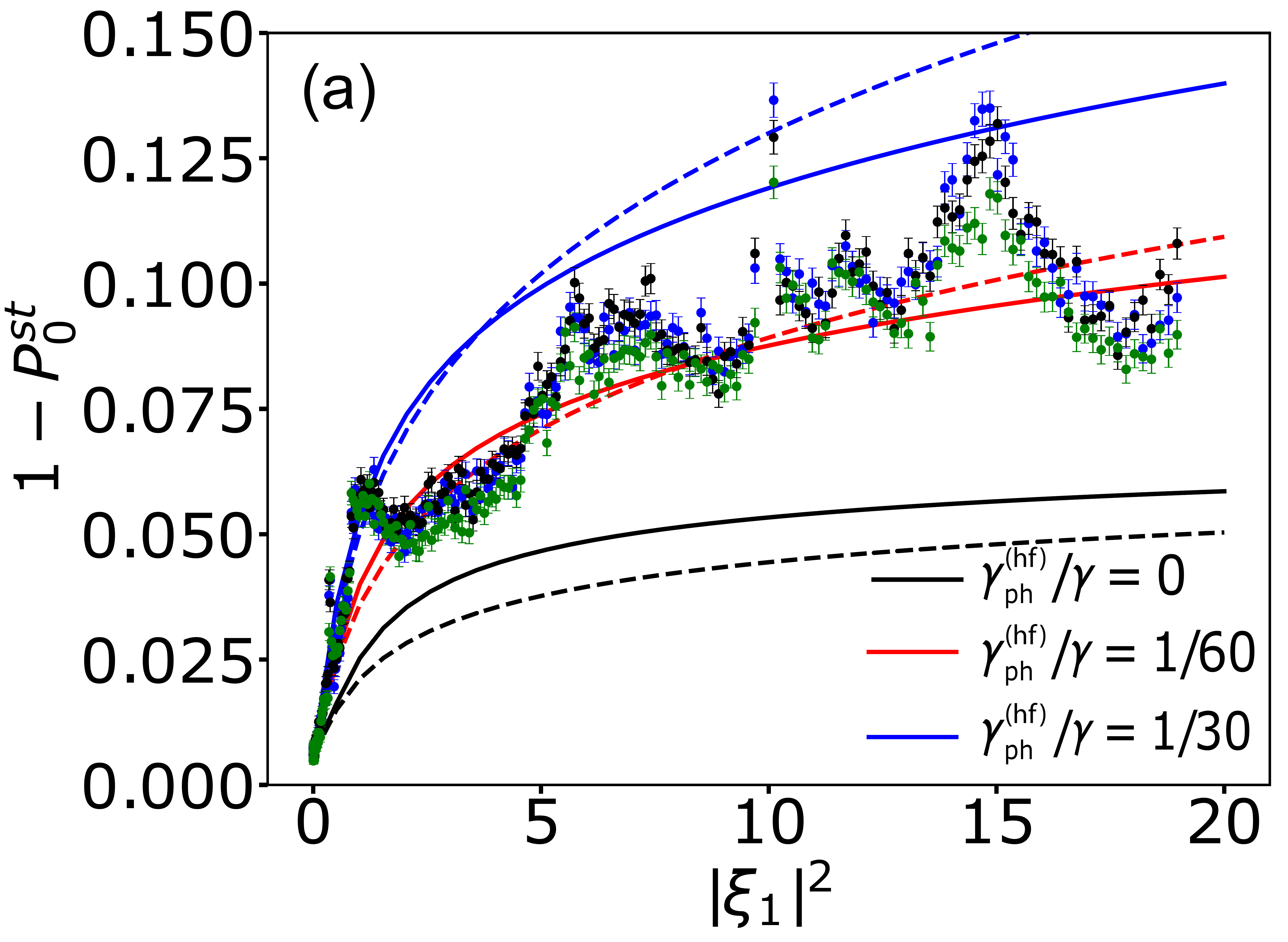}  \\
\includegraphics[width=5.5cm]{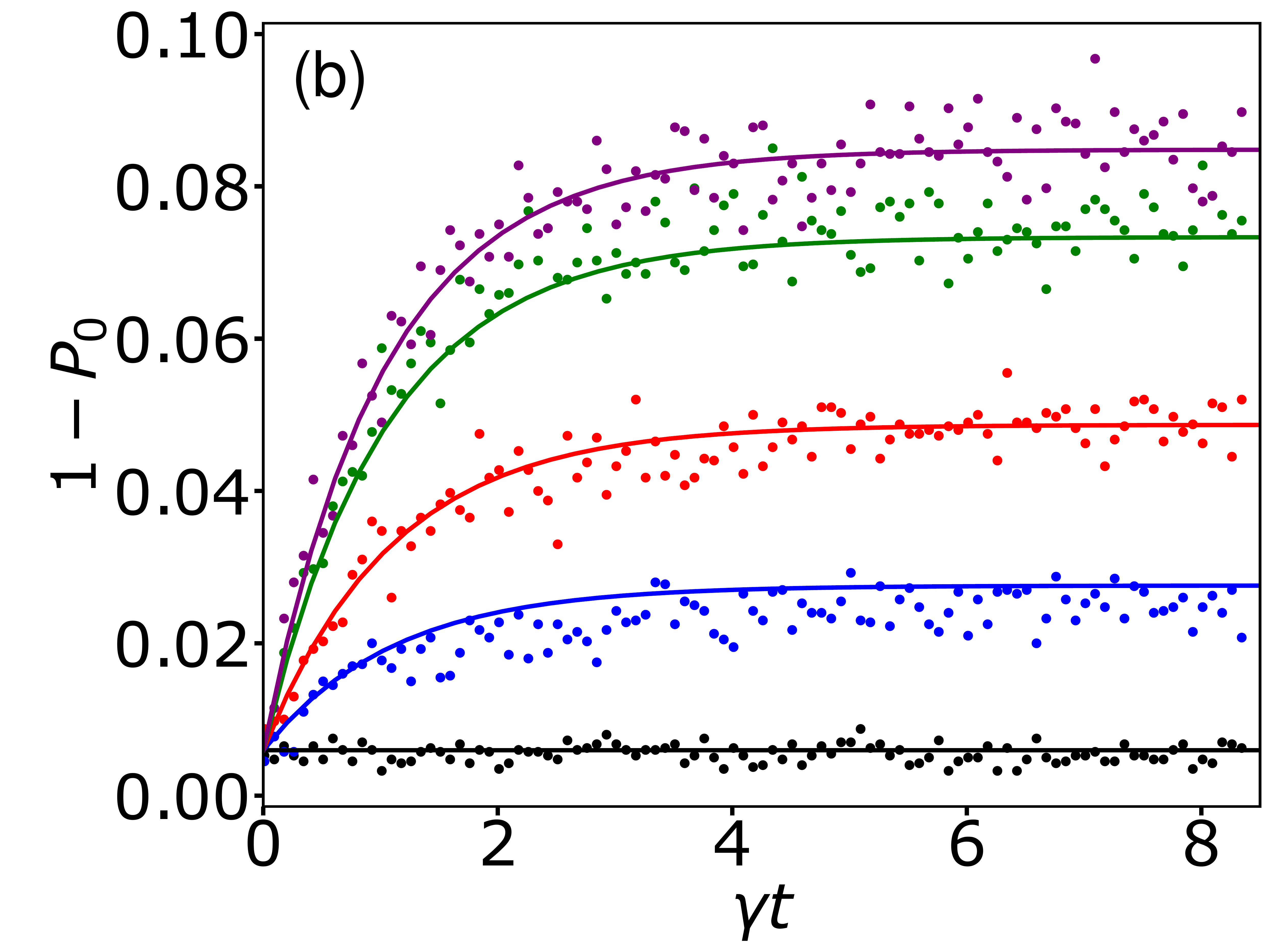} 
\caption{Comparison between the theory (solid lines) and experiment (dots) on the heating from the Floquet ground state $\Psi_0$ in the presence of one drive (drive-1). The detuning of the drive $\delta_1 = \alpha = 2\pi*71.68~\rm{MHz}$. (a) The steady-state population of not in the ground Floquet state as a function of the scaled drive power. The experimental data are taken for a pump duration $\sim 100~\rm{\mu s}$. The blue, black, green dots refer to 48,100, and 1000 ns ramping time, respectively. The spike around $|\xi_1|^2 = 0.5$ is likely due to leakage from the mixer that excites the ancilla when the ancilla frequency is Stark shifted to the frequency of this tone. 
The theoretical curve is a result of Eq.~(\ref{eq:diagonal}): $P_0 \equiv \rho_{00}$. We use a constant $\gamma$ and $\gamma_{\rm ph}^{\rm (hf)}$ whose ratio is taken to be $\gamma_{\rm ph}^{\rm (hf)}/\gamma =$ 0 (black), 1/60 (red) and 1/30 (blue); see the text for details. The dashed lines are results of semiclassical calculation (see Appendix~\ref{sec:semiclassics_heating}). (b) The transmon population not in the ground Floquet states as a function of the dimensionless time $\gamma t$ for various scaled drive powers. From bottom to top, the scaled drive powers are $|\xi_1|^2 $= 0 (black), 0.55 (blue), 1.52 (red), 4.85 (green) and 8.66 (purple). At $t=0$, when the drive is turned on, the ancilla is in an effective thermal equilibrium with its environment with a thermal population $n_{\rm th} \approx 0.006$. The independently measured decay rate of the ancilla is $\gamma = 90$ kHz. The solid lines refer to the results of simulation using Eq.~(\ref{eq:diagonal}) where we have neglected the finite time of ramping up and down the drive which is much shorter than the time scale set by $1/\gamma$ and assumed that the ancilla adiabatically evolves from Fock states $|n \rangle$ to the adiabatically-connected Floquet states $|\Psi_n \rangle$ during these times.}
\label{fig:heating}. 
\end{figure}

To corroborate the theory, we performed measurement on the drive-induced heating of the superconducting transmon qubit (the ancilla). The procedure of the experiment is similar to the AC Stark shift measurement in Sec.~\ref{sec:Stark_shift}. Before we turn on the drive, the ancilla is mostly in the ground state with a thermal population $n_{\rm th} \approx 0.006$. At time $t = 0$, we turn on the drive with a rise time 100 ns and then keep the drive on for various amount of time.  Finally, we measure the ancilla ground state population after we have turned off the drive. The drive envelope is symmetric with respect to ramping up and down each having a hyperbolic tangent shape. 
For zero drive amplitude, the ancilla remains in the ground state with a very small probability in the excited states due to thermal fluctuations; see the black dots in Fig.~\ref{fig:heating}(b). For a finite drive amplitude, the ancilla population in the excited (Floquet) states increases in time and then reaches a steady state during a time scale set by the relaxation rate $\gamma$ of the ancilla.   

To compare with the theory, we recorded the steady-state population for various drive amplitudes as shown Fig.~\ref{fig:heating}(a). In the presence of a single off-resonant drive, the rate of dephasing-induced hopping is determined by the spectral component of the ancilla frequency noise near the drive detuning frequency, as can be seen in Eq.~(\ref{eq:heating_rate_weakdrive}). For the range of drive power used in the experiment, the drive detunings from the AC Stark shifted ancilla transition frequencies range from tens to hundreds of MHz. We assumed that the dephasing rate $\gamma_{\rm ph}(\omega)$ is approximately flat in this frequency region and has an amplitude $\gamma_{\rm ph}^{\rm (hf)}$, to be differentiated from the dephasing rate obtained from Ramsey or spin echo experiment which is only sensitive to the low-frequency part of the noise spectrum; see below. We also assumed that the dissipation rate $\gamma(\omega)$ is approximately flat near the ancilla frequency and has a value $\gamma$.

As shown in Fig.~\ref{fig:heating}, the theory approximately matches the experiment for a dephasing rate $\gamma_{\rm ph}^{\rm (hf)} = \gamma/60$.
This is considerably smaller than the independently-measured dephasing rate of the undriven ancilla using Ramsey fringes: $\gamma_{\rm ph}^{(R)} = 88~\rm{kHz} \approx \gamma $. The latter is a measure of the noise spectrum near zero frequency over a frequency range set by the inverse length of the Ramsey experiment which is constrained by $\gamma_{\rm ph}^{(R)}$ itself~\cite{martinis2003a}; it is typically dominated by the low frequency component of the ancilla dephasing noise, including $1/f$ noise. The difference between $\gamma_{\rm ph}^{(R)}$ and $\gamma_{\rm ph}^{\rm (hf)}$ suggests a falling off of the dephasing noise spectrum over a range from tens of kHz to tens of MHz. Also, we emphasize that dissipation alone only accounts for half of the observed heating.

The observed heating due to dissipation can also be understood as a result of the drive-induced squeezing of the ancilla mode as can be seen from Eq.~(\ref{eq:H_D}). The lowering operator of the bare ancilla mode is a linear combination of both the lowering and raising operator of the squeezed ancilla mode. As a result, annihilation of the ancilla excitations due to coupling to the environment can lead to both creation and annihilation of the excitations of the squeezed mode. One can show using a semiclassical analysis that, even at zero temperature, there forms a finite-width Boltzman distribution over the states of the squeezed mode with an effective thermal population $\tilde n_{\rm th} = \sinh^2\phi$. Here, $\phi$ is the squeezing parameter and is controlled by a dimensionless parameter $\alpha |\xi_1|^2/\delta_1$. 

For a blue-detuned drive, the amount of squeezing saturates at strong drive due to the drive-induced frequency shift which pushes the ancilla frequency further away from the drive frequency and effectively constrains the maximum squeezing one can achieve. When the parameter $\alpha |\xi_1|^2/\delta_1$ becomes of the order ten, the squeezing saturates and the total population in the excited Floquet states approaches $\sim 7\%$. The black dashed line in Fig.~\ref{fig:heating}(a) shows the result of this analysis (see Appendix~\ref{sec:semiclassics_heating} for details), which qualitatively captures the behavior of the full Floquet analysis including the saturation of the heating at strong drive. 

The same semiclassical analysis shows that, on top of the dissipation-induced heating, ancilla dephasing leads to an additional effective thermal population $\tilde n_{\rm th} = 2|\overline \xi_1|^2 \gamma_{\rm ph}^{\rm (hf)}/\gamma$. 
The joint effects of the ancilla dissipation and dephasing are shown as the red and blue dashed lines in Fig.~\ref{fig:heating}(b) which also qualitatively match the full Floquet analysis.

To rule out the possibility that the observed heating is due to ramping up and down the drive too rapidly causing diabatic transitions (in particular for relatively strong drive), we performed the measurement for various ramping times ranging from 48 to 1000 ns. While we did observe slight variations in the steady-state population for different ramping times, the overall trend and the saturation value of the excited state population at large drive amplitude remains the same. For a very long ~5 $\rm{\mu}s$ ramping down time, we observed that the steady-state population in the excited states significantly reduces which is likely due to the ancilla re-equilibrates while the drive is turning off. We have also numerically verified that using a linear ramp in which the drive amplitude increases linearly in time from zero to its peak value ($|\xi_1|^2 = 10$) in 100 ns, the ancilla initially in the ground state has a 99.5\% overlap with the Floquet ground state $\Psi_0$ at the end of the ramp. This confirms that the drive is turned on adiabatically.

The exact source of the dephasing noise around the drive detuning frequency with a strength $\gamma_{\rm ph}^{\rm (hf)} \approx \gamma/60$ requires further investigation. The good agreement between theory and experiment over a wide range of the drive power (corresponding to an AC Stark shift $\delta E_{10}/2\pi\hbar$ up to $\sim 250$ MHz) suggests that the dephasing noise could be a broadband noise with a bandwidth larger than hundreds of MHz. 

We have ruled out the possibility that the observed heating comes from the tail of the spectrum of the dephasing noise induced by the photon number fluctuations of the low-$Q$ cavity mode for readout. Because of the dispersive coupling between the readout cavity and the transmon ancilla, thermal fluctuations of the readout cavity photon number become frequency fluctuations of the transmon; cf.~\cite{yan2016}. These frequency fluctuations have a Lorentzian spectrum $\propto \kappa n_{\rm th}(n_{\rm th}+1)/ (\kappa^2 + \omega^2)$ where $n_{\rm th}$ and $\kappa$ are the thermal photon number and relaxation rate of the readout cavity. Assuming that the measured Ramsey dephasing rate $\gamma^{(R)}_{\rm ph}$ all comes from the thermal fluctuations of the readout cavity mode, we obtained that $\gamma^{(R)}_{\rm ph} \approx  n_{\rm th}(n_{\rm th}+1)\chi^2/\kappa$ where $\chi$ is the dispersive coupling rate between the readout cavity and the ancilla; we have used the fact that $\kappa (\approx 1~\rm{MHz})\gg \gamma,\gamma^{(R)}_{\rm ph}$. Without knowing the values for $\chi$ and $n_{\rm th}$, we can deduce the dephasing rate at a higher frequency using the Lorentzian form of the noise spectrum and estimate an upper bound of the readout-cavity-induced dephasing rate at the drive detuning frequency ($\delta_1/2\pi = 71.68$~MHz) to be $\gamma^{(R)}_{\rm ph} (\kappa/\delta_1)^2 \sim 10^{-5}\gamma. $ This dephasing rate is two orders of magnitude smaller than what we found. However, we have not considered the thermal fluctuations of higher order modes of the readout cavity which usually have larger $\kappa$ and thus can potentially lead to larger transmon dephasing rate at high frequency.

\subsubsection{Decoherence of superpositions of Floquet states}
Because of ancilla dephasing and dissipation, superpositions of Floquet states decohere as described by Eq.~(\ref{eq:off_diagonal}). The decoherence rate $V_{mn}$ of a superposition of Floquet states $\Psi_m$ and $\Psi_n$ is given by 
\begin{align}
\label{eq:decoherence_rate}
V_{mn} &= V^{\gamma}_{mn} + V^{\gamma_{\rm ph}}_{mn},  \nonumber \\
V^{\gamma}_{mn}&=\frac{1}{2} \sum_{j\neq m} W^{\gamma}_{mj}+\frac{1}{2} \sum_{j\neq n} W^{\gamma}_{nj} + \sum_K\bigg[\frac {1}{2}\gamma(\omega_1+K\omega_{21}) \nonumber \\
&\times (2n_{\rm th}+1)|c_{mm,-K}-c_{nn,-K}|^2\bigg]  \nonumber \\
V^{\gamma_{\rm ph}}_{mn}&=\frac{1}{2} \sum_{j\neq m} W^{\gamma_{\rm ph}}_{mj}+\frac{1}{2} \sum_{j\neq n} W^{\gamma_{\rm ph}}_{nj} +\sum_K\big[\gamma_{\rm ph}(K\omega_{21}) \nonumber \\
&\times |(c^\dagger c)_{mm,K}-(c^\dagger c)_{nn,K}|^2 \big]
\end{align}
Here $V^{\gamma}$ and $V^{\gamma_{\rm ph}}$ are the rates of dissipation and dephasing-induced decoherence, respectively. In a spectroscopy measurement of the ancilla, the decoherence rate $V_{mn}$ sets the linewidth for the transition from the state $\Psi_m$ to $\Psi_n$; see Sec.~\ref{sec:inverse_Purcell}.

Equation~(\ref{eq:decoherence_rate}) shows that decoherence of superposition of two Floquet states $\Psi_m$ and $\Psi_n$ has two contributions: i) incoherent hopping from the two states to other states as given by the first two terms in both  $V^{\gamma}$ and $V^{\gamma_{\rm ph}}$; ii) ``pure dephasing" of the Floquet states with a rate given by the third terms in $V^{\gamma}$ and $V^{\gamma_{\rm ph}}$. This form of decoherence rate $V_{mn}$ is similar to the decoherence rates of superpositions of Fock states of an undriven oscillator; because the Floquet states are coherent admixture of Fock states, their effective dephasing and dissipation have mixed contribution from both dissipation and dephasing in the Fock basis.

\begin{figure}[ht]
\centering
\includegraphics[width=5 cm]{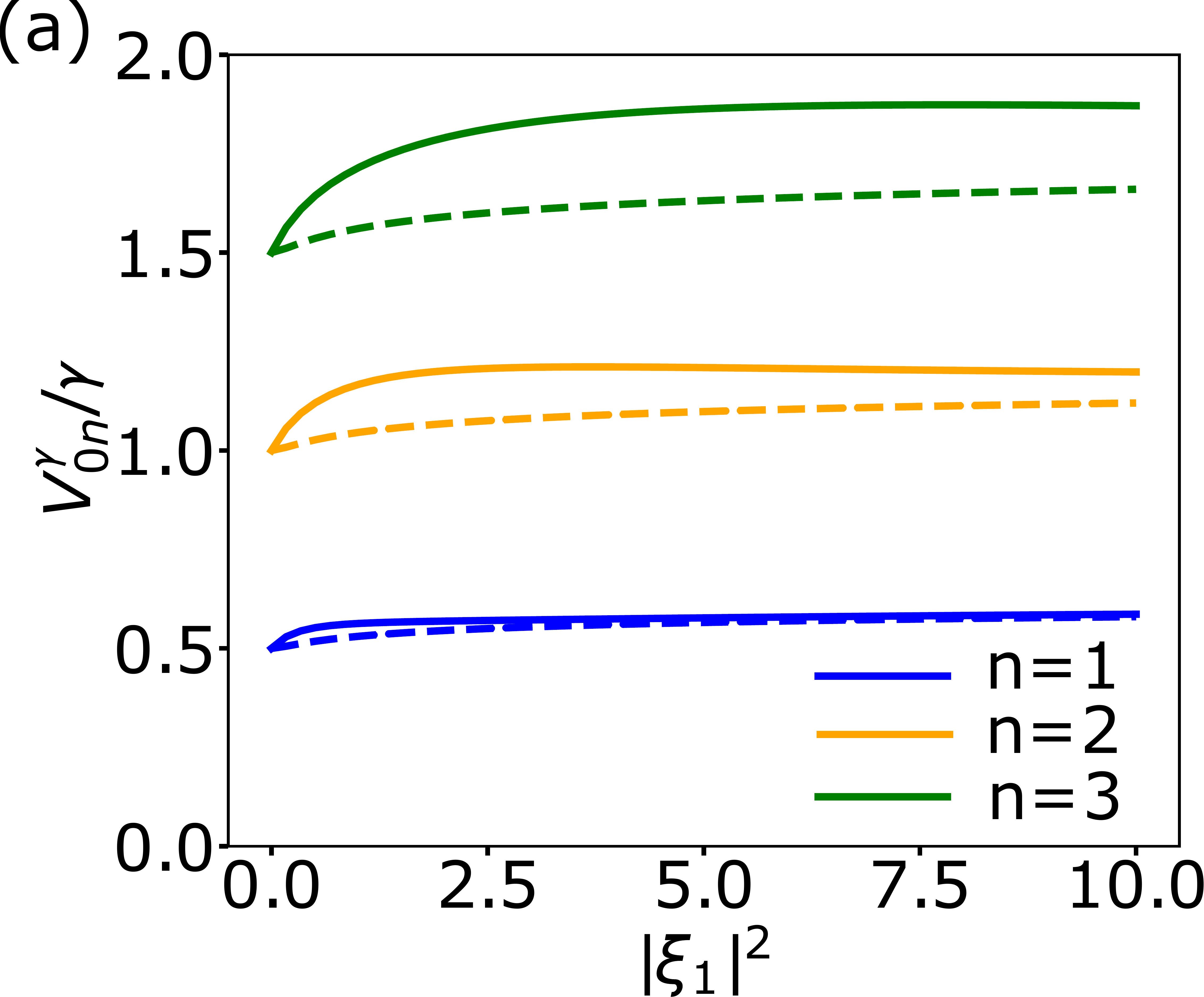} \\
\includegraphics[width=5 cm]{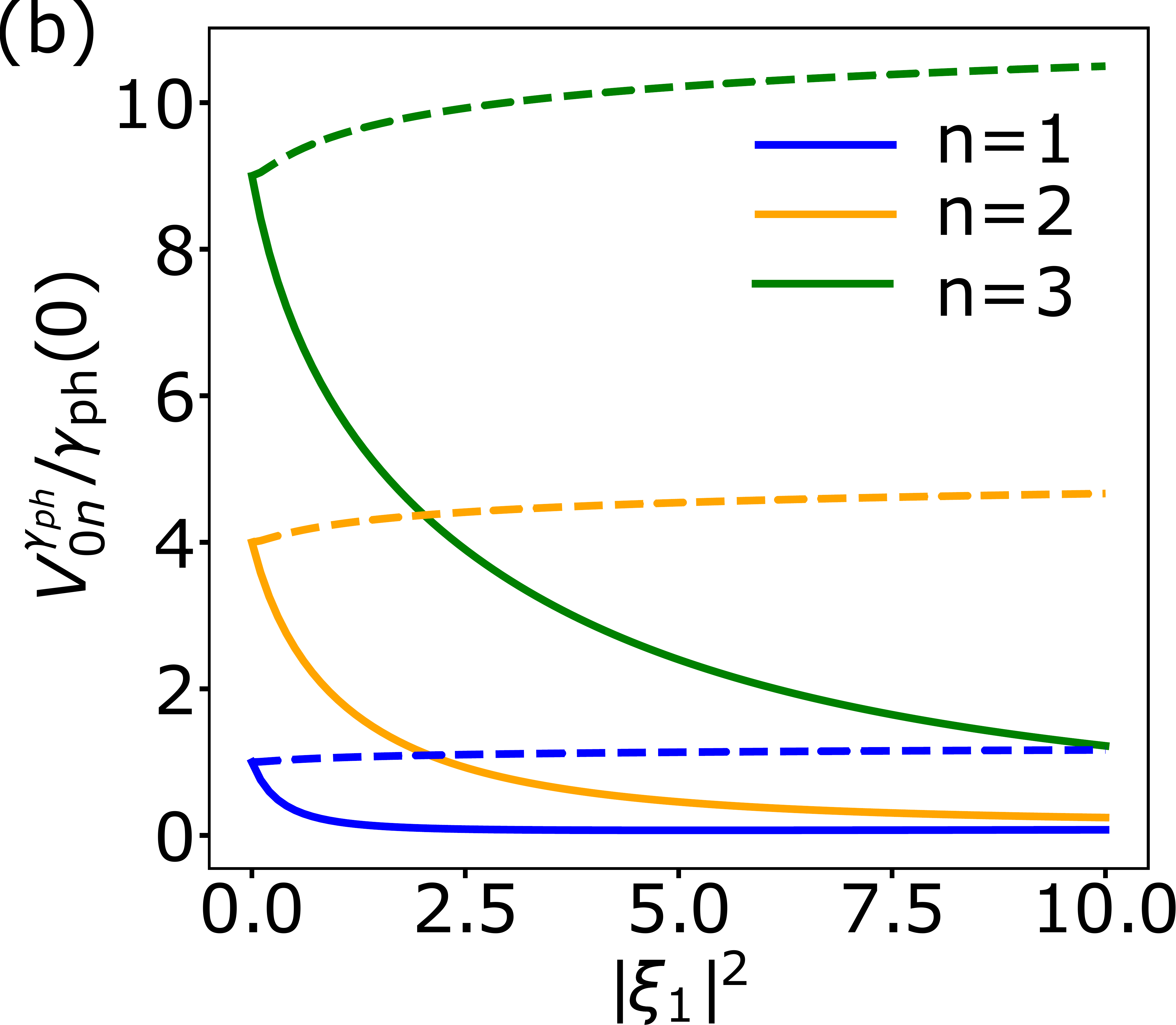}
\caption{The decoherence rate of a superposition of Floquet states $\Psi_0$ and $\Psi_n$ as a function of the scaled drive power for the case of one blue-detuned drive with detuning $\delta_1/\alpha = 1$. The solid and dashed lines refer to the full Floquet results and semiclassical results in Appendix~\ref{sec:semiclassics_decoherence_rate}, respectively. (a) The dissipation-induced decoherence rate $V^{\gamma}_{0n}$ in Eq.~(\ref{eq:V_gamma_simplified}). (b) The dephasing-induced decoherence rate $V^{\gamma_{\rm ph}}_{0n}$ in Eq.~(\ref{eq:V_gamma_ph_simplified}). The ratio $\gamma_{\rm ph}^{\rm (hf)}/\gamma_{\rm ph}(0)$ is chosen to be 1/60 as we found in Sec.~\ref{sec:heating_exp}.}
\label{fig:linewidth}
\end{figure}

Depending on the relative magnitude of ancilla dephasing and dissipation rates, the decoherence rates of Floquet states can have a non-trivial dependence on the drive parameters. In the following, we discuss the drive-dependence of the decoherence rate $V_{mn}$ for the case only drive-1 is present and $n_{\rm th} = 0$.

We first discuss the drive dependence of $V_{mn}^\gamma.$ Since $\Omega_2=0$, only $K=0$ term is needed in Eq.~(\ref{eq:decoherence_rate}). Assuming a constant dissipation rate $\gamma(\omega)=\gamma$, it follows from Eqs.~(\ref{eq:heating_rate_sum},\ref{eq:decoherence_rate}) that the expression for $V^{\gamma}_{mn}$ can be simplified to,
\begin{align}
\label{eq:V_gamma_simplified}
V^{\gamma}_{mn}=\frac{\gamma}{2}[\langle c^\dagger  c\rangle_m + \langle c^\dagger c\rangle_n - 2\Re (\langle c \rangle_m \langle c \rangle_n^*)],
\end{align}
where $\langle c^\dagger c\rangle_m, \langle c \rangle_m$ are the expectation values of the corresponding operators in the $m$-th stationary eigenstate of the Hamiltonian $\tilde H_c(t)$ in Eq.~(\ref{eq:rotated_Hc}) at $\Omega_2 = 0$.

At zero or weak drive-1 amplitude, the expression for $V^{\gamma}_{mn}$ reduces to the familiar form $V^{\gamma}_{mn} \approx  (m+n)\gamma/2$. For stronger drive, we find that the decoherence rate first increases as the drive power increases and then saturates at a large power; see Fig.~\ref{fig:linewidth}(a). This initial increase and saturation at large drive power can be understood similarly as the saturation of the heating rate shown in Fig.~\ref{fig:heating}. Within a semiclassical approximation, we obtain that $V_{mn}^\gamma \approx \gamma [(m+n) (\sinh^2\phi + 1/2)+\sinh^2\phi]$, where $\sinh^2\phi$ characterizes the drive-induced squeezing of the ancilla mode; see Appendix~\ref{sec:semiclassics_decoherence_rate}. The saturation in $V_{mn}^\gamma$ is due to the saturation of the squeezing as discussed in Sec.~\ref{sec:heating_exp}.

We now discuss the dephasing-induced decoherence of the Floquet states. To simplify the analysis, we assume that the dephasing noise spectrum is approximately flat near the drive detuning frequency and near zero frequency but can have different values in these two frequency regions. The smoothness of the noise spectrum near zero frequency is already used in the Markov approximation that leads to Eq.~(\ref{eq:reduced_density_matrix}). Under these assumptions, we obtain from Eqs.~(\ref{eq:heating_rate_sum},\ref{eq:decoherence_rate}) that
\begin{align}
\label{eq:V_gamma_ph_simplified}
V_{mn}^{\gamma_{\rm ph}} &= \gamma_{\rm ph}^{\rm (hf)} (\langle(c^\dagger c)^2\rangle_m-\langle c^\dagger c\rangle_m^2+\langle(c^\dagger c)^2\rangle_n-\langle c^\dagger c\rangle_n^2) \nonumber \\
&+ \gamma_{\rm ph}(0)(\langle c^\dagger c\rangle_m-\langle c^\dagger c\rangle_n)^2
\end{align} 
Here $\gamma_{\rm ph}^{\rm (hf)}$ denotes the high-frequency dephasing rate near the drive detuning frequency as we introduced in Sec.~\ref{sec:heating_exp}; $\gamma_{\rm ph}(0)$ denotes the low-frequency dephasing rate.

For the case $\gamma_{\rm ph}(0)\gg \gamma_{\rm ph}^{\rm (hf)}$, the decoherence rate $V^{\gamma_{\rm ph}}_{mn}$ is dominated by the second term on the right-hand side of Eq.~(\ref{eq:V_gamma_ph_simplified}) which describes the pure dephasing of the Floquet states. At zero drive amplitude, $V^{\gamma_{\rm ph}}_{mn}=\gamma_{\rm ph}(0) (m-n)^2$. 
As the drive power increases, the decoherence rate $V^{\gamma_{\rm ph}}_{mn}$ sharply decreases indicating that the difference in the expectation values of the ancilla number operator in different eigenstates decreases as a function of the drive power; see Fig.~\ref{fig:linewidth}(b). This is somewhat similar to a driven two-level system where the expectation values of the operator $\sigma_z$ in the two dressed eigenstates in the rotating frame of the drive $|\pm\rangle = \cos\theta |\uparrow\rangle \pm \sin\theta |\downarrow\rangle$ become the same at strong drive: $\langle + |\sigma_z|+\rangle = \langle -|\sigma_z|-\rangle\approx 0.$ Here $|\uparrow\rangle,|\downarrow\rangle$ are the bare states of the two level system and the drive-induced mixing angle $\theta$ between the two states becomes $\pi/4$ at strong drive. Quite interestingly, this behavior is not captured by the semiclassical analysis of the driven ancilla shown as the dashed lines suggesting the strongly quantum nature of this observation.  

When the ancilla is coupled to the cavities, the cavities can inherit finite decoherence rate from the ancilla via hybridizations with the ancilla states. In next section, we will discuss this effect in detail.

\subsection{Linear susceptibilities in the presence of ancilla dissipation and dephasing}
\label{sec:susceptibilities_dissipation}
In the presence of ancilla decoherence, the linear susceptibilities $\chi $ and $X $ can be found quite generally via Eqs.~(\ref{eq:chi1_formal}, \ref{eq:X1_formal}), where the two-time correlation functions can be calculated in a way similarly to that in Ref.~\cite{mollow1969a}. Equivalently, for the specific model of the bath considered in this section, one can calculate the susceptibilities by adding a perturbation $H_f$ to $H_c$ in Eq.~(\ref{eq:reduced_density_matrix}) and calculating $\rho_c$ and the expectation value of operator $c$ to leading order in $H_f$.

\subsubsection{Susceptibilities in the steady state}
In the limit of weak damping and dephasing where Eqs.~(\ref{eq:diagonal},\ref{eq:off_diagonal}) hold, we find that the ensemble-averaged susceptibility of the ancilla in the steady state is a sum of ``partial" susceptibilities corresponding to the ancilla being in a given Floquet state weighted by the population in this state,
\begin{align}
\label{eq:averaged_susceptibilities}
&\chi^{\rm st} (\omega,\omega+K\omega_{21}) = \sum_{m} P_m^{\rm st} \chi^{\rm st} _m(\omega,\omega+K\omega_{21}) \nonumber\\
&X ^{\rm st}(\omega,2\omega_1+K\omega_{21}-\omega)  \nonumber\\
&= \sum_m P_m^{\rm st} X ^{\rm st}_m(-\omega,2\omega_1+K\omega_{21}-\omega)
\end{align}
Here the population $P_m^{\rm st}$ is the steady-state solution of the rate equation~(\ref{eq:diagonal}) with $P_m\equiv \rho_{mm}$. The partial susceptibilities $\chi^{\rm st}_m, X^{\rm st}_m$ have the same form as in Eqs.~(\ref{eq:chi1}, \ref{eq:X1}) except that now the quasienergy differences acquire an imaginary part due to bath-induced broadening of quasienergy levels. More specifically, one needs to replace $\epsilon_{mn}$ with $\epsilon_{mn}+i V_{mn}$ in the first term of Eq.~(\ref{eq:chi1},\ref{eq:X1}) and by $\epsilon_{mn}-i V_{mn}$ in the second term. To be complete, we repeat the expressions here:
\begin{align}
\label{eq:chi1_decay}
& \chi^{\rm st} _m (\omega,\omega+K\omega_{21}) \nonumber\\
&=-\sum_{n\neq m,K'}\left[\frac{c_{mn,K'-K}(c^\dagger )_{nm,-K'}}{(\omega-\omega_1) + K'\omega_{21}+(\epsilon_{mn}/\hbar+i V_{mn}) }\right. \nonumber\\
&+\left.\frac{(c^\dagger )_{mn,-K'}c_{nm,K'-K}}{-(\omega-\omega_1+K\omega_{21})+(K-K')\omega_{21}+(\epsilon_{mn}/\hbar-i V_{mn})} \right] 
\end{align}
\begin{align}
\label{eq:X1_decay}
&X ^{\rm st}_m(-\omega,2\omega_1+K\omega_{21}-\omega) \nonumber \\
&= -\sum_{n\neq m,K'}\left[\frac{ c_{mn,K'-K} c_{nm,-K'}}{-(\omega-\omega_1)+K'\omega_{21}+(\epsilon_{mn}/\hbar+i V_{mn})}\right. \nonumber\\
&\left.+\frac{c_{mn,-K'}c_{nm,K'-K} }{-[K\omega_{21}-(\omega-\omega_1)]+(K-K')\omega_{21} +(\epsilon_{mn}/\hbar-i V_{mn})} \right]. 
\end{align}

Equation~(\ref{eq:averaged_susceptibilities}) shows that the effects of ancilla decoherence on its susceptibilities are two-fold. Firstly, due to the dissipation- and dephasing-induced random hopping between ancilla Floquet states, the unitary part of the susceptibilities is now a sum over the partial susceptibilities $\chi_m,X_m$ weighted by the probabilities of the ancilla in the state $\Psi_m$. Secondly, decoherence induces finite linewidth of the ancilla transitions; as a result, its susceptibilities obtain an imaginary part (or strictly speaking, a non-unitary part). 

The expressions for the susceptibilities in Eqs.~(\ref{eq:chi1_decay},\ref{eq:X1_decay}) apply when the probe frequency $\omega$ is relatively close to one of the ancilla resonances (corresponding to the values of $\omega$ where the real part of the denominators in the susceptibilities vanish). The reasons are twofold. Firstly, the detuning of the probe frequency $\omega$ from the ancilla resonance needs to be much smaller than the bandwidth of the dephasing noise near zero frequency and other frequencies corresponding to transitions between different quasienergy levels, so that the noise spectrum can be approximated as being flat over the frequency range set by this detuning and one can use here the expression for $V_{mn}^{\gamma_{\rm ph}}$ in Eq.~(\ref{eq:decoherence_rate}). As we have found previously, the noise spectrum can however be quite frequency-dependent and fall off rapidly over a relatively narrow frequency range. For large probe detuning, one needs to take into account the actual noise spectrum in calculating $V_{mn}^{\gamma_{\rm ph}}$ and the susceptibility spectrum is generally non-Lorentzian.  

Secondly, in arriving at Eqs.~(\ref{eq:chi1_decay},\ref{eq:X1_decay}), we have neglected the interference between different Lorentzian peaks (represented by the imaginary part of different terms in the summation in Eq.~(\ref{eq:chi1_decay},\ref{eq:X1_decay})). This corresponds to neglecting the coupling between different off-diagonal elements of the density matrix as assumed in Eq.~(\ref{eq:off_diagonal}). If the finite decoherence rate results in an overlap between two Lorentzian peaks or the probe frequency $\omega$ is in between two Lorentzian peaks with similar intensity,  the interference effect is non-negligible. 

Also of interest to us are the transient time-dependent susceptibilities where the ancilla 
is initially in one of the Floquet states and has not reached the steady state. A detailed discussion is given in Appendix~\ref{sec:transient_susceptibilities}.


\subsubsection{Inverse Purcell effect in the presence of drives}
\label{sec:inverse_Purcell}
As described by Eq.~(\ref{eq:delta_kappa_formal}) in Sec.~\ref{sec:linear_response}, the imaginary part of the susceptibility $\chi (\omega,\omega)$ is related to the ancilla-induced decay of the two cavity modes. This is sometimes referred to as the ``inverse Purcell effect" in cQED where microwave cavities inherit a finite decay rate from the artificial atom  \cite{reagor2016}. We will consider in this section the ancilla-induced decay where the ancilla has reached its steady state. 

In the absence of external drives on the ancilla, the inverse Purcell effect occurs via the mixing between the cavity excitations and the ancilla excitation from the ground to the first excited state. Because of this mixing, the cavity inherits a finite decay rate from the ancilla which adds to the intrinsic decay of the cavities. The rate of this inherited decay for a cavity at frequency $\omega$ with a coupling strength $g$ to the ancilla for the case $n_{\rm th} = 0$ reads,
\begin{align}
\label{eq:delta_kappa_zerodrive}
\delta\kappa(\omega) &= 2|g|^2 \Im\chi^{\rm st} (\omega,\omega) \nonumber \\
&=|g|^2 \frac{\gamma(\omega_c)+2\gamma_{\rm ph}(0)}{(\omega-\omega_c)^2+[\gamma(\omega_c)/2+\gamma_{\rm ph}(0)]^2}. 
\end{align}

We note that the ancilla dephasing also contributes to $\delta\kappa$. This can be understood as that the ancilla dephasing noise inducing incoherent hopping between the dressed cavity mode and the dressed ancilla with a rate approximately given by $2|g/(\omega-\omega_c)|^2\gamma_{\rm ph}$ in the limit of large $|\omega-\omega_c|$. 
If $\gamma \ll 2|g/(\omega-\omega_c)|^2\gamma_{\rm ph}$, the excitation can incoherently hop back and forth between the cavity and the ancilla many times.  In the opposite limit the excitation is lost after it hops into the ancilla before it can hop back.

For a detuning $|\omega-\omega_c|$ larger than the bandwidth of the dephasing noise, $\gamma_{\rm ph}(0)$ in Eq.~(\ref{eq:delta_kappa_zerodrive}) needs to be modified. A Fermi' s golden rule calculation shows that the rate of this incoherent hopping from the cavity to the ancilla is proportional to the dephasing rate $\gamma_{\rm ph}(\omega_c-\omega)$ at the detuning frequency between the cavity and the ancilla; see Appendix~\ref{sec:two_state_model}. 

\begin{figure}[!htbp]
\centering
\includegraphics[width=5.3 cm]{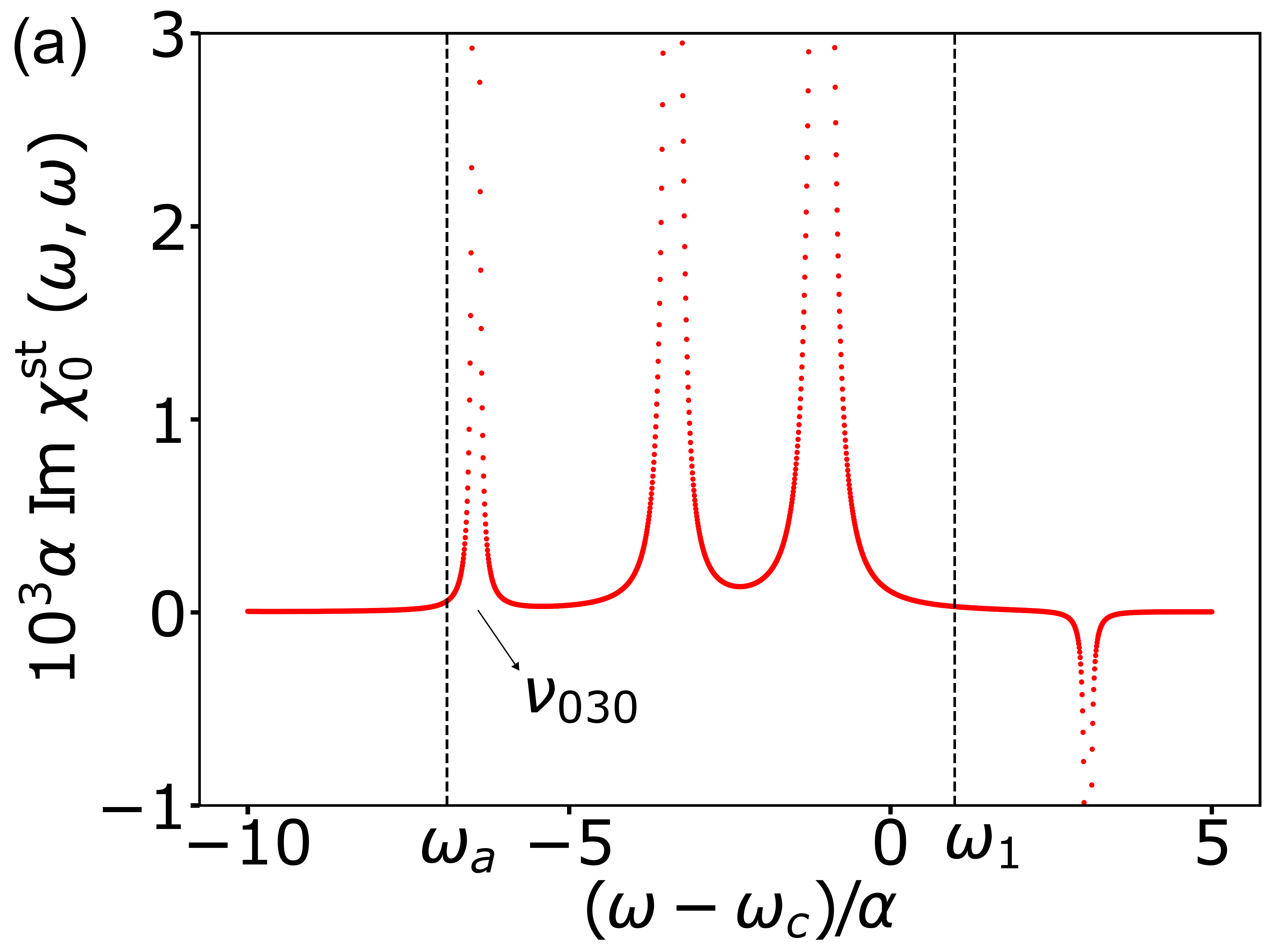} \\
\includegraphics[width = 5.5 cm]{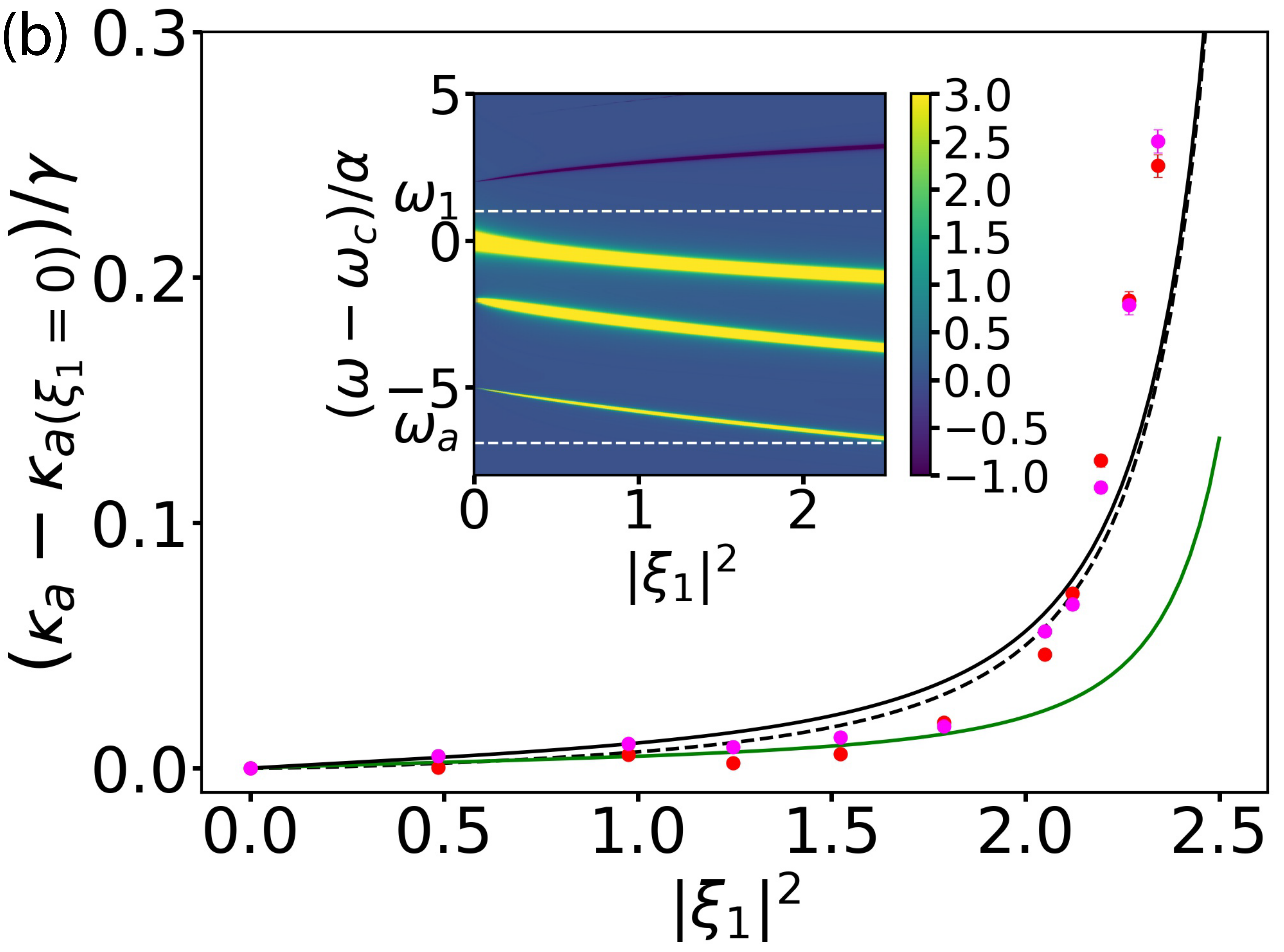} 
\caption{Illustration of the inverse Purcell effect in the presence of one drive (drive-1) on the ancilla. (a) An example of partial susceptibility spectrum $\Im \chi _0^{\rm st}(\omega,\omega)$ of the driven ancilla calculated using Eq.~(\ref{eq:chi1_decay}). The scaled drive power is $|\xi_1|^2 = 2$ and drive detuning is $\delta_1/\alpha = 1.$ (b) Comparison between the experiment (red dots) and theory (curves) on the rate of drive-induced decay of cavity $a$ as a function of the scaled drive power. As the drive power increases, the cavity becomes close to the resonance at frequency $\nu_{030}$ that excites the ancilla from the state $\Psi_0$ to $\Psi_3$ by absorbing two drive-1 photons and the cavity photon [The inset shows the susceptibility spectrum $10^3\alpha\Im \chi_0^{\rm st}(\omega,\omega)$ as a function of the drive power]. 
The vertical axis is the total decay rate of cavity $a$ minus the the same quantity in the absence of drive on the ancilla. 
The points with different colors (red and magenta) represent two sets of measurements with different cavity decay rates: $\kappa_a(\xi_1=0) = 6.13\pm0.08$~(kHz) (red) and $5.71\pm0.06$~(kHz) (magenta). In the theoretical calculation using Eq.~(\ref{eq:delta_kappa}) (black line), we used the previously-found dephasing rate $\gamma_{\rm ph}^{\rm (hf)} = \gamma/60$ to calculate the rate of dephasing-induced hopping among ancilla Floquet states, and the Ramsey dephasing rate $\gamma^{(R)}_{\rm ph}\approx \gamma$ to calculate the the rate of incoherent hopping between the cavity and ancilla. As explained in the text, this choice overestimates the latter hopping rate when the cavity is far detuned from the ancilla resonance, but matches the experiment well near the resonance which occurs at strong drive. As a comparison, the green line shows the result where an overall constant $\gamma_{\rm ph}(\omega) = \gamma_{\rm ph}^{\rm (hf)}$ was used. The dashed black line represents the partial contribution from the resonant peak at frequency $\nu_{030}$. The coupling strength $g_a$ and the detuning $\delta_a$ between cavity $a$ and the ancilla frequency $\omega_c$ is the same as in the beam-splitter experiment: $|g_a/\delta_a| = 0.047, \delta_a/\alpha = -6.9.$ }
\label{fig:absorption_spectrum}
\end{figure}

In the presence of drives on the ancilla, the cavity modes not only mix with ancilla transition from the first excited to the ground state, but also mix with transitions between higher levels of the ancilla.  Furthermore, there can also be amplification (negative damping) of the cavity modes at certain frequencies accompanied by absorption of drive excitations. In the limit of weak damping and dephasing, it follows from Eq.~(\ref{eq:averaged_susceptibilities}) that the total rate of the ancilla-induced decay (or amplification) is given by a sum of Lorentzians where each component in the sum corresponds to a resonance process that involves absorption or emission of a cavity mode excitation (see also Fig.~\ref{fig:absorption_spectrum}):
\begin{align}
\label{eq:delta_kappa}
&\delta \kappa(\omega) = \delta \kappa^\downarrow(\omega)- \delta \kappa^\uparrow(\omega) \nonumber \\
&\delta \kappa^\downarrow(\omega) = 2|g|^2 \sum_{m\neq n,K}P_m^{\rm st}\frac{|(c^\dagger )_{nm,-K}|^2 V_{mn}}{(\omega-\nu_{mnK})^2 + V_{mn}^2 } \nonumber \\ 
&\delta \kappa^\uparrow(\omega) =  2|g|^2 \sum_{m\neq n, K}P_m^{\rm st}\frac{|c_{nm,K}|^2 V_{mn}}{(\omega-\mu_{mnK})^2 + V_{mn}^2} 
\end{align}
Here $\delta\kappa^{\downarrow}$ and $\delta\kappa^{\uparrow}$ refer to the rates of transition down and up in the cavity Fock basis and they correspond to the term proportional to $\mathcal D[\sqrt{\delta \kappa^\downarrow}a] \rho $ and $\mathcal D[\sqrt{\delta \kappa^\uparrow}a^\dagger ] \rho$ in the master equation for the cavity, respectively; see Appendix~\ref{sec:full_eom}. The widths of the Lorentzians are given by $V_{mn}$ in Eq.~(\ref{eq:decoherence_rate}) and the locations of Lorentzians are given by $\nu_{mnK}$ and $\mu_{mnK}$; see below. 

Part of the linewidth in $V_{mn}$ in Eq.~(\ref{eq:delta_kappa}) comes from the dephasing of the Floquet states. Similar to the case without drive, this dephasing causes incoherent hopping between the cavity-like state and the ancilla-like state in the eigenbasis of the coupled cavity-ancilla system; see Eq.~(\ref{eq:inverse_Purcell_Rabi}). If the relaxation rate of the ancilla-like state exceeds the hopping rate from the ancilla to the cavity, the excitation that hops from the cavity to the ancilla will not have time to hop back, which effectively leads to photon loss from the cavity. A detailed analysis for this dephasing-induced hopping is given in Appendix~\ref{sec:two_state_model} based on a two-state approximation of the cavity-ancilla system. 

The set of Lorentzians in Eq.~(\ref{eq:delta_kappa}) centered at $\nu_{mnK}$ correspond to the absorption (therefore decay) of a cavity mode excitation and transition of the ancilla from Floquet state $\Psi_m$ to $\Psi_n$, accompanied by absorption or emission of $1-n+m$ drive-1 excitations and an exchange of $K$ excitations between the two drive reservoirs. The process becomes resonant when the mode frequency equals
\begin{align}
\label{eq:nu_nK}
\nu_{mnK} &= \omega_1 - K\omega_{21}+ \epsilon_{nm}/\hbar  \nonumber \\
&= (1-n+m)\omega_1 - K\omega_{21} + (E_{nm}+\delta E_{nm})/\hbar,
\end{align}
where in the second line of Eq.~(\ref{eq:nu_nK}), to better illustrate the resonance process, we have transformed from the quasienergy levels to the bare energy levels of the ancilla using Eq.~(\ref{eq:quasienergies}).
The strength of such a resonance process is given by the squared matrix element $|(c)^+_{nm,-K}|^2 \propto |\xi_1|^{|2(n-m-K+1)|}|\xi_2|^{2|K|}$ to leading order in the drive amplitudes. We emphasize that the absorption peaks centered at $\nu_{mnK}$ with the same $m$ and $n$ but different $K$ all share the same widths $V_{mn}$ because they all correspond to transitions between the same pair of Floquet states.

Similarly, the set of Lorentzians in Eq.~(\ref{eq:delta_kappa}) centered at $\mu_{mnK}$ correspond to the emission (therefore amplification) of a cavity mode excitation and transition of the ancilla from Floquet state $\Psi_m$ to $\Psi_n$, accompanied by the  absorption or emission of $1+n-m$ drive-1 photons and an exchange of $K$ excitations between the two drive reservoirs. The strength of such a resonance process is given by the squared matrix element $|c_{nm,K}|^2 \propto |\xi_1|^{|2(n-m+K+1)|}|\xi_2|^{2|K|}$ to leading order in the drive amplitudes. The process becomes resonant when the mode frequency equals
\begin{align}
\mu_{mnK}& = \omega_1  - K\omega_{21} - \epsilon_{nm}/\hbar \nonumber \\
& = (1+n-m)\omega_1 - K\omega_{21} - (E_{nm}+\delta E_{nm}) /\hbar
\end{align}

We show in Fig.~\ref{fig:absorption_spectrum} (a) an example of the spectrum $\Im \chi_0^{\rm st}(\omega,\omega)$ in the presence of one blue-detuned drive on the ancilla. The three positive Lorentzian peaks at negative $\omega-\omega_c$ from left to right represent the process of absorption of one probe photon at frequency $\omega$ and excitation of the ancilla from state $\Psi_0$ to $\Psi_1$, $\Psi_2$ and $\Psi_3$ by absorbing zero, one and two drive-1 excitations, respectively. The negative peak at negative $\omega-\omega_c$ corresponds to the emission of one probe photon and excitation of the ancilla from $\Psi_0$ to $\Psi_1$ by absorbing two drive-1 excitations. 

In an experiment, one needs to carefully choose the parameters of the drives and mode frequencies so as to avoid the aforementioned resonances. Even though the frequencies of the cavity modes may be far away from the resonance frequencies at weak drives, they may become close to resonance as the drive strengths increase because of the drive-induced AC Stark shift of the ancilla transition frequencies; see Fig.~\ref{fig:absorption_spectrum}. As a result, the ancilla-induced decay of the cavity mode that is initially away from any resonances very quickly increases as its frequency becomes close to one of the resonance frequencies. 

As the frequency of the cavity mode moves closer to the resonance, the linear response treatment of the ancilla-cavity coupling breaks down when the coupling strength becomes comparable to the detuning of the cavity mode from the resonance. Adiabatic elimination of the ancilla might also break down and one has to take into account the coherent Rabi oscillation that occurs between the cavity mode and the relevant ancilla quasienergy level. If cavity mode $a$ is close to, for instance, the resonance frequency $\nu_{0mK}$, one can neglect other resonance processes and the Hamiltonian that describes the Rabi oscillation reads,
\begin{align}
\label{eq:inverse_Purcell_Rabi}
H = (\omega_a - \nu_{0mK})a^\dagger a + g_a a  |u_m(0)\rangle \langle u_0(0)| (c^\dagger )_{m0,-K} + \rm{H.c.},
\end{align}
where we have made a unitary transformation $U = \exp[-i(c^\dagger c\omega_1t + a^\dagger a \nu_{0mK}t)]U_c(0,t) $ to the cavity-ancilla Hamiltonian~(\ref{eq:Hamiltonian}) to eliminate $H_c$ and transform into a frame that rotates at frequency $\nu_{0mK}$ for the mode $a$; $U_c$ is given in Eq.~(\ref{eq:Uc}). Here $g_a (c^\dagger )_{m0,-K}$ is the effective coupling strength of cavity mode $a$ to the transition of the ancilla from state $\Psi_0$ to $\Psi_m$, and $\omega_a - \nu_{0mK}$ is the relevant detuning. The rate of the Rabi oscillation due to the coupling depends non-trivially on the drive strengths as both the detuning $\omega_a-\nu_{0mK}$ from the resonance and the relevant Floquet matrix element $(c^\dagger )_{m0,-K}$ depend on the drive strengths. As a result, the exact condition for when the adiabaticity breaks down depends also non-trivially on the rate of turning on the drives.

\subsubsection{Comparison with experiment}
\label{sec:inverse_Purcell_exp}
To illustrate the above effect and confirm the theory, we performed an experiment where we measured the decay rate of cavity $a$ for increasing amplitude of drive-1 on the ancilla. The decay rate is measured by putting one photon in the cavity before turning on the drive on the ancilla. Then we keep the drive on for various durations of time and measure the population left in the cavity using another transmon ancilla dispersively coupled to the cavity \cite{gao2018a} after the drive has been turned off. For the chosen drive detuning, the frequency $\omega_a$ of the cavity is initially closest to the resonance frequency $\nu_{030}$ which corresponds to exciting the ancilla from state $\Psi_0$ to $\Psi_3$ by absorbing two drive photons and the photon from the cavity. As the drive amplitude increases and the resonance frequency $\nu_{030}$ moves closer to $\omega_a$ due to the AC Stark shift on the ancilla frequency, we observed a significant increase of the decay rate of the cavity; see Fig.~\ref{fig:absorption_spectrum}(b).

The generally good agreement between the theory and experiment over a wide range of drive power indicates that the observed enhanced cavity decay rate is dominated by the following two processes. Firstly, due to the hybridization between the cavity excitation and the ancilla excitation from the state $\Psi_0$ to $\Psi_3$, the cavity inherits finite decay rate from the ancilla. For this case, the inherited decay rate is close to $3|g_a (c^\dagger )_{30,0}/(\omega_a-\nu_{030})|^2 \gamma$ where $|g_a (c^\dagger )_{30,0}/(\omega_a-\nu_{030})|^2$ is the participation ratio of the ancilla excitation in the dressed cavity and $3\gamma$ is approximately the decay rate of the state $\Psi_3$ [see Fig.~\ref{fig:linewidth}(a)]. At $|\xi_1|^2  = 2.5$, the participation ratio is about 0.04.

Secondly, by the same hybridization mechanism, the ancilla dephasing induces incoherent hopping of excitation from the cavity to the ancilla which leads to cavity photon loss. To further confirm the role of this incoherent hopping, we show as the green line in Fig.~\ref{fig:absorption_spectrum}(b) the corresponding cavity decay rate where we calculated the incoherent hopping rate using the previously-found high-frequency dephasing rate $\gamma_{\rm ph}^{\rm (hf)}=\gamma/60$ in contrast to using the low-frequency Ramsey dephasing rate $\gamma^{(R)}_{\rm ph} \approx \gamma $ (shown as the black line).  We observed more than 50\% reduction in the calculated cavity decay rate at strong drive. We note that at weak drive, the green line agrees better with the experiment; this is consistent with the fact that the rate of this incoherent hopping depends on the noise spectrum at the detuning frequency between the cavity and the resonance frequency $\nu_{030}$ [see Appendix~\ref{sec:two_state_model}]. At weak drive, this detuning is of the order of $100$~MHz where the dephasing noise is relatively weak, as discussed in Sec.~\ref{sec:heating_exp}.

In the experiment, we found that the ancilla transition frequency $E_{32}/\hbar$ in the absence of drive is approximately 5~MHz lower than the value we expected from the Hamiltonian used to model the ancilla. This shift in the transition frequency would move the resonance frequency $\nu_{030}$ down by 5~MHz and thus the sharp rise in the inverse Purcell decay rate shown in Fig.~\ref{fig:absorption_spectrum}(b) would occur at a drive power smaller than what we expected.  This is consistent with what we observed in Fig.~\ref{fig:absorption_spectrum}(b). 

We also found that the decay rate $\kappa_a(\xi_1=0)$ of the cavity in the absence of the drive fluctuates by as large as 10\% from day to day. The data points with two different colors shown in Fig.~\ref{fig:absorption_spectrum}(b) represent two datasets each taken within an hour and separated in time by a day. They correspond to different decay rates $\kappa_a(\xi_1 = 0)$. The drive-induced decay rates $\kappa_a - \kappa_a(\xi_1=0)$ as shown in Fig.~\ref{fig:absorption_spectrum} are, however, very close between the two datasets and both agree qualitatively with the theory. 

Another complication is that the cavity decay rate is obtained by measuring its population over a time $\sim 1/\kappa_a$ that is much longer than the ancilla relaxation time $1/\gamma$ at least for weak drive. Immediately after the drive on the ancilla has been turned on, the ancilla is mostly in the state $\Psi_0$ (neglecting the very small thermal population $n_{\rm th} = 0.006$) and has not reached its steady-state yet; thus, the cavity decay inherited from the ancilla is not strictly speaking described by the steady-state decay rate in Eq.~(\ref{eq:delta_kappa}) but instead described by the transient susceptibility of the ancilla described in Appendix~\ref{sec:transient_susceptibilities}. However, for the range of drive power used in this measurement, the steady-state population of the excited Floquet states is less than 5\% as shown in Fig.~\ref{fig:heating}; also their contribution to the inverse Purcell decay rate is further suppressed by the large frequency detuning of the cavity from the relevant resonances. We verified numerically that the decay rate calculated from the transient susceptibility agrees within a few percent with the decay rate calculated from the steady-state susceptibility shown in Fig.~\ref{fig:absorption_spectrum}. Also, we verified that the inverse Purcell decay rate $\delta \kappa$ is dominated by $\delta \kappa^{\downarrow}$ for the same reasons as above.

To further confirm that the observed enhancement in the cavity decay rate is indeed due to the mixing of the cavity excitation with the ancilla excitation from the state $\Psi_0$ to $\Psi_3$, we simultaneously measured the population of the ancilla and the population of the cavity. We intentionally chose a relatively short ramping up and down time for the ancilla drive so that the one-photon state in the cavity  remains in the diabatic state as the drive is being turned on. As a function of the duration of the drive, we indeed observed anti-correlated oscillations of the population in the cavity one-photon state and population of the ancilla not in the ground state. Moreover, when we prepared the ancilla in the Fock state $|3\rangle$ and the cavity in the vacuum state, we also observed correlated oscillations in the ancilla ground state population and the cavity population in the one-photon state after turning on the drive.

\subsection{Ancilla-induced dephasing of the SWAP operation}
As discussed in Sec.~\ref{sec:heating}, due to the noise that accompanies the dissipation and dephasing, the driven ancilla randomly hops from one Floquet state to another. Because different Floquet state yields a different susceptibility, such hopping leads to fluctuations in the susceptibilities and thus fluctuations in the ancilla-induced interactions and frequency shift of the cavity modes. These fluctuations dephase the cavity modes as well as the quantum operations such as SWAP between them. We discuss this effect in this section and focus on the case of engineering a beam-splitter interaction between the two cavity modes. 

To capture the effect that different Floquet states yield different cavity frequency shifts and beam-splitter rates, we write down an effective Hamiltonian,
\begin{align}
\label{eq:H_eff}
H_{\rm eff} =& \sum_m \Big[\delta_{\rm {BS},m}a^\dagger a + g_{\rm {BS},m} a^\dagger b + g_{\rm {BS},m}^* a b^\dagger \Big]\otimes|\Psi_m\rangle \langle \Psi_m| \nonumber \\
\delta_{\rm {BS},m} & = (\delta\omega_{a,m}-\delta\omega_{a,0})-(\delta\omega_{b,m}-\delta\omega_{b,0})
\end{align}
Here $\delta\omega_{a,m}, \delta\omega_{b,m}$ denote the ancilla-induced frequency shift of cavity $a$, $b$, respectively, when the ancilla is in Floquet state $\Psi_m$, and $g_{\rm {BS},m}$ denotes the corresponding strength of the ancilla-induced beam-splitter interaction between them. $\delta_{\rm {BS},m}$ denotes the detuning from the beam-splitter resonance condition in Eq.~(\ref{eq:frequency_matching_BS}). Note that the definition of $\delta_{\rm{BS},m}$ is slightly different from that in Eq.~(\ref{eq:eom_BS}), as here we take into account explicitly the ancilla-induced cavity frequency shifts. We have assumed that the drive frequencies have been tuned so that when the ancilla is in Floquet state $\Psi_0$, the beam-splitter interaction is exactly on resonance, i.e., $\delta_{\rm{BS},0} = 0$. Importantly, since the frequencies of both cavities change as the transmon hops among Floquet states, it is only the difference in the frequency changes of the two cavities that matters. 

Formally, Eq.~(\ref{eq:H_eff}) can be derived by applying a unitary transformation to the original Hamiltonian in Eq.~(\ref{eq:Hamiltonian}) to eliminate the linear in $g_a,g_b$ terms and only keep slowly-rotating terms~\cite{cohen-tannoudji2004}. 

\subsubsection{Dispersion of the ancilla-induced cavity frequency shift and interaction}
The finite dispersion (difference) among different Floquet states of both the ancilla-induced frequency shift and the beam-splitter interaction strength is a consequence of the non-equidistance of the ancilla levels due to its finite anharmonicity. Level non-equidistance leads to different response to a probe when the ancilla is in different Floquet states, thereby yielding different susceptibilities. As we will show, the magnitude of this dispersion is controlled by the ratio of the ancilla anharmonicity $\alpha$ and the detunings of the drives and cavities from the ancilla frequency. In the following, we analyze this dispersion in the limit of small and large anharmonicity. 

We begin with the ancilla-induced frequency shift. In the absence of the drives on the ancilla, the ancilla-induced frequency shift of cavity $a$ when the ancilla is in the Fock state $|m\rangle$ can be found for arbitrary anharmonicity,
\begin{align}
\label{eq:frequency_shift_dispersion}
\delta\omega_{a,m} = |g_a|^2\frac{\delta_a-\alpha}{(\delta_a+m\alpha)[\delta_a+(m-1)\alpha]}
\end{align}
and similarly for $\delta\omega_{b,m}$. 

In the presence of weak drives, there is a small correction proportional to the drive power to the frequency shift in Eq.~(\ref{eq:frequency_shift_dispersion}). The result simplifies in the limit of small anharmonicity ($\alpha \ll |\delta_{1,2,a}|$):
\begin{align}
\label{eq:dw_smallalpha}
&\delta\omega_{a,m} \approx \delta\omega_{a,0}-2m\alpha|g_a/\delta_a|^2[1+\Delta_a(\xi_1,\xi_2)], \nonumber \\
& \Delta_a(\xi_1,\xi_2) = -2\alpha\sum_{j=1,2}|\xi_j|^2 [2\delta_a^{-1}+2\delta_j^{-1}+(\delta_a+\delta_j)^{-1}].
\end{align}
We note that the factor $2\alpha |g_a/\delta_a|^2$ corresponds to the cross-Kerr between cavity $a$ and the ancilla in the language of Ref.~\cite{Nigg2012} in the absence of the drives; $\Delta_a(\xi_1,\xi_2)$ can be understood as the drive-induced correction to this cross-Kerr. 
The expression for $\delta \omega_{a,0}$ is given in Eq.~(\ref{eq:chi0_weakdrive}) for weak drives but arbitrary ratio of $\alpha$ over $|\delta_{a,1,2}|$. 

In the opposite limit of large anharmonicity $\alpha \gg |\delta_{a,1,2}|$, we recover the Jaynes-Cummings model in the dispersive regime~\cite{blais2004a}:  
\begin{align}
\label{eq:dw_dispersion_largealpha}
&\delta\omega_{a,1} \approx -\delta\omega_{a,0} = -(|g_a|^2/\delta_a)[1-2\sum_{j=1,2}|\xi_j|^2(\delta_a+\delta_j)/\delta_a], \nonumber\\
&\delta\omega_{a,m>1}\approx 0
\end{align}

For the case $\omega_b-\omega_a \approx \omega_{21}$, the beam-splitter rate $g_{\rm{BS},m}$ among different Floquet states in the limit of weak drives and weak anharmonicity ($\alpha\ll |\delta_{1,2,a,b}|$) reads,
\begin{align}
\label{eq:gBS_dispersion_smallalpha}
g_{\rm{BS},m} =& g_{\rm{BS},0}\{1-2m\alpha [\delta_a^{-1}+\delta_b^{-1}+\delta_1^{-1}+\delta_2^{-1} \nonumber\\
&+(\delta_a+\delta_2)^{-1}]\}, \,\, g_{\rm{BS},0}=-2\alpha \xi_1\xi_2^* \frac{g_a^*g_b}{\delta_a\delta_b}.
\end{align} 
The result above can be conveniently found based on the four-wave mixing picture described in Appendix~\ref{sec:four_wave_mixing} and going to second order in $\alpha$. Importantly, the difference of the beam-splitter rates among different Floquet states goes as $\alpha^2$ and has the same dependence on the drive amplitudes as the beam-splitter rate itself in the considered limit. 

In the opposite limit of large anharmonicity where the ancilla is effectively a two-level system, we find that 
\begin{align}
\label{eq:gBS_dispersion_largealpha}
&g_{\rm {BS},1} = -g_{\rm {BS},0} = 2\xi_1\xi_2^*g_a^*g_b\frac{\delta_a+\delta_2}{\delta_a\delta_b},\nonumber \\
&g_{\rm{BS},m>1}=0.
\end{align}
The fact that $g_{\rm {BS},1}$ and $g_{\rm {BS},0}$ have the same magnitude but opposite sign follows from Eq.~(\ref{eq:chi1_formal}). This sign change in the beam-splitter rate $g_{\rm BS}$ from $\Psi_0$ to $\Psi_1$ is in a sense the maximal dispersion one can obtain; if the ancilla hops between these two Floquet states halfway through the SWAP operation, such hopping would completely nullify the SWAP operation.

\begin{figure}[ht]
\centering
\includegraphics[width=6cm]{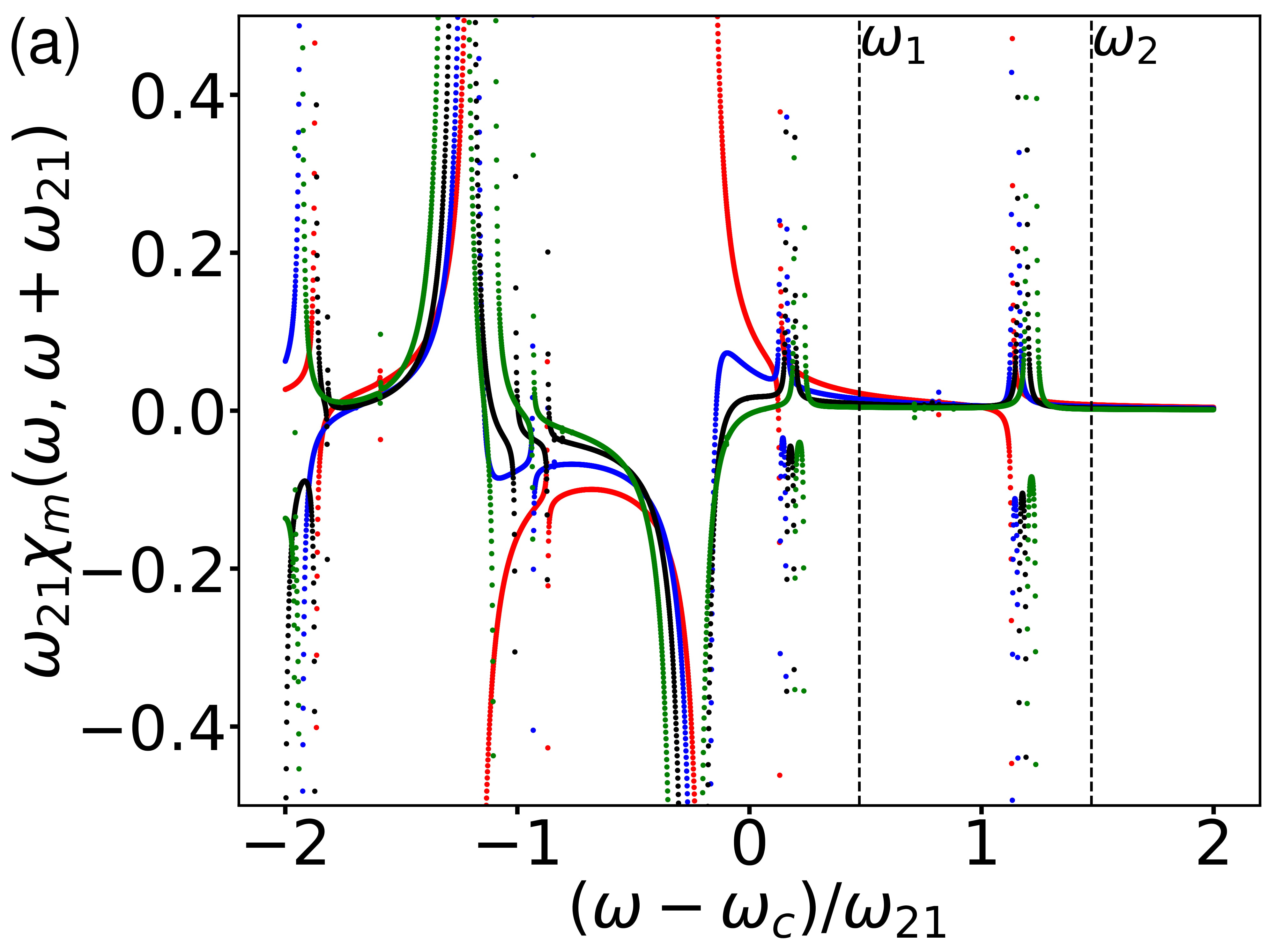} \hfill 
\includegraphics[width=6cm]{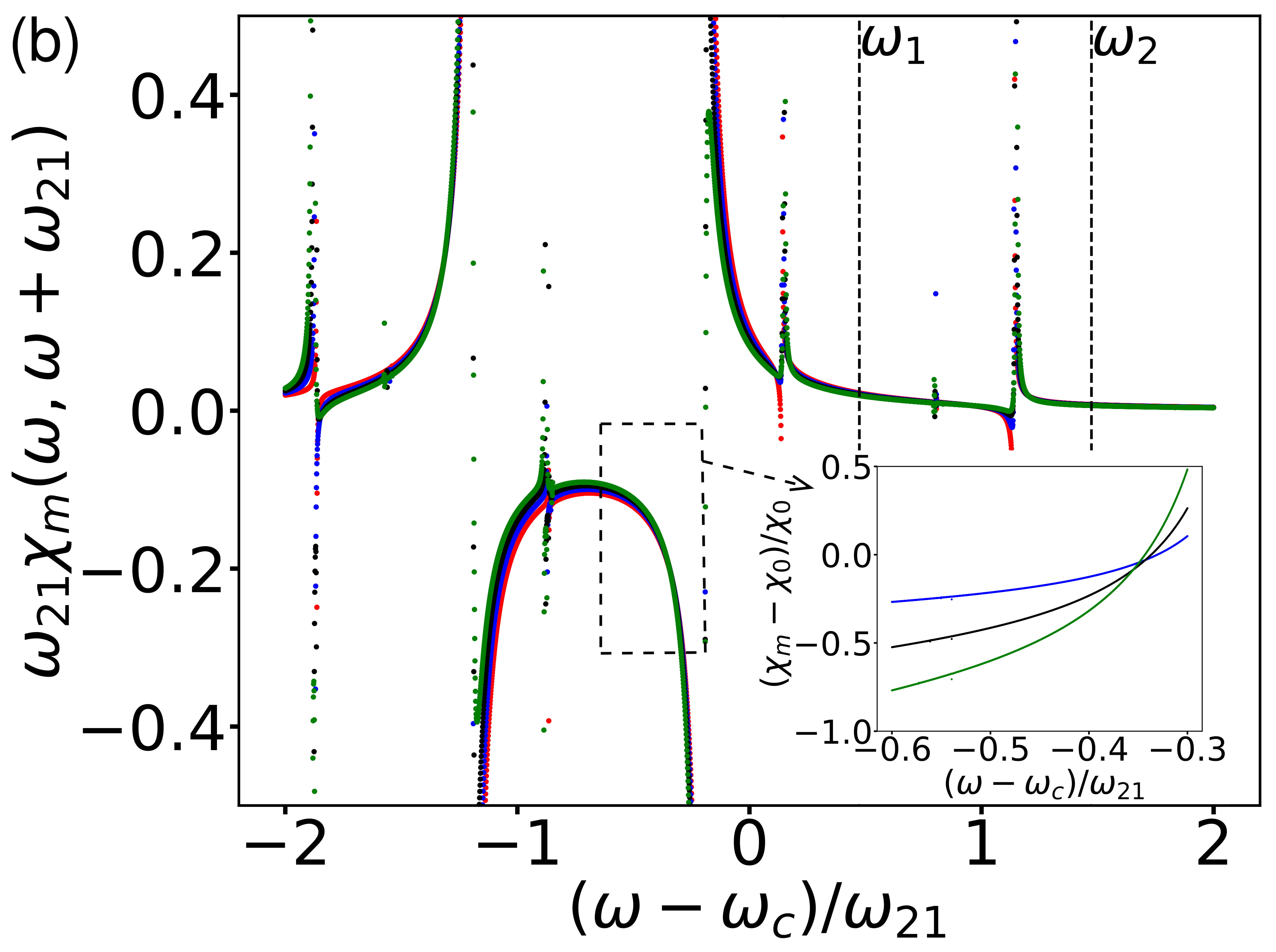} 
\caption{Comparison of the susceptibility spectra $\chi_m(\omega,\omega+\omega_{21})$ in Eq.~(\ref{eq:chi1}) for different $m$. The red, blue, black and green dots correspond to $m = 0,1,2,3$, respectively. (a) The scaled anharmonicity $\alpha/\delta_1 = 0.14$. The ratio of the drive detunings $\delta_2/\delta_1 = 3.1$. The scaled drive amplitudes $\xi_1 = 1.5, \xi_2 = 0.15.$ (b) $\alpha/\delta_1$ is reduced by a factor of 7 compared to panel (a), the drive amplitudes $\xi_1$ and $\xi_2$ are both increased by a factor of $\sqrt{7}$ so that the quantity $\alpha\xi_1\xi_2/\delta_1$ (which sets the beam-splitter rate) remains unchanged. The ratio $\delta_2/\delta_1$ remains the same. The inset shows the ratio $(\chi_m-\chi_0)/\chi_0$ for m = 1 (blue),2 (black), 3 (green) for the frequency region inside the dashed box. At particular probe frequencies, the difference between $\chi_{m\neq 0}$ and $\chi_0$ vanishes. The comparison between panels (a) and (b) clearly shows that the difference between the spectra $\chi_m(\omega,\omega+\omega_{21})$ for different $m$ goes down as the scaled anharmonicity $\alpha/\delta_1$ goes down. } 
\label{fig:dispersion_gBS}
\end{figure}

\begin{figure}[ht]
\centering
\includegraphics[width=6cm]{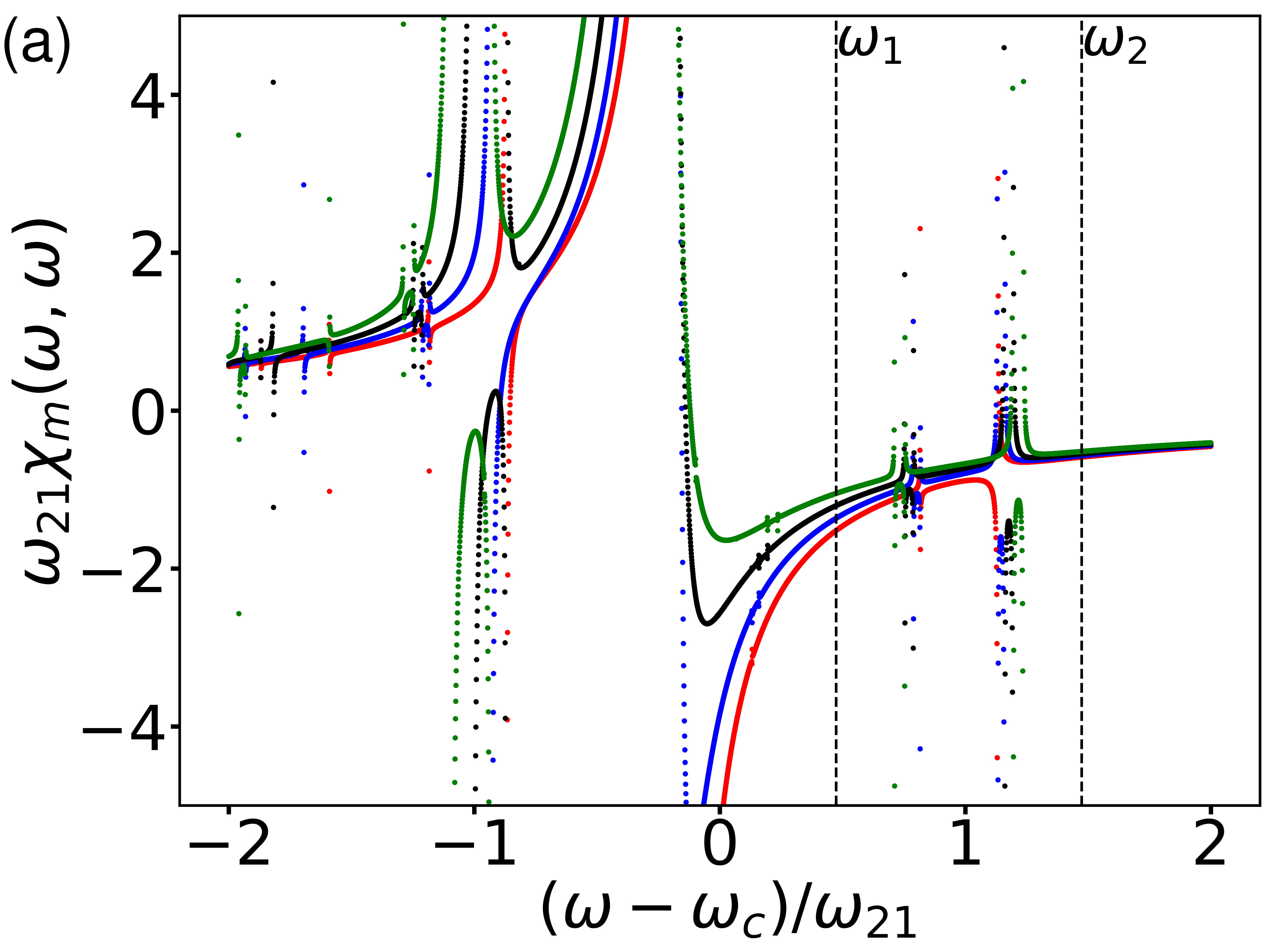} \hfill 
\includegraphics[width=6cm]{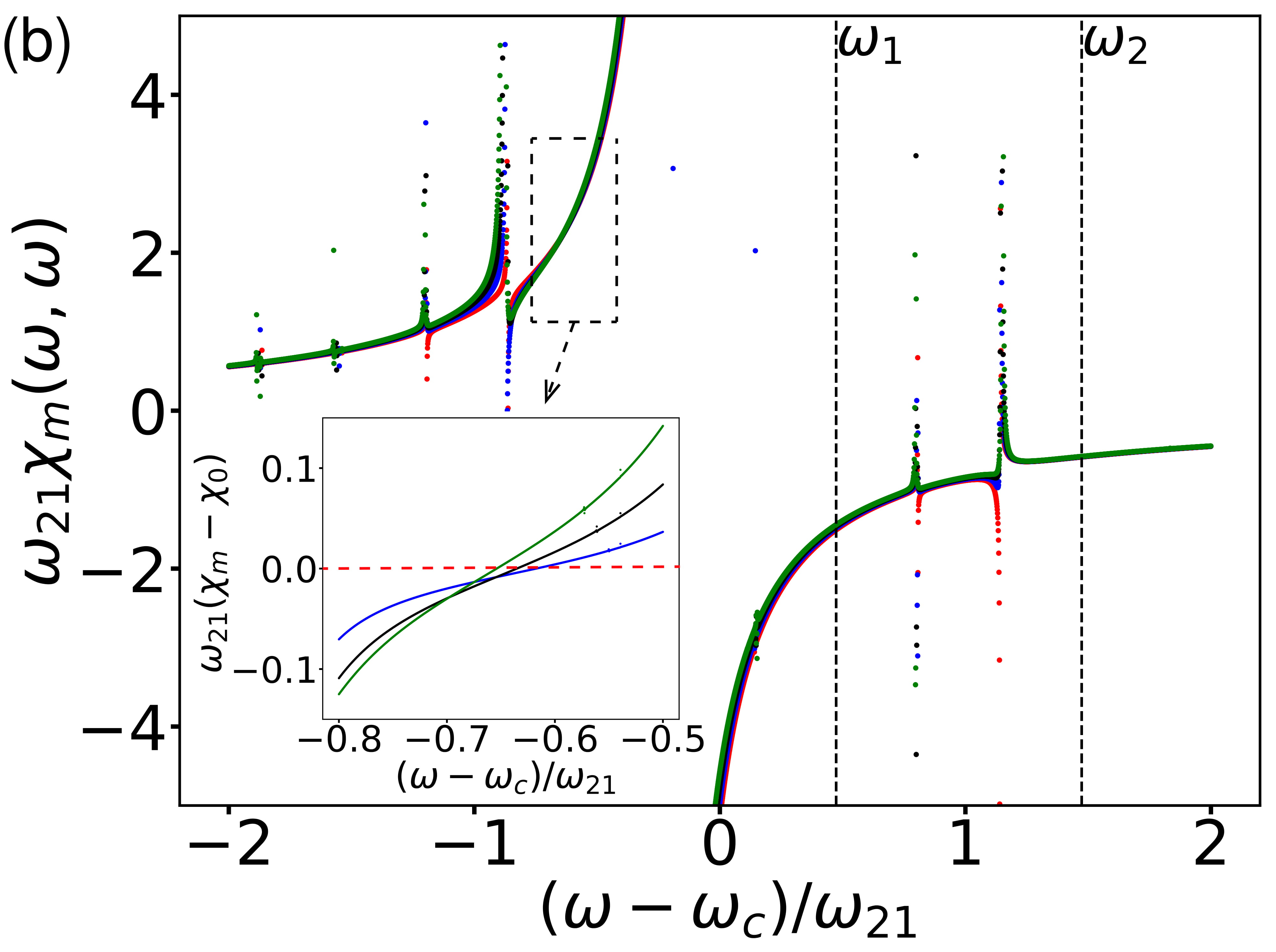} 
\caption{Comparison of the susceptibility spectra $\chi_m(\omega,\omega)$ in Eq.~(\ref{eq:chi1}) for different $m$. The red, blue, black and green dots correspond to $m = 0,1,2,3$, respectively. The parameters used in panels (a) and (b) are the same as those in Fig.~\ref{fig:dispersion_gBS}(a) and (b), respectively. The inset shows the difference $\chi_m - \chi_0$ for m = 1 (blue), 2 (black), 3 (green) for the part of the spectra inside the dashed box. It shows that $\chi_{m\neq 0}$ can become equal to $\chi_0$ at certain probe frequency $\omega$. The horizontal dashed red line is to guide the eye. The comparison between panels (a) and (b) clearly shows that the difference between the spectra $\chi_m(\omega,\omega)$ for different $m$ goes down as the scaled anharmonicity $\alpha/\delta_1$ goes down. We note that the seemingly isolated dots in the spectra including the two near zero frequency in panel (b) are a result of the finite sampling along the frequency axis which does not suffice to resolve the sharp change in the spectra in a narrow frequency range due to weak resonances.} 
\label{fig:dispersion_dw}
\end{figure}

Equations~(\ref{eq:dw_smallalpha}-\ref{eq:gBS_dispersion_largealpha}) show that the dispersion among different Floquet states of both the ancilla-induced frequency shift and the strength of the beam-splitter interaction goes down as ancilla anharmonicity decreases with respect to the drive detunings and cavity detunings from the ancilla and reaches a maximum as the anharmonicity goes to infinity. To further illustrate this result, we compare in Fig.~\ref{fig:dispersion_gBS} the dispersion of the susceptibility spectra $\chi_m(\omega,\omega+\omega_{21})$ [which relates to the ancilla-induced beam-splitter interaction via Eq.~(\ref{eq:gBS_simplified})] between the case of relatively large and small transmon anharmonicity. For smaller anharmonicity, we use stronger drives so that the strength of the beam-splitter interaction remains unchanged compared to the case of large anharmonicity. It is clearly shown that the spectra $\chi_m$ for different $m$ differ significantly from each other at large anharmonicity and become increasingly close to each other as $\alpha$ decreases. The spectra $\chi_m(\omega,\omega)$ [which relates to the ancilla-induced cavity frequency shift via Eq.~(\ref{eq:frequency_shift})] displays similiar behavior; see Fig.~\ref{fig:dispersion_dw}.

An interesting feature of the spectra $\chi_m(\omega,\omega)$ and $\chi_m(\omega,\omega+\omega_{21})$ shown in Figs.~\ref{fig:dispersion_gBS} and~\ref{fig:dispersion_dw} is that at particular probe frequencies, the difference between $\chi_{m\neq 0}$ and $\chi_0$ can vanish exactly, namely, the spectra $\chi_m$ with different $m$ can cross each other. Such crossings are a result of the interference between resonance processes canceling out the difference among ancilla Floquet states in their linear response. For instance, the spectrum $\chi_m(\omega,\omega+\omega_{21})$ in Fig.~\ref{fig:dispersion_gBS} has an ``inverted U" shape in the frequency range $\omega_c-\omega_{21}< \omega < \omega_c$ as a result of the interference between the resonance near $\omega_c$ and $\omega_c-\omega_{21}$. Due to the non-equidistance of the ancilla transition frequencies, different $\chi_m$ are shifted with respect to each other, necessarily leading to crossings of the spectra in this frequency range. For weak drives and weak anharmonicity, the location of this crossing can be found by setting the terms in the square bracket of Eq.~(\ref{eq:gBS_dispersion_smallalpha}) to zero. 

To reduce the infidelity due to ancilla hopping among Floquet states, it is desirable to minimize the magnitude of both $g_{\rm{BS},m \neq 0}-g_{\rm{BS},0}$ and $\delta_{\rm {BS},m}$ in Eq.~(\ref{eq:H_eff}). The former can be reduced by designing the system parameters and choosing the drive parameters so that the frequency of the lower-frequency cavity (cavity $a$) is close to or at the point where the spectrum $\chi_{m \neq 0}$ crosses with $\chi_0$ in Fig.~\ref{fig:dispersion_gBS}. To reduce $|\delta_{\rm{BS},m}|$, we need the change in the frequency shift $\delta\omega_{a,m}-\delta \omega_{a,0}$ of cavity $a$ due to transmon jumps to match that of cavity $b$. This requires not simply designing the cavity frequencies to be close to the crossing point in Fig.~\ref{fig:dispersion_dw}(b), but rather to be close to where $|g_a|^2 [\chi_m(\omega_a,\omega_a) -\chi_0(\omega_a,\omega_a)]$ matches the corresponding quantity for cavity $b$. For the frequency configuration used in the current experiment where the ancilla frequency is in between the two cavity frequencies, it is preferable to place the frequency of cavity $a$ to the right of the crossing point where $\chi_{m\neq 0} > \chi_0$ which is also the case for cavity $b$.


\subsubsection{Estimating the SWAP infidelity due to the fluctuating beam-splitter rate and cavity frequencies}
Random transitions of the ancilla among Floquet states lead to fluctuations in $g_{\rm BS}$ and $\delta_{\rm BS}$ in Eq.~(\ref{eq:H_eff}), which dephase the quantum operations that rely on the engineered beam-splitter interaction. In this section, we give a simple guideline on estimating the infidelity of the SWAP operation due to these fluctuations in various limits. In Appendix~\ref{sec:infidelity_scaling}, based on the scaling of this infidelity with respect to transmon anharmonicity $\alpha$, we argue that a transmon ancilla with small anharmonicity is favored in reducing the infidelity.


To quantify the effect of these fluctuations, we consider the simplest case where there is initially one photon in cavity $a$ and no photon in cavity $b$, and the ancilla is in state $\Psi_0$ at $t=0$. We are interested in the probability of finding this photon in cavity $b$ after a pre-selected SWAP time $t_{\rm SWAP} = \pi/2 |g_{\rm BS,0}|$. Neglecting the decay of the cavities, their dynamics in the one-photon manifold can be mapped to that of a spin $1/2$ subject to a Rabi drive,
\begin{align}
H = \delta_{\rm BS}(t)\sigma_z/2+g_{\rm BS}(t)\sigma_x  \nonumber
\end{align}
The eigenstates of $\sigma_z$ represent the states where the photon is in cavity $a$ or $b$. Here, we explicitly write the beam-splitter rate $g_{\rm BS}$ and the detuning $\delta_{\rm BS}$ as being time-dependent due to the random hopping of the ancilla among different Floquet states. 

One can think of the processes represented by $\delta_{\rm BS}(t)$ and $g_{\rm BS}(t)$ as Markov chains that take discrete values  $\delta_{\rm{BS},m}$ and $g_{\rm {BS},m}$, respectively. The probability to take each value is governed by the balance equation~(\ref{eq:diagonal}). Importantly, fluctuations in $\delta_{\rm BS}$ are correlated with those in $g_{\rm BS}$ as they come from the same source. In the following, we consider all $g_{\rm {BS},m}$ to be real (so $g_{\rm BS}(t)$ is also real) which can always be made so by choosing a proper gauge for the cavity modes. This is possible because all $g_{\rm {BS},m}$ share the same phase as determined by the relative phase between the two drives and the phases in the coupling constants $g_a,g_b$; see the discussion below Eq.~(\ref{eq:sol_BS}). $g_{\rm{BS},m}$ can however be positive or negative for different $m$.

To understand the effect of the fluctuations in $g_{\rm BS}$ and $\delta_{\rm BS}$, we consider two different limits. The first limit is that the correlation time of these fluctuations set by the inverse $\gamma^{-1}$ of the ancilla relaxation rate is much larger than the duration of the quantum operation: $\gamma t_{\rm SWAP}\ll1 $. In this limit, the ancilla rarely makes a transition from state $\Psi_0$ to other states during the operation. Assuming that the ancilla at most makes one transition which occurs with almost uniform probability across the operation time, one can estimate the SWAP infidelity using the following formula, 
\begin{align}
1-\mathcal F_{\rm SWAP}\approx &\sum_m \int_0^{t_{\rm SWAP}}[1-P_b^{(m)}(t_{\rm jump},t_{\rm SWAP})] \nonumber \\ &\times W_{0m}dt_{\rm jump},
\end{align}
where $P_b^{(m)}(t_{\rm jump},t_{\rm SWAP})$ denotes the probability of finding a photon in cavity $b$ at $t = t_{\rm SWAP}$ conditioned on there being one jump from Floquet state $\Psi_0$ to $\Psi_m$ at time $t = t_{\rm jump}$. After some algebra, we find that
\begin{align}
\label{eq:infidelity}
1-\mathcal F_{\rm SWAP} = &\sum_m \frac{W_{0m}t_{\rm SWAP}}{2}\Big[1+\sin(2\tilde g_{\rm BS,m} t_{\rm SWAP}) \nonumber \\ 
&\times \frac{|g_{\rm{BS},0}|}{\pi}\frac{g_{\rm BS,m}}{\tilde g_{\rm BS,m}}\frac{g_{\rm BS,m}+g_{\rm BS,0}}{\tilde g_{\rm BS,m}^2-g_{\rm BS,0}^2}\Big]
\end{align}
where $\tilde g_{\rm BS,m}$ is the Rabi rate when the ancilla is in state $\Psi_m$:  $\tilde g_{\rm BS,m}=\sqrt{\delta_{\rm BS,m}^2/4+g_{\rm BS,m}^2}$.

In the case where $g_{\rm {BS},m}$ is not very different from $g_{\rm {BS},0}$ and the detuning $\delta_{\rm{BS},m}$ is also much smaller than $g_{\rm{BS},0}$, i.e.,  $|\delta_{\rm{BS},m}|, |g_{\rm {BS},m} - g_{\rm{BS},0}|\ll |g_{\rm{BS},0}|$, the infidelity in Eq.~(\ref{eq:infidelity}) can be expanded in terms of these differences,
\begin{align}
\label{eq:infidelity_smallalpha}
1-\mathcal F_{\rm SWAP} \approx& \sum_m \frac{W_{0m}t_{\rm{SWAP}}}{4}\Bigg[\frac{3}{2}\left( \frac{\delta_{\rm{BS},m}}{2g_{\rm{BS},0}}\right)^2 \nonumber \\
&+\frac{\pi^2}{3}\left( \frac{g_{\rm {BS},m}-g_{\rm {BS},0}}{g_{\rm {BS},0}} \right)^2\Bigg]
\end{align}
Interestingly, to leading order in the expansion, the infidelity does not have terms linear in $(g_{\rm {BS},m}-g_{\rm {BS},0})/g_{\rm {BS},0}$. We verified that such quadratic dependence in fact holds for any operations including 50/50 beam-splitter, not just SWAP. 

The expression above explicitly shows that the infidelity goes down as the differences among Floquet states in the detuning $\delta_{\rm{BS},m}$ and the beam-splitter strength $g_{\rm{BS},m}$ go down. As shown in Eqs.~(\ref{eq:dw_smallalpha},\ref{eq:gBS_dispersion_smallalpha}), these differences generally decrease as the transmon anharmonicity decreases. In Appendix~\ref{sec:infidelity_scaling}, we show that for fixed SWAP rate, the rate of dephasing-induced heating goes up as the anharmonicity decreases whereas the rate of dissipation-induced heating stays the same. Overall, the infidelity decreases linearly in the ancilla anharmonicity. 

In the opposite limit $\gamma t_{\rm SWAP} \gg 1$, the ancilla undergoes many transitions among different Floquet states during the operation and quickly reaches the steady state after the drives have been turned on. Assuming that the fluctuations in $g_{\rm BS}(t)$ and $\delta_{\rm BS}(t)$ are weak and frequent over the slow time scale $\sim t\gg \gamma^{-1}$, they can be treated as white Gaussian noise. In the case where the fluctuations in $g_{\rm BS}$ dominate over those in $\delta_{\rm BS}$, we find that the probability to find a photon in cavity $b$ at time $t$ reads
\begin{align}
\label{eq:transfer_prob_fast}
P_b(t)\approx \frac{1}{2}[1-\exp(- 4\sigma_{g_{\rm BS}} t)\cos( 2|\langle g_{\rm{BS}} \rangle| t) ],
\end{align}
where $\sigma_{g_{\rm BS}} = \int_0^\infty dt \langle g_{\rm BS}(t) [g_{\rm BS}(0) - \langle g_{\rm BS} \rangle] \rangle$. Here we have neglected the initial transient dynamics of the ancilla immediately after the drives have been turned on. Note that to maximize the SWAP fidelity in this case, one needs to choose $t_{\rm SWAP}$ to be $\pi/2|\langle g_{\rm BS} \rangle|$ and the SWAP infidelity is given by $1-\mathcal F_{\rm SWAP}= [1-\exp(2\pi\sigma_{g_{\rm BS}}/|\langle g_{\rm {BS}}\rangle|)]/2$. In the limit of small $\alpha$ where Eq.~(\ref{eq:gBS_dispersion_smallalpha}) applies, one can write $g_{\rm BS}(t)$ as $g_{\rm BS}(t) = g_{\rm{BS},0} + \delta g_{\rm BS} \hat m(t)$, where $\hat m(t)$ represents the fluctuating occupation number of the ancilla and $\delta g_{\rm BS} = - g_{\rm{BS},0} \{2\alpha [\delta_a^{-1}+\delta_b^{-1}+\delta_1^{-1}+\delta_2^{-1}+(\delta_a+\delta_2)^{-1}]\}$. Assuming that the ancilla is in an effective thermal equilibrium with thermal occupation $\tilde n_{\rm th}$ (see Appendix~\ref{sec:semiclassics_heating}), we find that 
\begin{align}
\langle g_{\rm{BS}} \rangle = g_{\rm {BS},0} + \delta g_{\rm BS} \tilde n_{\rm th},\,\, \sigma_{g_{\rm {BS}}}= \delta g_{\rm BS}^2 \tilde n_{\rm th}(\tilde n_{\rm th}+1)/\gamma.
\end{align}
An interesting feature of the above result is that $\sigma_{g_{\rm BS}}$ goes down as the ancilla relaxation rate $\gamma$ goes up. This is in direct analogy to the motional narrowing effect in nuclear magnetic resonance. 

In the case where the fluctuations in $\delta_{\rm BS}$ dominate over those in $g_{\rm BS}$, we obtain that
\begin{align}
P_b(t) \approx &\frac{1}{2}\{1-\exp(-\sigma_{\delta_{\rm BS}} t/2)[\cos( 2 \tilde g_{\rm {BS}}t) \nonumber\\
&+ (\sigma_{\delta_{\rm BS}}/4\tilde g_{\rm {BS}})\sin(2\tilde g_{\rm {BS}}t)]\},
\end{align}
where  $\sigma_{\delta_{\rm BS}} =\int_0^\infty dt \langle \delta_{\rm BS}(t) \delta_{\rm BS}(0)\rangle$ and $\tilde g_{\rm {BS}}=\sqrt{ \langle g_{\rm BS}\rangle^2-(\sigma_{\delta_{\rm BS}}/4)^2}$. Here we have assumed that the drive frequencies have been tuned such that $\langle \delta_{\rm BS}\rangle = 0$. In the limit of small $\alpha$ where Eq.~(\ref{eq:dw_smallalpha}) applies, we find that the decay rate of the probability reads $\sigma_{\delta_{\rm BS}} =4 \tilde n_{\rm th}(\tilde n_{\rm th}+1)\alpha^2 [|g_a/\delta_a|^2 (1+\Delta_a) - |g_b/\delta_b|^2(1+\Delta_b)]^2/\gamma$. In the case where fluctuations in $\delta_{\rm BS}$ are comparable to those in $g_{\rm BS}$, one needs to take into account the correlations between these fluctuations and we will not go into detail here.

\section{Conclusions}
\label{sec:conclusion}
We have presented a theoretical framework supported by experiments on generating tunable bilinear interaction between bosonic modes based on the superconducting circuit architecture that consists of long-lived microwave cavities (used to store the encoded quantum information) and transmon ancillas. We showed that for a system of two off-resonant cavities coupled to a common transmon ancilla, applying two periodic drives with properly-chosen  frequencies on the ancilla allows inducing resonant beam-splitter type and two-mode squeezing interactions between the two cavities. These interactions are the essential ingredients for implementing various entangling gates between the encoded cavities such as controlled SWAP and exponential SWAP and for quantum simulations of complex bosonic systems.
The agreement between theory and experiment is excellent paving the way for using the theory for designing and optimizing future experimental devices.

The rates of the engineered bilinear interactions are shown to be related to the linear response of the driven ancilla to the coupling to the cavities. Unlike a static system, the linear response of the two-tone driven nonlinear transmon is characterized by a susceptibility matrix that relates the probe (one cavity) at one frequency to the response (the other cavity) at a generally different frequency. For a weakly nonlinear oscillator such as a transmon, the linear susceptibilities as a function of the probe frequency have distinct resonant structures at frequencies that are in resonance with ancilla transitions not only between neighboring levels but also non-neighboring levels due to the presence of external drives.

We developed a Floquet theory for the two-tone driven nonlinear ancilla that allows calculating the rate of the ancilla-mediated interaction beyond the perturbative regime. Interference of the two drives leads to a non-trivial periodic modulation of the ancilla Floquet states. These modulations, on one hand, provide the frequency mixing capability of the ancilla, on the other hand, strongly modify the spectrum of the ancilla via effects such as AC Stark shift and multi-photon resonance. The effects become particularly strong when the drive frequency difference becomes comparable to their strengths.  

We identify two major sources of the beam-splitter and SWAP infidelity due to finite coherence time of the ancilla. Even when the environment that leads to ancilla decoherence is at zero temperature, the quantum noise that accompanies ancilla dissipation and the frequency noise that leads to ancilla dephasing can cause the ancilla to incoherently hop from the Floquet ``ground" state (that adiabatically connects to the ancilla vacuum state) to other Floquet states. This is a result of the interplay of the drive on the ancilla and the noise. This noise-induced hopping leads to decoherence of the ancilla-mediated interaction between the cavities. For a relatively strong drive blue-detuned from a typical transmon ancilla by hundreds of linewidths, we found that the steady-state population in the Floquet excited states can be as large as 10\% even when the transmon only has a thermal population as low as 0.6\% in the absence of drive. 

A second important source of infidelity is that the cavity inherits finite decay rate from the typically lossier transmon ancilla via the hybridization of the cavity with the ancilla. In the presence of drives on the ancilla, the rate of this inherited decay could be strongly enhanced due to the hybridization of the cavity with transitions between higher levels of the ancilla assisted by the drives. Equally important is that ancilla dephasing causes incoherent hopping of excitations between the cavity and the ancilla via the same hybridization mechanism. Such hopping effectively cause the cavity to lose photons which we found is the dominant cavity loss mechanism when the cavity frequency is close to the drive-induced resonances that excite the ancilla.

\section{Acknowledgments}
This research was supported by NSF DMR-1609326 and by the Army Research Office (ARO) under Grant Numbers ARO W911NF-14-1-0011 and W911NF-18-1-0212. B. J. L. is supported by Yale QIMP Fellowship; Y. Y. G. was supported by an A*STAR NSS Fellowship. The views and conclusions contained in this document are those of the authors and should not be  interpreted as representing the official policies, either expressed or implied, of the Army Research Office (ARO), or the U.S. Government. The U.S. Government is authorized to reproduce and distribute reprints for Government purposes notwithstanding any copyright notation herein. Y. Z. would like to thank M. I. Dykman, Shruti Puri and Connor Hann for helpful discussions.

\appendix
\section{Expansion of the Cooper pair box Hamiltonian}
\label{sec:Cooper_pair_box}
The transmon superconducting qubit is described by a Cooper pair box Hamiltonian \cite{koch2007,girvin2014} 
\begin{align}
\label{eq:Cooperbox}
H = 4E_C \hat n^2 - E_J {\cos \hat \phi}
\end{align} 
where $E_C$ is the charging energy, and $E_J$ is the  Josephson energy. Given that the transmon qubit is operated in the regime $E_J\gg E_C$, we have neglected the offset charge and treated the phase variable $\phi$ as being compact.  The goal of this section is to show that in the limit $E_J\gg E_C$ and for not extremely strong drive, it is sufficient to expand the cosine potential and keep up to the quartic term. 

We now write the phase and number operator in terms of the creation and annihilation operator $c^\dagger$ and $c$,
\begin{align*}
\hat \phi &=\frac{1}{\sqrt{2}} \left(\frac{8E_C}{E_J}\right)^{1/4}(c+c^\dagger ) \\
\hat n &= \frac{1}{\sqrt{2}}i\left(\frac{8E_C}{E_J}\right)^{-1/4}(c-c^\dagger )
\end{align*}
We then expand the cosine potential in Eq.~(\ref{eq:Cooperbox}), and keep up to $\phi^6$. After neglecting the non-RWA terms, we obtain in terms of $c^\dagger ,c$
\begin{align}
\label{eq:Cooperbox_truncated}
&H/\hbar  \approx \omega_c c^\dagger  c - \frac{\alpha}{2} c^{\dagger2}c^2 + \frac{1}{6}\beta c^{+3}c^3 \nonumber\\
&\hbar\omega_c = \sqrt{8E_C E_J}, \,\hbar \alpha = E_C,\,\hbar\beta = \frac{1}{3}E_C\sqrt{2E_C/E_J }
\end{align}
We have neglected the correction $\mathcal O(\sqrt{E_C/E_J})$ in $\omega_c$ and $\alpha$.

Neglecting the six-th order term in Eq.~(\ref{eq:Cooperbox_truncated}) requires that 
\begin{align}
\label{eq:expansion_condition}
 \sqrt{E_C/E_J}\langle c^\dagger c\rangle \sim \frac{\alpha}{\omega_c}\langle c^\dagger c\rangle \ll 1
\end{align}
For the transmon used in the current experiment ($\omega_c/\alpha \approx 80$; see Sec.~\ref{sec:Stark_shift_exp}), the above condition is satisfied as long as the drives are not exceedingly strong, i.e. $\langle c^\dagger c \rangle \sim |\xi_{1,2}|^2 \ll \omega_c/\alpha$, where $\xi_{1,2}$ are the drive amplitudes scaled by the corresponding drive detunings. The condition to neglect the non-RWA terms in the expansion such as $c^{+3} c$ is the same as the above condition. The regime of a very strong drive where the full cosine potential needs to be taken into account was considered in a recent paper \cite{verney2018}. 

We emphasize that the nonlinear dependence of the AC Stark shift on the drive power we observed in Sec.~\ref{sec:Stark_shift} occurs at a much smaller drive power than that required for the condition in Eq.~(\ref{eq:expansion_condition}) to break down. More precisely, it occurs when $(\alpha/\delta_{1,2}) |\xi_{1,2}|^2 \sim 1$ where $\delta_{1,2}$ is the drive detuning from the ancilla frequency $\omega_c$ which is typtically much smaller than the ancilla frequency itself.

\section{Deriving the effective beam-splitter and two-mode squeezing interactions between cavity modes based on the four wave mixing picture}
\label{sec:four_wave_mixing}
In this section, we derive the ancilla-induced bilinear interaction between two cavity modes assuming weak ancilla anharmonicity. We follow the method of ``black box quantization'' presented in Ref.~\cite{Nigg2012}.

\begin{figure}
\includegraphics[width=6cm]{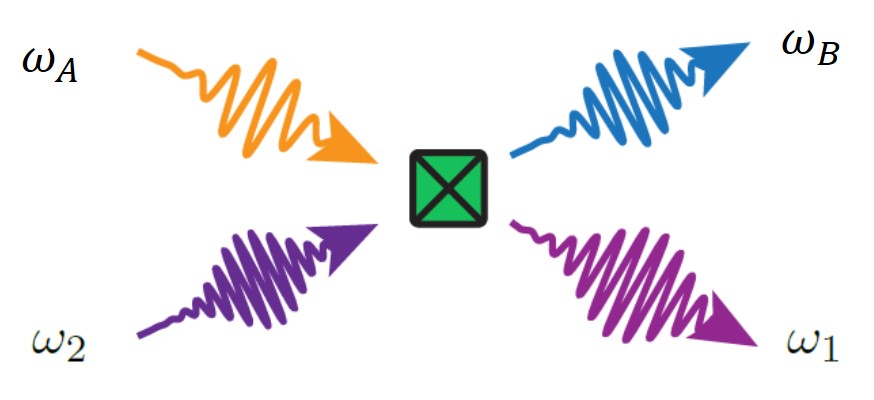}  
\caption{A schematic of the four-wave mixing process between the two cavity modes and two drives that leads to the beam-splitter interaction between the dressed cavity modes $A$ and $B$. The process becomes resonant when $\omega_A+\omega_2=\omega_B+\omega_1$. The green box represents the transmon ancilla as a frequency mixer.}
\label{fig:4_wave_mixing}
\end{figure}

The idea is to first diagonalize the quadratic part of the Hamiltonian Eq.~(\ref{eq:Hamiltonian}) excluding the drive terms. Eigenmodes $A,B,C$ of the quadratic Hamiltonian are linear combinations of bare modes $a,b,c$. In particular, for the considered regime of large detuning between the ancilla and the cavity modes, eigenmode $A$ is mostly bare mode $a$ and has a small mixing with mode $b$ and $c$, similarly for eigenmodes $B$ and $C$. We now express the annihilation operator of the bare ancilla in terms of that of the eigenmodes:
\[ c = \xi_A A+\xi_B B+\xi_C C \]
where $\xi_{A,B,C}$ characterize the mixings between bare mode $c$ and eigenmodes $A,B,C$. For weak couplings, $\xi_{A,B} \approx g_{a,b}/(\omega_{a,b}-\omega_c)$, $\xi_C\approx 1$. Because of such mixings, the nonlinear term $-\alpha c^{\dagger 2}c^2/2$ in the Hamiltonian (\ref{eq:Hamiltonian}) naturally provides a four-wave mixing between the eigenmodes. For the purpose of engineering beam-splitter and two-mode squeezing interactions between eigenmodes $A$ and $B$, we keep the following terms of interest:
\begin{align}
\label{eq:mixing}
&H_{4-wave}/\hbar \nonumber\\
&= -\frac{\alpha}{2}\left(|\xi_C|^2\xi_A^*\xi_B C^\dagger C A^\dagger B + \xi_C^2\xi_A^*\xi_B^* C^2A^\dagger B^\dagger \right)+\rm{H.c.}.
\end{align}

In terms of the eigenmodes, the drive term in Eq.~(\ref{eq:Hamiltonian}) becomes $\sum_{X=A,B,C}\xi_X X (\Omega_1^* e^{i\omega_1 t}+\Omega_2^* e^{i\omega_2 t}) +\rm{H.c.} $. Now we perform a displacement transformation on (dressed) ancilla $C$ to eliminate the drives on $C$:
\begin{align}
\label{eq:displacement_dressed}
\hat D^\dagger C \hat D &= C + \xi_1e^{-i\omega_1 t}+\xi_2e^{-i\omega_2 t}, \nonumber \\ 
\xi_{1,2}&=\Omega_{1,2}/ (\omega_{1,2} - \omega_C).
\end{align}
After such a displacement, the nonlinear mixing terms in Eq.~(\ref{eq:mixing}) provide mixing between the classical drives and (dressed) cavity modes $A$ and $B$. 

When the frequency difference of modes $A$ and $B$ match the frequency difference of the two drives, $\omega_2-\omega_1 = \omega_B-\omega_A$(see Fig.~\ref{fig:4_wave_mixing}) , there arises a resonant beam-splitter interaction between modes A and B: $g_{\rm BS} A^\dagger B + \rm{H.c.}$ where 
\begin{align}
\label{eq:g_BS_four_wave}
g_{\rm BS}\approx -\alpha \xi_A^*\xi_B\xi_1\xi_2^*.
\end{align}
When the frequency sum of the two drives matches with the frequency sum of modes $A$ and $B$, $\omega_1+\omega_2=\omega_A+\omega_B$, one obtains a two-mode squeezing interaction between $A$ and $B$: $g_{\rm TMS}A^\dagger B^\dagger  + \rm{H.c.}$ where 
\begin{align}
\label{eq:g_TMS_four_wave}
g_{\rm TMS} \approx -\alpha\xi_A^*\xi_B^* \xi_1\xi_2.
\end{align}

We emphasize that the above formulas for $g_{\rm BS}$ and $g_{\rm TMS}$ are valid for weak anharmonicity and weak drives. It is not hard to see that Eqs.~(\ref{eq:BS_weakdrive},\ref{eq:TMS_weakdrive}) obtained for weak drives reduce to Eqs.~(\ref{eq:g_BS_four_wave},\ref{eq:g_TMS_four_wave}) to leading order in the anharmonicity. The correction in  Eqs.~(\ref{eq:BS_weakdrive},\ref{eq:TMS_weakdrive}) comes from the terms we have neglected in the four wave mixing which is linear in both the cavity operator and the drive field such as $\alpha \xi_C^{*2}\xi_B \xi_1e^{-i\omega_1t}C^{\dagger 2}B$ and $\alpha \xi_2^*\xi_A^* \xi_C^2 e^{i\omega_2t} A^\dagger C^2$. These terms are off-resonant individually but together can yield a beam-splitter (or two-mode squeezing) interaction term between eigenmodes $A$ and $B$ to second order in the anharmonicity. 

The four wave mixing also naturally gives the AC Stark shift on the ancilla. Again, neglecting the off-resonant terms, we obtain
\begin{align}
\label{eq:Stark_shift_small_alpha}
H_{ss}/\hbar = -2\alpha (|\xi_1|^2 + |\xi_2|^2) C^\dagger C.
\end{align}
Eq.~(\ref{eq:Stark_shift_weakdrive}) in the main text reduces to the above expression to leading order in $\alpha$.

Lastly, we comment that by modifying the displacement transformation in Eq.~(\ref{eq:displacement_dressed}) so that operator C is displaced by the full classical response taking into account the nonlinearity, the higher-order effect of the nonlinearity and the drive can be partially captured; see Appendix~\ref{sec:semiclassical} for the classical analysis of the two-tone driven ancilla.

\section{Heisenberg-Langevin equations of motion and the equivalent quantum master equation}
\label{sec:full_eom}
Ancilla-induced dissipation on the cavities is accompanied by quantum noise. We give in this section the explicit expressions for the noise and also present the quantum master equation equivalent to the Heisenberg-Langevin equation.

We repeat the equations of motion~(\ref{eq:eom_BS}) including the noise that accompanies the dissipative part $\delta\kappa$ and $\kappa_{\rm BS}$,
\begin{align}
\label{eq:full_eom_BS}
\dot {\tilde a} &= -\frac{\delta\kappa_a}{2} \tilde a -  (ig_{\rm BS}+\kappa_{\rm BS})\tilde b+ \hat \xi_a(t) + \hat \eta_a(t) , \nonumber\\ 
\dot {\tilde b} &= - \frac{\delta\kappa_b}{2} \tilde b - (ig_{\rm BS}^* + \kappa^*_{\rm BS})\tilde a + \hat \xi_b(t) + \hat \eta_b(t).
\end{align} 
For simplicity, we have neglected the intrinsic dissipation of the cavities which simply renormalizes $\delta\kappa_{a,b}$. The Gaussian quantum noise $\hat \xi_a$ (and similarly for $\hat \xi_b$) associated with $\delta\kappa_a$ has the following properties,
\begin{align}
&[\hat \xi_a(t),\hat \xi_a^\dagger(0)] = \delta \kappa_a \delta(t), \nonumber \\
&\langle \hat \xi_a(t)\hat \xi_a^\dagger(0)\rangle = \delta\kappa_a^\downarrow \delta(t),\,
\langle \hat \xi_a^\dagger(t)\hat \xi_a(0)\rangle = \delta\kappa_a^\uparrow \delta(t),
\end{align}
where $\delta \kappa_a^{\downarrow}$ and $\delta \kappa_a^{\uparrow}$ are the ancilla-induced cavity transition down and up rates. The decay rate $\delta\kappa_a$ in the equation of motion is the difference between them: $\delta\kappa_a = \delta \kappa_a^\downarrow - \delta \kappa_a^\uparrow$. $\delta \kappa_a^{\downarrow}$ and $\delta \kappa_a^{\uparrow}$ can be calculated using Eqs.~(\ref{eq:chi1_formal},\ref{eq:delta_kappa_formal}), and they correspond to the term $\langle c^{(0)}(t)c^{(0)\dagger}(t') \rangle$ and $\langle c^{(0)\dagger}(t)c^{(0)}(t') \rangle$ in the commutator of Eq.~(\ref{eq:chi1_formal}), respectively.  Note that $\delta\kappa_a$ can become negative when the transition up rate dominates, corresponding to anti-damping of the mode; see Sec.~\ref{sec:inverse_Purcell}.

The Gaussian quantum noises $\hat \eta_a$ and $\hat \eta_b$ associated with $\kappa_{\rm BS}$ have the following properties,
\begin{align}
&[\hat \eta_a(t),\hat \eta_b^\dagger(0)] = 2\kappa_{\rm BS} \delta(t), [\hat \eta_a(t),\hat \eta_a^\dagger(0)] = 0, \nonumber \\
&\langle \hat \eta_a(t)\hat \eta_b^\dagger(0)\rangle = 2\kappa_{\rm BS}^\downarrow \delta(t),\,
\langle \hat \eta_b^\dagger(t)\hat \eta_a(0)\rangle = 2\kappa_{\rm BS}^\uparrow \delta(t),
\end{align}
where $\kappa_{\rm BS} = \kappa_{\rm BS}^\downarrow - \kappa_{\rm BS}^\uparrow$. Similar to $\delta\kappa^{\downarrow,\uparrow}$, $\kappa_{\rm BS}^\downarrow$ and $\kappa_{\rm BS}^\uparrow$, which can be calculated using Eqs.~(\ref{eq:chi1_formal},\ref{eq:kappa_BS}), correspond to the term $\langle c^{(0)}(t)c^{(0)\dagger}(t') \rangle$ and $\langle c^{(0)\dagger}(t)c^{(0)}(t') \rangle$ in the commutator of Eq.~(\ref{eq:chi1_formal}), respectively.

The quantum master equation equivalent to Eq.~(\ref{eq:full_eom_BS}) reads, 
\begin{align}
\label{eq:full_master}
\dot \rho = &-i[g_{\rm BS}\tilde a^\dagger \tilde b+g_{\rm BS}^*\tilde a \tilde b^\dagger, \rho]/\hbar \nonumber \\
&+\delta\kappa_a^\downarrow\mathcal D[\tilde a]\rho + \delta\kappa_a^\uparrow\mathcal D[\tilde a^\dagger]\rho +\delta\kappa_b^\downarrow\mathcal D[\tilde b]\rho + \delta\kappa_b^\uparrow\mathcal D[\tilde b^\dagger]\rho \nonumber \\
&-\kappa_{\rm BS}^\downarrow (\tilde a^\dagger\tilde b \rho + \rho\tilde a^\dagger \tilde b - 2\tilde b\rho \tilde a^\dagger) + \rm{H.c.} \nonumber \\
&-\kappa_{\rm BS}^\uparrow (\tilde a^\dagger\tilde b \rho + \rho\tilde a^\dagger \tilde b - 2\tilde a^\dagger \rho \tilde b) + \rm{H.c.},
\end{align}
where $\mathcal D[a]\rho \equiv a\rho a^\dagger - \{a^\dagger a,\rho\}/2.$ The equation above can be derived starting from the equation of motion for the full density matrix of cavity-ancilla system and tracing over the ancilla degree of freedom using the Born-Markov approximation.

Equation~(\ref{eq:full_master}) can be simplified and written in the following Linbladian form,
\begin{align}
\label{eq:full_Linbladian_simplified}
&\dot \rho = -i[g_{\rm BS}\tilde a^\dagger \tilde b+g_{\rm BS}^*\tilde a \tilde b^\dagger, \rho]/\hbar \nonumber \\
&+\delta\kappa_a^\downarrow \mathcal D[\tilde a + (2\kappa_{\rm BS}^\downarrow/\delta\kappa_a^\downarrow)\tilde b]\rho + (\delta\kappa_b^\downarrow- |2\kappa_{\rm BS}^\downarrow|^2/\delta\kappa_a^\downarrow)\mathcal D[\tilde b]\rho \nonumber \\
&+\delta\kappa_a^\uparrow \mathcal D[\tilde a^\dagger + (2\kappa_{\rm BS}^\uparrow/\delta\kappa_a^\uparrow)\tilde b^\dagger]\rho + (\delta\kappa_b^\uparrow- |2\kappa_{\rm BS}^\uparrow|^2/\delta\kappa_a^\uparrow)\mathcal D[\tilde b^\dagger]\rho.
\end{align}

One can interpret Eq.~(\ref{eq:full_Linbladian_simplified}) in the following way. Because the frequency of cavity $b$ is converted by the ancilla to the frequency of cavity $a$, the interference of the two effectively-degenerate modes causes a certain linear combination of the two modes as determined by the ratio of $\kappa_{\rm BS}^{\downarrow,\uparrow}$ and $\delta\kappa_a^{\downarrow,\uparrow}$ to be subject to damping (or anti-damping) whereas the orthogonal combination is immune to such loss (or gain). This is manifested in the first term of the second and third line on the right-hand side of Eq.~(\ref{eq:full_Linbladian_simplified}). The second term of the second and third line on the right-hand side of Eq.~(\ref{eq:full_Linbladian_simplified}) accounts for the fact that cavity $b$ can lose or gain photons via the ancilla without being converted into the frequency of cavity $a$. 

In the special case where the two cavity modes are degenerate in frequency [$K=0$ in Eq.~(\ref{eq:frequency_matching_BS})], we have the identity $\delta\kappa_a^{\downarrow,\uparrow}\delta\kappa_b^{\downarrow,\uparrow} = |2\kappa_{\rm BS}^{\downarrow,\uparrow}|^2$. Then the second term of the second and third line on the right-hand side of Eq.~(\ref{eq:full_Linbladian_simplified}) vanish. We have a ``bright mode"  given by the linear combination $g_a \tilde a + g_b \tilde b$ that is damped or anti-damped by the ancilla and a ``dark mode" $g_b \tilde a - g_a \tilde b$ that is neither damped nor anti-damped by the ancilla. In the case of $K\neq 0$, we generally have $\delta\kappa_a^{\downarrow,\uparrow}\delta\kappa_b^{\downarrow,\uparrow} \geq |2\kappa_{\rm BS}^{\downarrow,\uparrow}|^2$.

A similar analysis can be applied to the case where there is a non-unitary two-mode squeezing interaction between the two cavity modes as described in Eq.~(\ref{eq:eom_TMS}) of the main text. We will not discuss this in detail here.

\section{Nonlinear susceptibilities of the driven nonlinear ancilla}
\label{sec:nonlinear_susceptibilities}
Analogous to the linear susceptibilities, nonlinear susceptibilities of the driven ancilla relate to the ancilla-induced ``nonlinear frequency shift" and ``nonlinear decay" of the two cavity modes. To the leading order in the coupling between the ancilla and cavity modes, the nonlinear frequency shift corresponds to the ancilla-induced self-Kerr of each cavity mode and the cross-Kerr between them. The nonlinear decay corresponds to a decay channel where the rate of decay depends on the instantaneous energy of the modes. We will explore in this section their relation with the nonlinear susceptibilities of the ancilla. 

Let us consider two weak probes on the ancilla, one at frequency $\omega$ and the other at $\omega'$: $H_f/\hbar = -f_\omega c^\dagger  e^{-i\omega t}-f_{\omega'} c^\dagger  e^{-i\omega' t}+h.c.$. Each of them represents a cavity mode. Relevant to the ancilla-induced Kerr on the cavity modes is the third order nonlinear response of the ancilla to the probes,
\begin{align}
\label{eq:nonlinear_response}
\langle c^{(3)}\rangle = & f_\omega |f_\omega|^2 \chi^{(3)}(\omega,\omega,-\omega,\omega) e^{-i\omega t} \nonumber \\
&+f_\omega |f_{\omega'}|^2 \chi^{(3)}(\omega,\omega',-\omega',\omega) e^{-i\omega t}
\end{align}
Here we have introduced the third-order nonlinear susceptibilities $\chi^{(3)}$. In the language of nonlinear optics \cite{boyd}, the first three arguments of $\chi^{(3)}$ indicate the frequencies of probe (incident) photons, and the last argument indicates the frequency of the outgoing photon. We have adopted the convention that the positive frequency corresponds to a field with complex amplitude $f_\omega$ and the negative frequency corresponds to a field with complex amplitude $f^*_\omega$.
By construction, one can permute the positions of the first three arguments without changing the value of $\chi^{(3)}$. 

Similar to Sec.~\ref{sec:equations_of_motion}, we now substitute Eq.~(\ref{eq:nonlinear_response}) into the equations of motion of the two cavity modes Eq.~(\ref{eq:Heisenberg}), and replace $f_\omega,f_{\omega'}$ with annihilation operators $-g_a a,-g_b b$, and $\omega,\omega'$ with $\omega_a,\omega_b$, respectively.  After disregarding non-resonant terms, we obtain for cavity mode $a$
\begin{align}
\dot a &= i[H_{\rm Kerr},a] -\kappa_{aa}^{\rm nl} (a^\dagger a)a -\kappa_{ab}^{\rm nl}(b^\dagger b) a + ...\nonumber \\
H_{\rm Kerr} &=-\chi_{aa}a^{\dagger 2}a^2/2 - \chi_{bb}b^{\dagger 2}b^2/2-\chi_{ab}a^\dagger b^\dagger ab
\end{align}
where ``...'' represents other linear in $a$ terms. The ancilla-induced Kerr effects characterized by $\chi_{aa},\chi_{bb}$ and $\chi_{ab}$ and rates of nonlinear decay are related to the susceptibility $\chi^{(3)}$ via the following relations:
\begin{align}
\chi_{aa} &= |g_{a}|^4 \Re\chi^{(3)}(\omega_{a},\omega_{a},-\omega_{a},\omega_{a}), \nonumber \\
\chi_{ab} &= |g_{a}g_{b}|^2 \Re \chi^{(3)}(\omega_{a},\omega_{b},-\omega_{b},\omega_{a}) 
\end{align}
\begin{align}
\kappa_{aa}^{\rm nl} &= |g_a|^4 \Im\chi^{(3)}(\omega_a,\omega_a,-\omega_a,\omega_a),  \nonumber\\
\kappa_{ab}^{\rm nl} &= |g_ag_b|^2 \Im\chi^{(3)}(\omega_a,\omega_b,-\omega_b,\omega_a).
\end{align}

The nonlinear susceptibility $\chi^{(3)}$ can be calculated by going to higher order perturbation in $H_f$ using Eq.~(\ref{eq:Liouville}). In the absence of ancilla decoherence where $\Im \chi^{(3)}$ vanishes, the ancilla-induced Kerr can be calculated simply using time-independent perturbation theory based on the mapping to the static tight-binding Hamiltonian as described in Appendix~\ref{sec:tight-binding}. After this mapping, the total Hamiltonian including the cavity modes can be written as 
\begin{align}
\label{eq:total_tb}
H &= H_{\rm tb} + \tilde H_0 + H_I, \nonumber \\
\tilde H_0 &= (\omega_a-\omega_1)a^\dagger a + (\omega_b-\omega_1)b^\dagger b  , \nonumber \\
H_I &= (g_a a + g_b b)c^\dagger  + \rm{H.c.},
\end{align}
where we have transformed to the rotating frame of drive-1 and $H_{\rm tb}$ is given by Eq.~(\ref{eq:tight-binding}).

In order to find the ancilla-induced Kerr while the ancilla is in a tight-binding eigenstate $\Phi_{\epsilon_m}$ (which corresponds to a Floquet state $\Psi_m$ with quasienergy $\epsilon_m$; see Sec.~\ref{sec:Floquet_formulation}), one only needs to calculate the energy shift $\delta E$ of the state $|\Phi_{\epsilon_m},N_a,N_b\rangle$ due to the perturbation $H_I$ in Eq.~(\ref{eq:total_tb}). Here $N_a,N_b$ indicate the occupation number in mode $a$ and $b$. The fourth order correction $\delta E^{(4)}$ immediately gives the Kerr. More precisely, if we group terms in $\delta E^{(4)}$ according to their power in $N_a$ and $N_b$, the terms proportional to $N_a^2, N_b^2$, and $N_a N_b$ give the ancilla-induced self-Kerr and cross-Kerr, respectively:
\[ \delta E^{(4)} = -(\chi_{aa,m} /2)N_a^2- (\chi_{bb,m}/2) N_b^2 -\chi_{ab,m}N_aN_b + ...,\]
where $\chi_{aa,m}$ denotes the ancilla-induced self-Kerr of the mode-$a$ when it is in the state $\Phi_{\epsilon_m}$, and similarly for $\chi_{bb,m}, \chi_{ab,m}$.

A more rigorous way of finding the ancilla-induced Kerr is to apply a unitary transformation to the Hamiltonian in Eq.~(\ref{eq:total_tb}) that eliminates the off-resonant ancilla-cavity coupling to a given order in the coupling rate $g_{a,b}$. To the fourth order in the coupling rate, the resulting Hamiltonian contains the following terms,
\begin{align}
H_{\rm Kerr} =&-\sum_m |\Phi_{\epsilon_m}\rangle \langle \Phi_{\epsilon_m}|\bigg[ \chi_{aa,m}a^{\dagger 2}a^2/2 + \chi_{bb,m}b^{\dagger 2}b^2/2
\nonumber \\
&+\chi_{ab,m}a^\dagger b^\dagger ab\bigg]
\end{align}
 A systematic procedure to find this unitary transformation can be found in Ref.~\cite{cohen-tannoudji2004}.


\section{Mapping the time-dependent Hamiltonian $\tilde H_c(t)$ to a time-independent tight-binding model}
\label{sec:tight-binding}

\begin{figure}[ht]
\centering
\includegraphics[width=6cm]{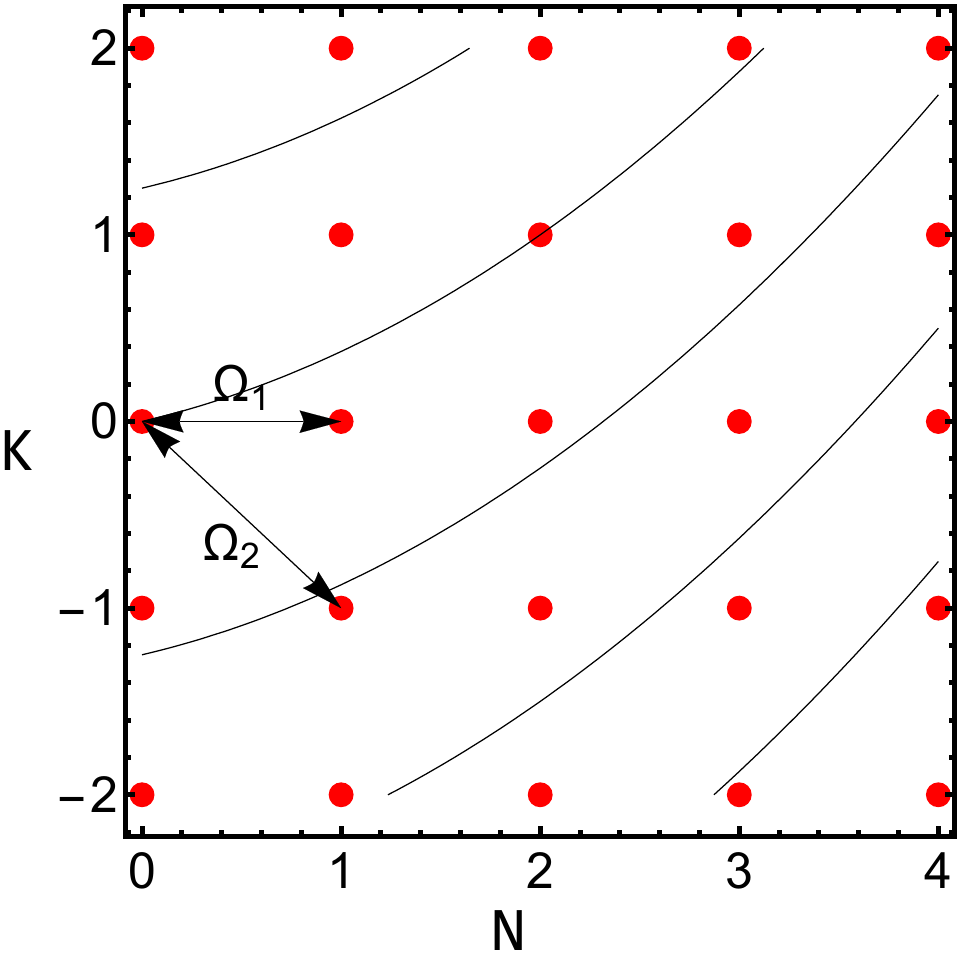} 
\caption{A 2D lattice that represents the tight-binding Hamiltonian $H_{\rm tb}$ in  Eq.~(\ref{eq:tight-binding}). Quantum number $N$ indicates the level of the ancilla in the Fock space. $K$ indicates the change in the number of excitations in drive-2 reservoir. Direct transfer of excitations between the drives and the ancilla corresponds to hopping on the 2D lattice: hopping along the horizontal direction (indicated by $\Omega_1$) means exchange of excitations between drive-1 and ancilla, and along one of diagonal directions (indicated by $\Omega_2$) means exchange of excitations between drive-2 and the ancilla. Indirect hopping along vertical direction can happen via two direct hopping along the diagonal and horizontal directions meaning exchange of excitations between drive-1 and drive-2. The solid lines are contours of constant on-site energy obtained by treating N and K as continuous variables for the parameters: $\delta_1/\alpha = 1.5, \omega_{21}/\alpha = 4$. For these parameters, site $(0,0)$ and $(2,1)$ have the same onsite energy indicating multi-photon resonance as discussed in Sec.~\ref{sec:multi-photon_resonance}.}
\label{fig:schematic_tight-binding}
\end{figure}

To find the Floquet states and quasienergies, one can write the function $u_m(t)$ in terms of its Fourier components in the Fock basis $|N\rangle$, 
\begin{align*}
u_m(t) = \sum_{N,K} f_{NK}^{(\epsilon_m)} e^{iK\omega_{21}t} |N\rangle, 
\end{align*}
One can think of the Fourier index $K$ as a second quantum number, and rewrite the state $u_m$ in a time-independent form: $u_m \rightarrow  \Phi_{\epsilon_m}=\sum_{N,K} f_{NK}^{(\epsilon_m)}  |N,K\rangle$. Then it follows from Eq.~(\ref{eq:Schrodinger_Floquet}) that the amplitudes $f_{NK}$ can be found by solving a time-independent Schr\"odinger equation with a tight-binding Hamiltonian on a 2D lattice:
\begin{align}
\label{eq:tight-binding}
& H_{\rm tb}\Phi_\epsilon = \epsilon \Phi_\epsilon,\, \Phi_\epsilon = \sum_{N,K} f_{NK}^{(\epsilon)} |N,K\rangle \nonumber \\
&H_{\rm tb}/\hbar =\sum_{NK}( \mathcal E_N/\hbar + K\omega_{21})|N,K\rangle\langle N,K| + \sqrt{N+1}  \nonumber\\
&\times\left(\Omega_1|N+1,K\rangle\langle N,K| + \Omega_2|N+1,K-1\rangle\langle N,K| + \rm{H.c.}\right)
\end{align}
where $\mathcal E_N/\hbar = -\delta_1 N - \alpha N(N-1)/2.$ Hamiltonian $H_{\rm tb}$ has on-site energy $\mathcal E_N+K\omega_{21}$ on the site $(N,K)$ and hopping between neighboring sites along certain directions; see Fig.\ref{fig:schematic_tight-binding}. Its eigenenergies are the quasienergies $\epsilon$ in the extended Brillouin zone scheme and its eigenstates $\Phi_\epsilon$ have a one-to-one correspondence to the states $u(t)$. Such a mapping to time-independent Hamiltonian by going to the Fourier basis has been studied in Ref.~\cite{Shirley1965}.

The form of tight-binding Hamiltonian $H_{\rm tb}$ can be understood by thinking of the drive fields quantum mechanically. One can think of $|N,K\rangle$ as the state in which there are $N$ excitations in the ancilla, and there has been an increase of $K$ excitations in the reservoir of drive-2 which contains a large number of excitations. Because we are in the rotating frame of drive-1, the onsite energy on site $(N,K)$ is counted from the energy of $N+K$ drive-1 excitations. The interaction between the drives and the ancilla leads to hopping between different sites. More precisely, exchange of an excitation between drive-1 reservoir and the ancilla leads to hopping between site $|N-1,K\rangle$ and $|N,K\rangle$, for each $N,K$; exchange of an excitation between drive-2 reservoir and the ancilla leads to hopping between site $|N-1,K+1\rangle$ and $|N,K\rangle$, for each $N,K$. The two types of hopping combined allow one to start from one lattice site and get to any site on the lattice. One can derive the same tight-binding Hamiltonian if one treats the driving fields quantum mechanically and expands about a large excitation number in the fields \cite{Shirley1965}. 

Hamiltonian $H_{\rm tb}$ is invariant up to a constant with respect to translation along the $K$ direction. Because it has a linear potential along K direction and the hopping strength is independent of K, translation along K direction by any number of sites simply amounts to changing the energy by integer multiples of $\hbar \omega_{21}$. One can show that eigenstates of $H_{\rm tb}$ whose eigenenergies differ by integer multiples of $\hbar \omega_{21}$ satisfy the relation: $f_{NK}^{(\epsilon)} = f_{N(K+\delta K)}^{(\epsilon+\delta K\omega_{21})}$. It follows that those eigenstates with eigenenergies differing by integer multiples of $\hbar\omega_{21}$ yield the same Floquet state $\Psi$. This explains why there are only N independent Floquet states $\Psi$ if the system Hilbert space has dimension N. In practice, one can choose any set of inequivalent states $\Phi_{\epsilon_m}$ (or $u_m$) for the analysis and understand that there are infinitely many replicas of them whose wavefunctions on the lattice are simply shifted along the $K$ direction by an integer number of sites. 

Hamiltonian $H_{\rm tb}$ allows us to calculate the rate of the ancilla-induced interaction between the two cavity modes using time-independent perturbation theory similar to how we calculate the ancilla-induced Kerr in Appendix~\ref{sec:nonlinear_susceptibilities}. Starting from Eq.~(\ref{eq:total_tb}), after a unitary transformation to eliminate the linear in $g_a,g_b$ terms, one arrives at an effective Hamiltonian in the interaction picture
\begin{align}
H_{\rm BS} = \sum_m g_{\rm BS,m} |\Phi_{\epsilon_m}\rangle \langle \Phi_{\epsilon_m-K\hbar\omega_{21}}|  a^\dagger  b + \rm {h.c.}
\end{align}
when the frequency matching condition for the beam-splitter interaction is satisfied: $\omega_b - \omega_a = K\omega_{21}$. Here the superscript $m$ in $g_{\rm BS,m}$ indicates the corresponding beam-splitter rate when the ancilla is in the state $\Phi_{\epsilon_m}$. 

Without knowing the exact form of the unitary transformation leading to $H_{\rm BS}$, one can calculate $g_{\rm BS,m}$ using degenerate perturbation theory. When the appropriate frequency matching condition is satisfied, states $|\Phi_ {\epsilon_m-K\hbar\omega_{21}},0_a,1_b\rangle$ and $|\Phi_{\epsilon_m},1_a,0_b\rangle$ are degenerate in the absence of interaction between the cavity modes and the ancilla. The beam-splitter rate $g_{\rm BS}^{(m)}$ is given by the energy splitting between the two states caused by the interaction to leading order in $g_a,g_b$. To see that one indeed arrives at the same expression for $g_{\rm BS}$ as that obtained from time-dependent Floquet theory, one just needs to replace the Fourier component $c_{mn,K}$ in $\chi_m(\omega,\omega+K\omega_{21})$ in Eq.~(\ref{eq:chi1}) with corresponding matrix element in the basis of tight-binding wavefunctions:
\[ c_{mn,K} = \langle \Phi_{\epsilon_m+K\omega_{21}}|c|\Phi_{\epsilon_n}\rangle, \]
After this replacement, Eq.~(\ref{eq:chi1}) has exactly  the same form as the energy splitting in the second-order degenerate perturbation theory.

The rate of ancilla-induced two-mode squeezing interaction can also be derived in the same way. The corresponding effective Hamiltonian in the interaction picture has the form
\begin{align}
H_{\rm TMS} = \sum_m g_{\rm TMS,m} |\Phi_{\epsilon_m}\rangle \langle \Phi_{\epsilon_m+K\hbar\omega_{21}}|  a^\dagger  b^\dagger + \rm {h.c.},
\end{align}
when the cavity mode frequencies satisfy $\omega_a+\omega_b = 2\omega_1+K\omega_{21}$. The rate $g_{\rm TMS,m}$ can be calculated as the energy splitting between the two states $|\Phi_ {\epsilon_m+K\hbar\omega_{21}},0_a,0_b\rangle$ and $|\Phi_{\epsilon_m},1_a,1_b\rangle$ using degenerate perturbation theory.

\section{Semiclassical analysis of the two-tone driven ancilla}
\label{sec:semiclassical}
In this section, we analyze the semiclassical dynamics of the driven ancilla starting from the Hamiltonian $\tilde H_c$ in Eq.~(\ref{eq:rotated_Hc}) in the rotating frame of drive-1.  The analysis applies to the regime where the drives on the ancilla are relatively strong or the drive detuning is much larger than the ancilla anharmonicity. We note that the experiments presented in the main text are not quite in the semiclassical regime yet, but the qualitative features of the results are already captured by the semiclassical analysis. 

For the purpose of semiclassical analysis, it is convenient to transform from the creation and annihilation operators $c^\dagger ,c$ to the quadratures $P,Q$ in the rotating frame,
\begin{equation}
Q = \sqrt{\frac{\lambda}{2}}(c+c^\dagger ), P = -i\sqrt{\frac{\lambda}{2}}(c-c^\dagger ),
\end{equation}
where we have introduced the dimensionless Planck constant \[ \lambda = \alpha/2|\delta_1|.  \] Note that $\alpha \propto \hbar$.

In terms of $P$ and $Q$, Hamiltonian $\tilde H_c(t)$ can be expressed as a dimensionless Hamiltonian $g(Q,P,t)$\cite{dykman2012},
\begin{align}
\label{eq:g}
\tilde H_c/\hbar &= \frac{2|\delta_1|^2}{\alpha} g(Q,P,t), \nonumber\\
g(Q,P,t) &= -\frac{1}{4}[Q^2+P^2+\sgn(\delta_1)]^2+ \overline \Omega_1 Q 
 \nonumber \\
&+ \overline \Omega_2 \cos(\omega_{21}t+\phi_{21})Q - \overline \Omega_2\sin(\omega_{21}t+\phi_{21})P,
\end{align} 
where the dimensionless driving amplitudes $\overline \Omega_{1,2} = \sqrt{\alpha}|\Omega_{1,2}|/|\delta_1|^{3/2}$. Without loss of generality, we have chosen a gauge for the ancilla $c$ such that $\Omega_1$ is real and positive, and $\Omega_2 = |\Omega_2|e^{-i\phi_{21}}$. 

Here we emphasize that the squared dimensionless drive amplitude $|\overline \Omega_{1,2}|^2 = \alpha|\Omega_{1,2}/\delta_{1,2}|^2/|\delta_{1,2}|$ can be understood as the ratio of the drive-induced frequency shift and the drive detuning (see Eq.~(\ref{eq:Stark_shift_weakdrive})). In the presence of only drive-1, $\overline \Omega_1$ is the only parameter that controls the classical dynamics; see below.

\subsection{Classical equations of motion}
The Hamilton equations of motion of the Hamiltonian $g(Q,P,t)$ read,
\begin{align}
\frac{dQ}{d\bar t} = \frac{\partial g}{\partial P} - \frac{\bar\gamma}{2}Q, \nonumber \\
\frac{dP}{d\bar t}= -\frac{\partial g}{\partial Q} -\frac{\bar\gamma}{2}P,
\end{align}
where the dimensionless time $\bar t = t/|\delta_1|$. We have also added a dissipation with a scaled rate $\bar \gamma = \gamma/|\delta_1|$. Note that here $\gamma$ is the energy decay rate in accordance with the definition used in the main text; see Sec.~\ref{sec:dissipation}. 

On a time scale set by the decay rate $\gamma$, the nonlinear oscillator reaches a steady state which is a periodic orbit with period $\tau = 2\pi/\omega_{21}$ in the phase space. Such a periodic steady-state solution can be found by substituting the Fourier decomposition of $Q$ and $P$, $Q(t) = \sum_K Q_K e^{iK\omega_{21}t}, P(t) = \sum_K P_K e^{iK\omega_{21}t}$ into the equations of motion and solving for the Fourier components. 

\begin{figure}[ht]
\centering
\includegraphics[width=6cm,height=6cm]{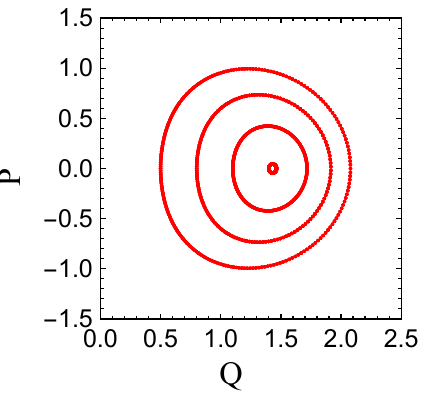} 
\caption{Spectroscopic Poincar\'e section of the Hamiltonian $g(Q,P,t)$. The scaled driving amplitudes $\overline \Omega_1 = 0.89, \overline \Omega_2 = 4.5,$. The scaled frequency $\omega_{21}/|\delta_1| = 15.$}
\label{fig:Poincare_section}
\end{figure}

In the absence of dissipation, the oscillator generally performs quasi-periodic motion in the phase space in the presence of the two drives. Such quasi-periodicity can be understood as follows. When there is only drive-1 ($\Omega_2 = 0$), given an initial condition, the oscillator occupies a periodic orbit of the Hamiltonian $g$ at $\Omega_2 = 0$. Turning on drive-2, the orbit is modulated by the second drive at frequency $\omega_{21}$ and becomes quasiperiodic unless frequency $\omega_{21}$ is commensurate with the period of this orbit. A convenient way to present such quasi-periodic motion is via the stroboscopic Poincar\'e section where the position of the oscillator in the phase space is recorded every period $\tau$; see Fig.~\ref{fig:Poincare_section}. Quasi-periodic orbits in the phase space fill up a full loop in the Poincar\'e section and form an island~\cite{ketzmerick2010}. The center of the island is the periodic steady-state to which the system flows to in the presence of a weak damping \footnote{There can also form other islands of regular loops in the Poincar\'e  section due to nonlinear resonance; not shown in Fig.~\ref{fig:Poincare_section}}.

\subsection{Semiclassical quantization in the presence of one drive}
\label{sec:one_drive_quantization}
When there is only drive-1, Hamiltonian $g$ is time-independent; its semiclassical eigenergies can be found using the Bohr-Sommerfeld quantization rule. For a relatively strong drive-1 or large detuning ($\lambda \ll 1$), one can also expand the Hamiltonian $g$ about one of the stable equilibrium positions of the oscillator in the phase space, and then quantize the fluctuations around the equilibrium position; cf. \cite{dykman2012}. The equilibrium position corresponds to the classical steady state in the limit $\gamma\rightarrow 0$. We will focus here on the case of positive detuning $\delta_1>0$ where there is only one stable equilibrium position but will keep the formulation general so that it also applies to negative detuning. 

\begin{figure}
\includegraphics[width = 6 cm]{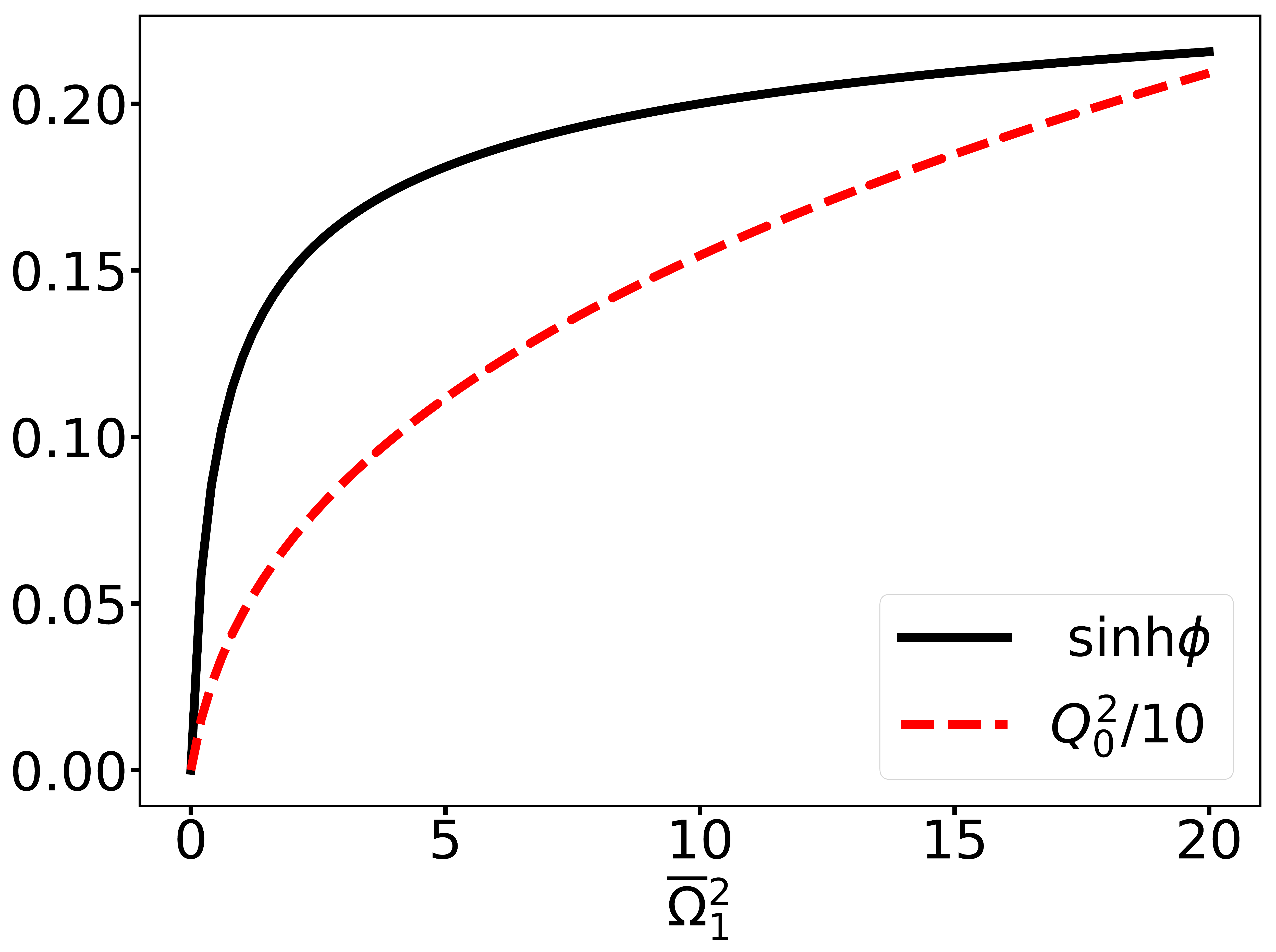}
\caption{The squeezing parameter $\sinh\phi$  and the squared classical response $Q_0^2$ as a function of the scaled drive power $\overline \Omega_1^2$. In terms of $\xi_1$ defined in the main text, $\overline \Omega_1^2 = \alpha|\xi_1|^2/\delta_1.$}
\label{fig:squeezing}
\end{figure}

A fully equivalent way of finding the semiclassical eigenstates and eigenenergies near the equilibrium position is to start from the Hamiltonian $\tilde H_c$ (at $\Omega_2 = 0$) in terms of $c,c^\dagger $; see Eq.~(\ref{eq:rotated_Hc}). One first makes a displacement transformation to eliminate the linear in $c$ term: 
\begin{align}
\label{eq:displacement_semiclassical}
c \rightarrow c + Q_0/\sqrt{2\lambda}.
\end{align}
$Q_0$ is the value of $Q$ at the equilibrium position of the oscillator and is given by one of the real solutions to the cubic equation: 
\begin{align}
\label{eq:equation_for_Q0}
Q_0(Q_0^2+\sgn(\delta_1)) = \overline \Omega_1. 
\end{align}
Note that the displacement $Q_0/\sqrt{2\lambda}$ is denoted as $\overline \xi_1$ in the main text (neglecting the phase in the drive amplitude $\Omega_1$). In terms of the scaled drive strength $\xi_1$ used in the main text, the above equation becomes $\overline \xi_1   (\alpha|\overline \xi_1|^2/\delta_1 + 1) = \xi_1$. The dependence of $Q_0^2$ on the scaled drive power $\overline \Omega_1^2$ is shown in Fig.~\ref{fig:squeezing} for $\delta_1>0$.

For negative detuning, there can be three real roots where two of them correspond to stable equilibrium positions. For the considered case of positive detuning, there is only one real root. The value of P at the equilibrium position is $P_0 = 0$ since the Hamiltonian is even in $P$. Note that such a displacement differs from the displacement discussed in Sec.~\ref{sec:weak_anharmonicity} as it takes into account the nonlinearity of the oscillator. 

After the displacement transformation, the Hamiltonian $\tilde H_c$ becomes,
\begin{align}
\label{eq:displaced_Hamiltonian}
\tilde H_c/\hbar = -\delta_1c^\dagger c -2\delta_1 Q_0^2 c^\dagger c - \delta_1 Q_0^2 (c^{\dagger2}+c^2)/2 + ...
\end{align}
where $...$ represent terms that are non-quadratic in $c$ and $c^\dagger$. We note that at the quadratic level, the drive induces frequency shift and squeezing of the ancilla mode. This Hamiltonian has a similar form as the Hamiltonian in Eq.~(\ref{eq:H_D}) of the main text but does not have linear in $c$ or $c^\dagger$ terms and has $\xi$ replaced by $Q_0/\sqrt{2\lambda}$. 

Next, we diagonalize the quadratic part of the above Hamiltonian via a squeezing transformation
\begin{align}
\label{eq:squeezing}
c =  c_{\rm aux} \cosh\phi - c_{\rm aux}^+\sinh\phi,
\end{align}
where $c_{\rm aux},c^\dagger _{\rm aux}$ can be thought of as the annihilation and creation operators of an auxiliary mode that corresponds to the small vibrations about the equilibrium position of the oscillator. The squeezing angle $\phi$ in Eq.~(\ref{eq:squeezing}) expressed in terms of the frequency $\omega_{\rm aux}$ (see below) of the auxiliary mode obeys
\begin{align}
\sinh \phi  = \sgn(Q_0) \left [ \frac{(2Q_0^2+\sgn(\delta_1))\sgn(Q_0)-\omega_{\rm aux}/|\delta_1|}{2\omega_{\rm aux}/|\delta_1|} \right]^{1/2}
\end{align}
The squeezing parameter $\sinh\phi$ is only a function of $Q_0$ which is controlled by the dimensionless drive amplitude $\overline \Omega_1$ through Eq.~(\ref{eq:equation_for_Q0}). 

The dependence of the squeezing parameter $\sinh\phi$ on the scaled drive power for positive detuning $\delta_1>0$ is shown in Fig.~\ref{fig:squeezing}. For small $\overline \Omega_1$, it is linear in $\overline \Omega_1^2$; for large $\overline \Omega_1$, it saturates to a value equal to $[(2-\sqrt{3})/2\sqrt{3}]^{1/2}$. This saturation can be understood as a result of the competition between the drive-induced frequency shift [the second term in Eq.~(\ref{eq:displaced_Hamiltonian})] and squeezing [the third term in Eq.~(\ref{eq:displaced_Hamiltonian})]. For positive detuning ($\delta_1>0$), the drive-induced frequency shift pushes the effective ancilla frequency further away from the drive frequency which in turn limits the amount of squeezing. Since both the frequency shift term and squeezing term in Eq.~(\ref{eq:displaced_Hamiltonian}) grow linearly in $Q_0^2$, the amount of squeezing one can achieve saturates at large drive power. 

For a large detuning over anharmonicity ($|\delta_1|\gg \alpha$) or a relatively strong drive, one can neglect the nonlinear in $c_{\rm aux},c_{\rm aux}^+$ terms in $\tilde H_c$. The resulting Hamiltonian in terms of $c_{\rm aux}$ and $c^\dagger _{\rm aux}$ reads,
\begin{align}
\label{eq:Hamiltonian_auxiliary}
\tilde H_c/\hbar \approx -\sgn(Q_0)\omega_{\rm aux} c_{\rm aux}^+c_{\rm aux},
\end{align}
where the frequency of the auxiliary mode is
\begin{align}
\label{eq:omega_aux}
\omega_{\rm aux} &= |\delta_1|\sqrt{( Q_0^2 + \sgn(\delta_1))(3Q_0^2+\sgn(\delta_1))}.
\end{align}
We note that eigenenergies of $\tilde H_c$ can be negative, and excited states can have lower eigenenergies than the ground state, a consequence of being in the rotating frame of the drive. The frequency $\omega_{\rm aux}$ depends on the drive amplitude through the drive dependence in $Q_0$. For the considered case of positive detuning, $Q_0 > 0$; $\omega_{\rm aux}$ monotonically increases as the drive amplitude increases. 

Equation~(\ref{eq:omega_aux}) allows us to calculate the approximate AC Stark shift of the transition frequency $E_{10}$ of the ancilla in the regime where the semiclassical analysis applies.  $\omega_{\rm aux}$ relates to the AC Stark shift $\delta E_{10}$  througth the relation 
\begin{align}
\delta E_{10}/\hbar \approx \delta_1 - \omega_{\rm aux}. 
\end{align}
We note that $\delta E_{10}$ starts off being linear in the drive power [see also Eq.~(\ref{eq:Stark_shift_weakdrive})], then becomes sublinear, and for a relatively strong drive, it becomes proportional to $\Omega_1^{2/3}$.

In the presence of two drives, in principle, one can follow the same procedure to find semiclassical quasienergy states and quasienergies of Hamiltonian $\tilde H_c(t)$ or $g(Q,P,t)$ in Eq.~(\ref{eq:g}) by expanding the Hamiltonian about the ``equilibrium position" (now a periodic orbit in the phase space, or a point in the Poincar\'e section; see Fig.~\ref{fig:Poincare_section}). The resulting Hamiltonian does not have terms linear in $c,c^\dagger$ but is still periodic in time with periodicity $\tau$. Similar to the case of one drive, one can find the approximate quasienergies by diagonalizing the quadratic part of the resulting Hamiltonian. One can take the point of view that the quasiperiodic orbits surrounding such a equilibrium position now become quantized quasienergy levels just like the case of static Hamiltonian where periodic orbits surrounding a stable equilibrium position form quantized energy levels \cite{gutzwiller1971}. Going away from the equilibrium position, there also exists a general semiclassical quantization scheme for Floquet systems \cite{breuer1991,bensch1992}; such an analysis is beyond the scope of this paper. 

\subsubsection{Dissipation- and dephasing-induced heating in the semiclassical regime}
\label{sec:semiclassics_heating}
Because of the squeezing transformation from $c$ to $c_{\rm aux}$, transition down in the ladder of Fock states of operator $c$ (due to dissipation) can correspond to both transition up and down in the ladder of Fock states of operator $c_{\rm aux}$. This leads to an effective ``heating'' of the auxiliary mode as discussed in Sec.~\ref{sec:dissipation} even when the bath that leads to the damping of the ancilla is at zero temperature. Importantly, the squeezing vanishes if $\alpha = 0$ where the system is linear. 

Likewise, due to the displacement transformation in Eq.~(\ref{eq:displacement_semiclassical}), dephasing noise leads to transitions between different eigenstates of the auxiliary modes. To see the effects of ancilla dissipation and dephasing on the auxiliary mode, we substitute $c$ in Eq.~(\ref{eq:master_equation_Markovian}) with $c_{\rm aux} \cosh\phi - c_{\rm aux}^+\sinh\phi+Q_0/\sqrt{2\lambda}$  and $\tilde H_c$ with Eq.~(\ref{eq:Hamiltonian_auxiliary}). Only keeping the terms in the Lindbladian that contain equal number of $c_{\rm aux}$ and $c^\dagger _{\rm aux}$, we obtain the quantum master equation for the auxiliary mode to be~\cite{Dykman2011},
\begin{align}
\label{eq:master_equation_auxiliary}
&\dot \rho = -i[\tilde H_c,\rho]/\hbar + \bigg\{\mathcal D[\sqrt{\gamma^\downarrow_{\rm aux}} c_{\rm aux}] + \mathcal D[\sqrt{\gamma^\uparrow_{\rm aux}}c^\dagger _{\rm aux}] \nonumber \\
&+\mathcal D[\sqrt{\gamma_2}c^2_{\rm aux}] + \mathcal D[\sqrt{\gamma_2}c^{\dagger 2}_{\rm aux}]+\mathcal D[\sqrt{2\tilde \gamma_{\rm ph}}c^\dagger _{\rm aux}c_{\rm aux}]\bigg\}\rho, \nonumber \\
\end{align}
The transition down and up rates between neighboring levels of the auxiliary mode read,
\begin{align}
\label{eq:semiclassical_gamma_up}
\gamma^{\downarrow}_{\rm aux} = (n_{\rm th}+1)\gamma \cosh^2\phi + n_{\rm th}\gamma \sinh^2\phi + \gamma_{\rm ph}^{\rm (hf)} Q_0^2/\lambda, \nonumber \\
\gamma^{\uparrow}_{\rm aux} = (n_{\rm th}+1)\gamma \sinh^2\phi + n_{\rm th}\gamma \cosh^2\phi +  \gamma_{\rm ph}^{\rm (hf)} Q_0^2/\lambda
\end{align}
We emphasize that even at $n_{\rm th}=0$, there is a finite transition up rate $\gamma^\uparrow_{\rm aux}$. The transition up and down rates induced by the dephasing noise are the same as a result of the assumed symmetric noise spectrum in Eq.~(\ref{eq:master_equation_Markovian}). Note that we have replaced in Eq.~(\ref{eq:semiclassical_gamma_up}) $\gamma_{\rm ph}$ with $\gamma_{\rm ph}^{\rm (hf)}$ as introduced in Sec.~\ref{sec:heating_exp} of the main text to emphasize that the rates of transitions caused by dephasing is determined by the dephasing noise at the drive detuning frequency which as we found is much smaller than the dephasing rate obtained from the Ramsey or spin echo measurement.

In addition, the dephasing noise also induces two-photon transitions with a rate $\gamma_2 = 2\gamma_{\rm ph}^{\rm (hf)}\sinh^2\phi  \cosh^2\phi$. However, this rate is small compared to the transition rates between neighboring levels in the considered semiclassical limit of $\lambda \ll 1.$ The dephasing rate of the auxiliary mode $\tilde \gamma_{\rm ph}$ is proportional to the original ancilla dephasing rate $\gamma_{\rm ph}$ and becomes equal to $\gamma_{\rm ph}$ in the limit of weak drive. 

Neglecting the two-photon transitions, the steady-state distribution among the 
eigenstates of the auxiliary mode is of the Boltzmann form and can be written as follows
\begin{align}
P_n^{\rm st} = \left(\frac{\tilde n_{\rm th}}{\tilde n_{\rm th}+1}\right)^n/(\tilde n_{\rm th}+1)
\end{align}
where the effective thermal population is given by
\begin{align}
\label{eq:tilde_n_th}
\tilde n_{\rm th} = n_{\rm th} + (2n_{\rm th}+1)\sinh^2\phi + (Q_0^2/\lambda)\gamma_{\rm ph}^{\rm (hf)}/\gamma
\end{align}
In the absence of ancilla dephasing, the above equation reduces to that found in Ref.~\cite{Dykman2011}.

The total population in the excited eigenstates of the auxiliary mode $1-P^{\rm st}_0 = \tilde n_{\rm th}/(\tilde n_{\rm th}+1)$ as a function of the scaled drive power is shown in Fig.~\ref{fig:heating} of the main text. For the case $\gamma_{\rm ph}^{\rm (hf)}/\gamma = 0$, the effective thermal population $\tilde n_{\rm th}$ saturates at strong drive due to the saturation of $\sinh^2\phi$ in Eq.~(\ref{eq:tilde_n_th}) (see also Fig.~\ref{fig:squeezing}). For a non-zero $\gamma_{\rm ph}^{\rm (hf)}/\gamma$, the effective temperature $\tilde n_{\rm th}$ would rise indefinitely according to Eq.~(\ref{eq:tilde_n_th}) but eventually will be constrained by the finite bandwidth of the noise that leads to ancilla dephasing. We note that for not very small $\lambda$, the semiclassical analysis already captures qualitatively the behavior of the results from the full Floquet analysis. 

\subsubsection{Decoherence rates of the driven ancilla}
\label{sec:semiclassics_decoherence_rate}
We discuss in this section the drive-dependence of the decoherence rates $V_{mn}$ of superpositions of Floquet states in Eq.~(\ref{eq:decoherence_rate}) in the semiclassical approximation. 

We approximate the Floquet states as the eigenstates of the auxiliary mode described by the Hamiltonian in Eq.~(\ref{eq:Hamiltonian_auxiliary}). Using Eq.~(\ref{eq:V_gamma_simplified}) and the squeezing transformation in Eq.~(\ref{eq:squeezing}), we find the dissipation-induced decoherence rate to be
\begin{align}
V^{\gamma}_{mn} \approx \gamma [ (\sinh^2 \phi +1/2)(m+n)+\sinh^2 \phi].
\end{align}
The result above is shown in Fig.~\ref{fig:linewidth}(a) and qualitatively matches the full Floquet results. 

Similarly, using Eq.~(\ref{eq:V_gamma_ph_simplified}), we find that the dephasing-induced decoherence rate to be
\begin{align}
\label{eq:semiclassics_V_gamma_ph}
V^{\gamma_{\rm ph}}_{mn} &\approx \gamma_{\rm ph}^{\rm (hf)}\sinh^2\phi\cosh^2\phi(m^2+n^2+m+n+2) \nonumber \\
&+\gamma_{\rm ph}(0)(2\sinh^2 \phi + 1)^2(m-n)^2.
\end{align}
In the case $\gamma_{\rm ph}^{\rm (hf)}\ll \gamma_{\rm ph}(0)$, the above result shows that $V_{mn}^{\gamma_{\rm ph}}$ increases as a function of drive power and saturates to a value slightly larger than its value in the absence of the drive. However, it does not capture the sharp decrease of $V^{\gamma_{\rm ph}}_{mn}$ as the drive power increases found using the full Floquet analysis shown in Fig.~\ref{fig:linewidth}(b). This indicates the significance of the quantum correction from the non-quadratic terms in Eq.~(\ref{eq:displaced_Hamiltonian}) which are neglected in obtaining Eq.~(\ref{eq:Hamiltonian_auxiliary}).

\subsubsection{Nonlinearities of the auxiliary mode}
\label{sec:auxiliary_nonlinearity}
We discuss now the effects of the terms nonlinear in $c_{\rm aux},c^\dagger _{\rm aux}$ that we have neglected in arriving at Eq.~(\ref{eq:Hamiltonian_auxiliary}). Those nonlinear terms come from the original ancilla nonlinearity $-\alpha c^{\dagger 2}c^2/2$. For clarification, we list those terms below:
\begin{align}
\label{eq:nonlinear_c_aux}
-\frac{\alpha}{2}c^{\dagger 2}c^2 \rightarrow -\frac{\alpha}{2}(c^\dagger _{\rm aux}\cosh\phi - c_{\rm aux}\sinh \phi + Q_0/\sqrt{2\lambda})^2 \nonumber \\
\times(c_{\rm aux}\cosh\phi - c^\dagger _{\rm aux}\sinh \phi + Q_0/\sqrt{2\lambda})^2
\end{align}

On the one hand, nonlinear terms in $c_{\rm aux},c^\dagger _{\rm aux}$ make the levels of the auxiliary mode slightly non-equidistant. Collecting all the resonant (i.e. non-rotating) nonlinear terms, we find that the effective self-Kerr of the auxiliary mode (corresponding to a term $-\hbar\alpha_{\rm aux}c^{\dagger 2}_{\rm aux}c_{\rm aux}^2/2$ in the Hamiltonian) has the following form
\begin{align}
\label{eq:alpha_aux_general}
\frac{\alpha_{\rm aux}}{\alpha} &= C_4(\phi) - C_3(\phi)\frac{Q_0^2}{(\omega_{\rm aux}/|\delta_1|)}\sgn(Q_0)
\end{align}
where $C_4, C_3$ are positive functions of the squeezing angle $\phi$. The term proportional to $C_4$ comes from the quartic terms in $c_{\rm aux},c^\dagger _{\rm aux}$ in Eq.~(\ref{eq:nonlinear_c_aux}) whereas the term proportional to $C_3$ comes from the cubic terms in $c_{\rm aux},c^\dagger _{\rm aux}$ taken to second order. In general, both terms are of the same order of magnitude and can have different signs depending on the sign of $Q_0$. 

For the case $Q_0>0$, we find after some algebra that
\begin{align}
\label{eq:alpha_aux}
\frac{\alpha_{\rm aux}}{\alpha} &= \frac{-3Q_0^4 + 2}{2[3Q_0^2+\sgn(\delta_1)]^2}.
\end{align}
The above result applies to both positive detuning ($\delta_1>0$) and negative detuning ($\delta_1<0$) as long as $Q_0>0$. Interestingly, for a strong drive-1 where $\overline \Omega_1 \gg 1$ such that $Q_0 \gg 1$, the effective Kerr of the auxiliary mode changes from being positive to negative (with respect to the sign of $\alpha$) and approaches $-\alpha/6$ in the limit $\overline\Omega_1 \rightarrow \infty.$ This  sign change in $\alpha_{\rm aux}$ is a consequence of the $C_3$ term dominating over the $C_4$ term in Eq.~(\ref{eq:alpha_aux_general}).

The change in the sign of the effective anharmonicity $\alpha_{\rm aux}$ has an interesting consequence. If one now turns on the second drive on the ancilla, the AC Stark shift to the transition frequencies of the auxiliary mode due to this drive can also change sign depending on the sign of $\alpha_{\rm aux}$. Indeed, one finds that,  to leading order in the drive amplitude $\Omega_2$, the AC Stark shift to the frequency $\epsilon_{10}/\hbar$ of transition between the ground and excited state of the auxiliary mode reads
\begin{align}
&\delta\epsilon_{10}/\hbar \nonumber \\
&= -2\alpha_{\rm aux} |\Omega_2|^2 \frac{2\sinh^2\phi (\omega_{21}^2+\omega_{\rm aux}^2)+(\omega_{21}-\omega_{\rm aux})^2}{(\omega_{21}^2-\omega_{\rm aux}^2)^2}.
\end{align}
Clearly, $\epsilon_{10}/\hbar$ changes sign when $\alpha_{\rm aux}$ changes sign.  This sign change in the AC Stark shift due to drive-2 was also observed in the experiment where the strength of drive-1 has not yet reached the semiclassical regime yet; see Sec.~\ref{sec:Stark_shift}. Quite interestingly, for stronger drive-2, the AC Stark shift becomes non-monotonic in its amplitude as shown in the main text.  

We point out that the expression for the effective anharmonicity $\alpha_{\rm aux}$ in Eq.~(\ref{eq:alpha_aux}) can also be found using the semiclassical quantization rule: $g_n - g_{n-1} = -\lambda \nu(g_{n-1})$ where $g_n$ is the eigenenergy of the $n$-th excited state of the Hamiltonian $g$ about the equilibrium position and $\nu(g_{n-1})$ is the frequency of the orbit for the classical Hamiltonian at energy $g_{n-1}$. Expanding $\nu(g_n)$ about the ground state energy $g_0$ and keeping to leading order term in $\lambda$, one finds that \cite{dykman2012} \[\frac{\alpha_{\rm aux}}{\alpha} = -\frac{1}{2}\frac{d\nu}{dg}\bigg|_{g=g_0}\nu(g_0).\] The expression above is invariant with respect to the choice of coordinate system. 

In addition to renormalizing the effective Kerr $\alpha_{\rm aux}$, the cubic terms in $c_{\rm aux}, c^\dagger _{\rm aux}$  in Eq.~(\ref{eq:nonlinear_c_aux}) also provide a capability of three-wave mixing. If one now turns on the second drive on the ancilla and couples the ancilla to two off-resonant cavity modes, the three-wave mixing among the two cavities and drive-2 can lead to beam-splitter or two-mode squeezing interaction between the two cavities. We note that the strength of such three-wave mixing is proportional to $Q_0$ which follows Eq.~(\ref{eq:equation_for_Q0}). When $\overline \Omega_1$ becomes of order one, $Q_0$ becomes sublinear in $\overline \Omega_1$ as can be seen in Fig.~\ref{fig:beam-splitter} of the main text. 

\section{Quasienergy level anti-crossing and experimental evidence}
\label{sec:level_anti-crossing}
In this section, we present experimental evidence of quasienergy level anti-crossing due to multi-photon resonance and show that the Floquet theory accurately predicts the locations of the level anti-crossings. 

As we discussed in Sec.~\ref{sec:multi-photon_resonance}, if the ancilla comes close to, or passes through, a quasienergy level anti-crossing near the peak of the drive envelopes where the drive amplitudes change slowly, the probability of diabatic vs. adiabatic transition between the two levels can become comparable. In an AC Stark shift measurement of the ancilla transition frequency $E_{10}/\hbar$ as described in Sec.~\ref{sec:Stark_shift_exp}, if the above situation occurs, we would observe that the ancilla has a finite probablity of not remaining in the ground state regardless of the frequency of the spectrosopy tone. 

\begin{figure}[ht]
\centering
\includegraphics[width=7cm]{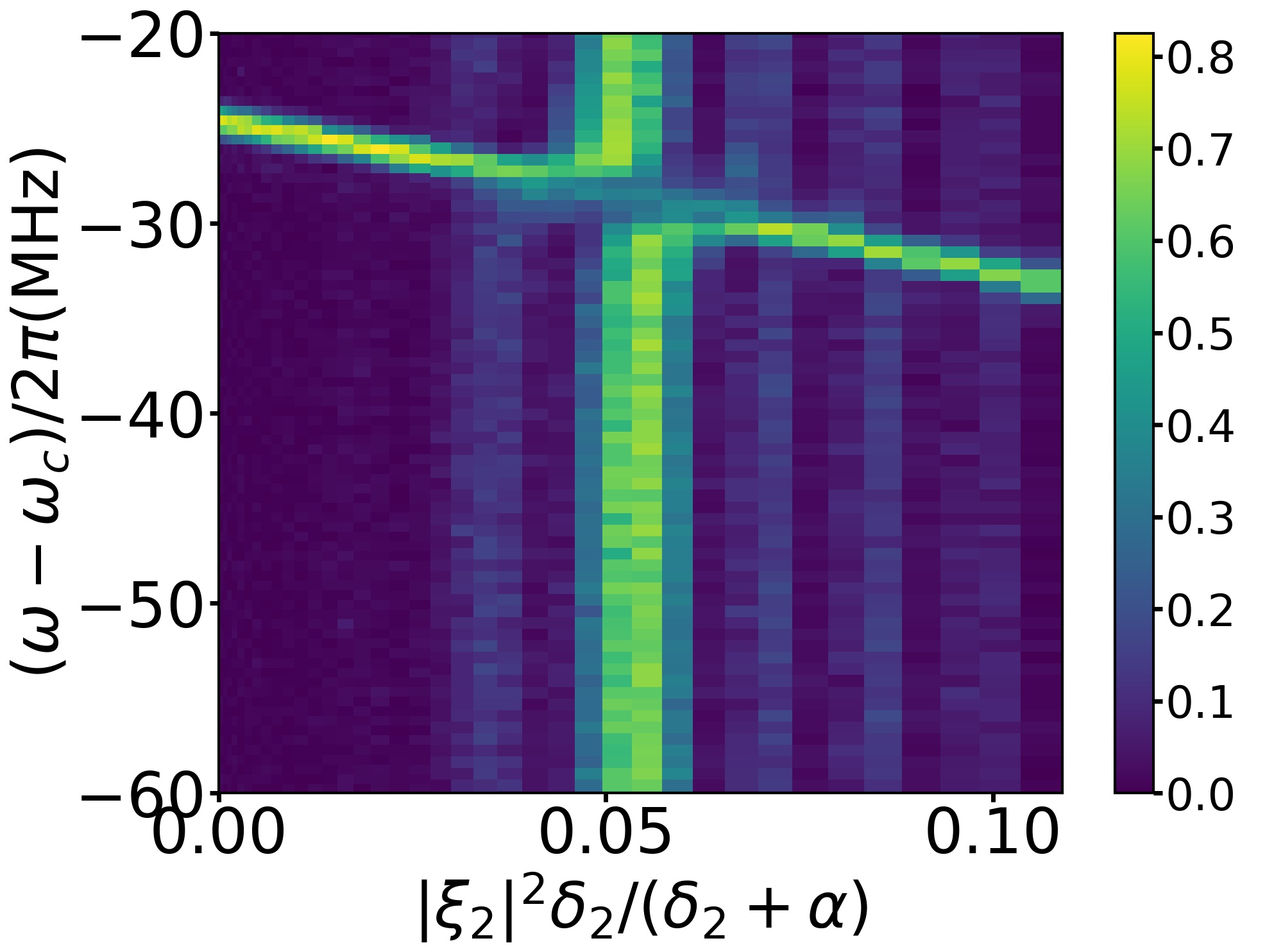} \hfill
\includegraphics[width=6.5cm]{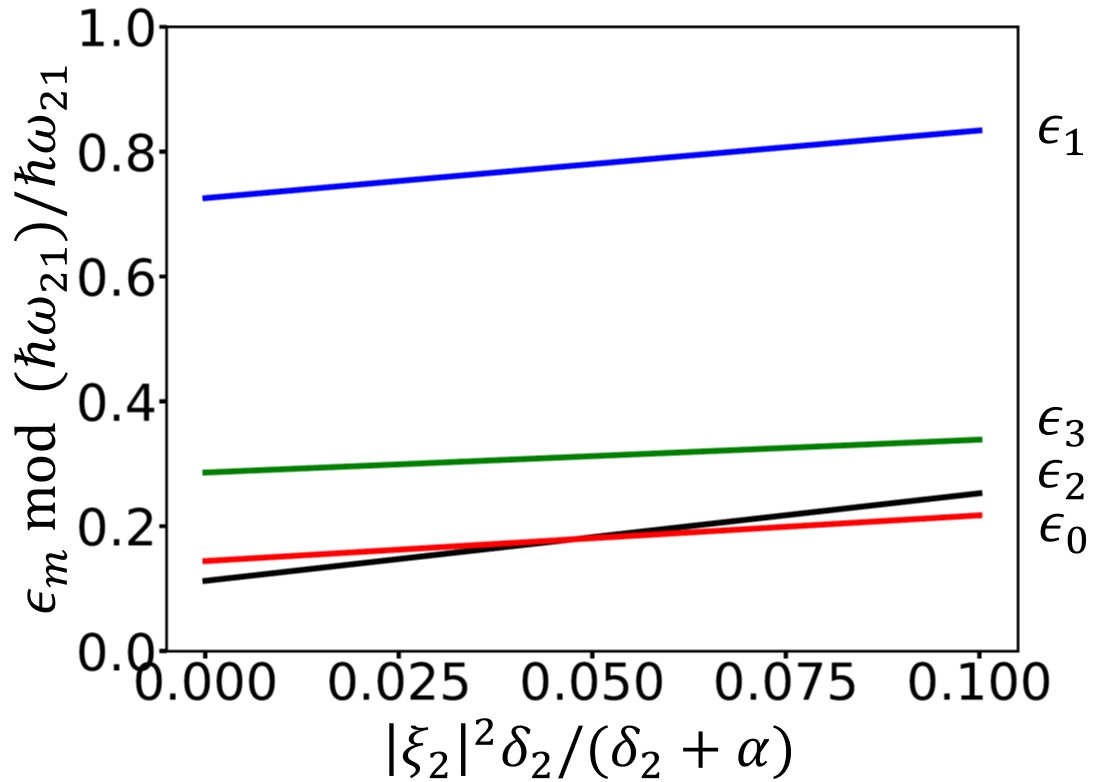} 
\caption{Quasienergy level anti-crossing due to multi-photon resonance. Top: Spectroscopy of the two-tone driven ancilla as a function of the scaled drive-2 power. The power of drive-1 is fixed at $|\xi_1|^2\delta_1/(\delta_1+\alpha) = 0.22$. The scaled drive detunings $\delta_1/\alpha = 1,\delta_2/\alpha =4.5.$ The vertical axis is the frequency of the spectroscopy tone (a $\pi$ pulse) counted from the ancilla transition frequency $\omega_c$. The color indicates the population of the ancilla not in the ground state. Bottom: the quasienergy spectrum of the driven ancilla for the same range of drive strengths as in the top panel. From top to down at $\xi_2 = 0$, the blue, green, red and black lines refer to the quasienergy levels $\epsilon_1, \epsilon_3, \epsilon_2, \epsilon_0$, respectively, projected into the same Brillouin zone. The anti-crossing between $\epsilon_0$ and $\epsilon_2$ indicates a multi-photon resonance where the ancilla is excited from the ground to the second excited state by absorbing three drive-1 photons and emitting one drive-2 photon. The gap of the anti-crossing is too weak to be seen on the scale of the plot. In the top panel, the dip at around $-30$~MHz at the drive strength where the anti-crossing occurs is likely due to that the ancilla decays from the second  to the first excited state and then be de-excited to the ground state by the spectroscopy tone.}
\label{fig:level_anti-crossing_exp}
\end{figure}

We show in Fig.~\ref{fig:level_anti-crossing_exp} an example of this situation where we vary the strength of drive-2 and keep the strength of drive-1 fixed. At a particular strength of drive-2, there is a sharp vertical line in the spectrum indicating a high probability of the ancilla not being in the ground state regardless of the frequency of the spectroscopy tone. A comparison with the theoretical quasienergy spectrum shows that there is an anti-crossing between quasienergy levels $\epsilon_0$ and $\epsilon_2$ at that particular drive strength; as a result, the ancilla has a significant probablity of transitioning from the state $\Psi_0$ to $\Psi_2$ during the ramping up and down of the drives. Near this resonance, we observed weak oscillations in the probability of the ancilla not being in the ground state, a typical situation in a Landau-Zener transition. Far away from the resonance, these oscillations damp out indicating a (mostly) diabatic transition when the ancilla passes through the resonance. 

Lastly, we emphasize that the process of multi-photon resonance is strongly suppressed when the frequency difference of the two drive tones is much larger than their drive amplitudes. In particular, this is the case for the beam-splitter experiment presented in Fig.~\ref{fig:beam-splitter}, where the frequency difference of the two drives $\omega_{21}/\alpha \approx 15$, was much larger than what we used here. 

\section{Transient susceptibilities}
\label{sec:transient_susceptibilities}
Also of interest to us are the transient susceptibilities where the ancilla is initialized in a given Floquet state. The analysis of the transient susceptibilities greatly simplifies in the limit of weak damping and dephasing where Eqs.~(\ref{eq:diagonal},\ref{eq:off_diagonal}) hold and the probe detunings from the ancilla resonances are much larger than the corresponding linewidths. Assuming that the ancilla is initially in a pure Floquet state $\Psi_j$ at $t=0$, we find that the ensemble-averaged susceptibilities read 
\begin{align}
\label{eq:chi1_transient}
&\chi (\omega,\omega+K\omega_{21},t) = \sum_{m} P_m (t) \chi _m(\omega,\omega+K\omega_{21})  \nonumber\\
&+ i \sum_{mnK'}[P_m(t) V_{mn} + \dot P_m(t)]\left( \frac{\mathcal  F_{mnKK'}}{\delta_{mnK'}^2(\omega)} - \frac{\mathcal F_{nmKK'}}{\delta_{nmK'}^2(\omega)}\right)
\end{align}
\begin{align}
\label{eq:X1_transient}
&X (-\omega,2\omega_1+K\omega_{21}-\omega,t)  \nonumber \\
&= \sum_m P_m(t) X _m(-\omega,2\omega_1+K\omega_{21}-\omega)+  i \sum_{mnK'}[P_m(t) V_{mn} \nonumber \\
& + \dot P_m(t)]\left( \frac{\tilde{\mathcal  F}_{mnKK'}}{\delta_{mnK'}^2(-\omega)} - \frac{ \tilde{\mathcal F}_{nmKK'}}{\delta_{nmK'}^2(-\omega)}\right) ,\nonumber \\ 
\end{align}
Here $P_m(t)$ denotes the time-dependent population of the ancilla in the state $\Psi_m$ and it satisfies the rate equation~(\ref{eq:diagonal}) with $P_m(t)\equiv \rho_{mm}(t)$ and the initial condition $\rho_{mm}(0) = \delta_{mj}.$ The (unitary) partial susceptibilities $\chi_m$ and $X_m$ are given in Eqs.~(\ref{eq:chi1}, \ref{eq:X1}).  We have introduced a shorthand notation for the squared Floquet matrix elements and the detuning of the probe frequency from the corresponding Floquet resonances 
\begin{align*}
&\mathcal F_{mn KK'} = c_{mn,K'-K}(c^\dagger )_{nm,-K'}, \\
&\tilde{\mathcal F}_{mn KK'} = c_{mn,K'-K}c_{nm,-K'}, \\
&\delta_{mnK'}(\omega) = (\omega-\omega_1) -\epsilon_{nm}/\hbar + K'\omega_{21}.
\end{align*}
Eqs.~(\ref{eq:chi1_transient},\ref{eq:X1_transient}) apply when the detunings $\delta_{mnK'}$ are much larger than the linewidths $V_{mn}$ and the changing rate of $P_m(t)$.

After a time set by the inverse relaxation time of the ancilla, the transient susceptibilities become the steady-state susceptibilities. This can be seen by setting $P_m(t)$ to the steady-state population $P_m^{\rm st}$ and $\dot P_m(0)$ to zero in Eqs.~(\ref{eq:chi1_transient},\ref{eq:X1_transient}). A more general result for the steady-state susceptibilities beyond the regime $|\delta_{mnK'}|\gg V_{mn}$ is given in Eq.~(\ref{eq:averaged_susceptibilities}) of the main text.


One can think of the ensemble-averaged susceptibilities in Eqs.~(\ref{eq:chi1_transient},\ref{eq:X1_transient}) in a quantum-trajectory-like picture where the ancilla randomly jumps from one Floquet state to another. In between the jumps, the ancilla remains in a given Floquet state, say, $\Psi_j$. Then the ``instantaneous" susceptibilities of the ancilla are given by the short-time limit of Eqs.~(\ref{eq:chi1_transient},\ref{eq:X1_transient}), where $t$ is much shorter than the relaxation time of the ancilla. Specifically, we need to set all  probabilities $P_{m\neq j}(t)$ in Eqs.~(\ref{eq:chi1_transient},\ref{eq:X1_transient}) to zero, $P_j$ to 1, and the time derivative of the probabilities to $\dot P_{m\neq j} = W_{jm}$, $\dot P_j = - \sum_{j'\neq j} W_{jj'}$ based on Eq.~(\ref{eq:diagonal}). We denote the resulting susceptibility as $\chi_j^{\rm tr}$ which is given by,
\begin{align}
\label{eq:chi1_j}
&\chi_j^{\rm tr}(\omega,\omega+K\omega_{21}) =\chi _j(\omega,\omega+K\omega_{21})  \nonumber\\
&+ i \sum_{nK'}[V_{jn} - \sum_{j'\neq j} W_{jj'}]\left( \frac{\mathcal  F_{jnKK'}}{\delta_{jnK'}^2(\omega)} - \frac{\mathcal F_{njKK'}}{\delta_{njK'}^2(\omega)}\right) \nonumber \\
& + i \sum_{nK',j'\neq j} W_{jj'}\left( \frac{\mathcal  F_{j'nKK'}}{\delta_{j'nK'}^2(\omega)} - \frac{\mathcal F_{nj'KK'}}{\delta_{nj'K'}^2(\omega)}\right) 
\end{align}
Note that the expression above is different from the partial susceptibility $\chi_j^{\rm st}$ in Eq.~(\ref{eq:chi1_decay}). One can show that summing the instantaneous susceptibility $\chi_j^{\rm tr}$ over the steady-state distribution $P_j^{\rm st}$ recovers the ensemble-averaged steady-state susceptibility $\chi^{\rm st}$ in Eq.~(\ref{eq:averaged_susceptibilities}). The instantaneous susceptibility $X_j^{\rm tr}$ can be found similarly.

\section{Incoherent hopping between the cavity and ancilla induced by ancilla dephasing}
\label{sec:two_state_model}
In this section, we study in detail the dephasing-induced incoherent hopping between the cavity and ancilla using a two-state approximation.  To be concrete, we consider the situation described by the Hamiltonian in Eq.~(\ref{eq:inverse_Purcell_Rabi}) where cavity mode $a$ is close to the resonance $\nu_{0mK}$ and other resonance processes can be neglected. 

Implicitly assumed in the formula for the inverse Purcell decay rate in Eq.~(\ref{eq:delta_kappa}) is that the drives are turned on relatively slowly so that any initial state of the cavity mode stays in the adiabatic state of the coupled ancilla-cavity system described by Eq.~(\ref{eq:inverse_Purcell_Rabi}) as the drives are being turned on. Here, we make the same assumption and show that ancilla dissipation and dephasing lead to incoherent hopping between the adiabatic states of the Hamiltonian in Eq.~(\ref{eq:inverse_Purcell_Rabi}), and the hopping rate reduces to the inverse Purcell decay rate in the weak coupling limit.

\begin{figure}[ht]
\centering
\includegraphics[width=7cm]{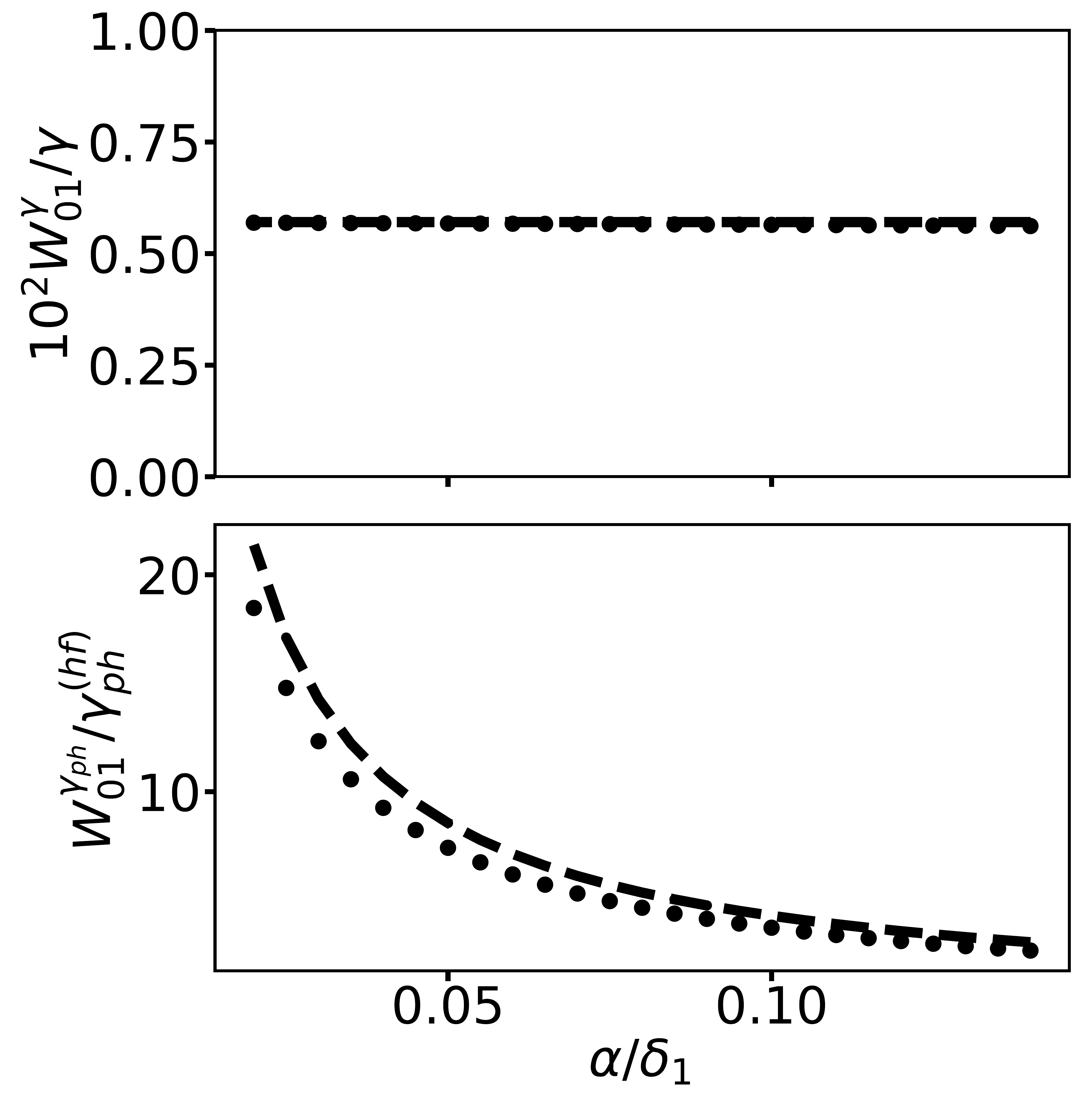} 
\caption{The scaling of the rates $W^{\gamma}_{01}$ and $W^{\gamma_{\rm ph}}_{01}$ with respect to $\alpha/\delta_1$ for fixed $\alpha|\xi_1|^2/\delta_1 = 0.315 $ and $\alpha |\xi_2|^2/\delta_2 = 0.001 $. The ratio $\delta_2/\delta_1 = 3.1.$ and $n_{\rm th} = 0$. The dots show the Floquet calculation using Eq.~(\ref{eq:hopping_rate}). The dashed lines show the semiclassical result in Eq.~(\ref{eq:semiclassical_gamma_up}). For the chosen parameters, the transition rates are primarily due to the much stronger drive-1. We clarify that the rate $W^{\gamma_{\rm ph}}_{01}$ is non-zero even when the ancilla is linear ($\alpha = 0$) and is proportional to the scaled drive power $|\xi_1|^2+|\xi_2|^2$ for weak drives; see Eq.~(\ref{eq:heating_rate_weakdrive}). The dependence on $\alpha/\delta_1$ shown in the figure is a result of decreasing $\alpha/\delta_1$ while
keeping $\alpha|\xi_1|^2/\delta_1$ and $\alpha |\xi_2|^2/\delta_2$ fixed, which requires increasing $\xi_1,\xi_2$ proportionally.}
\label{fig:heating_rates_scaling}
\end{figure}

To relate to the experiment in Sec.~\ref{sec:inverse_Purcell_exp}, we consider that cavity mode $a$ is initially in the Fock state $|1_a\rangle$ and the ancilla is in the vacuum state $|0_c\rangle$ before the drives are turned on. After the drives have been turned on adiabatically, the state $|1_a, 0_c\rangle$ becomes an eigenstate of the the Hamiltonian in Eq.~(\ref{eq:inverse_Purcell_Rabi}): $\psi_1 = \alpha_1 |1_a,u_0\rangle + \beta_1 |0_a, u_m\rangle$. We denote another eigenstate of the Hamiltonian in Eq.~(\ref{eq:inverse_Purcell_Rabi}) in the same subspace as $\psi_2 = \alpha_2 |1_a,u_0\rangle + \beta_2 |0_a, u_m\rangle$. In the weak coupling limit, $\psi_1$ will be mostly $|1_a,u_0\rangle$ and $\psi_2$ will be mostly $|0_a,u_m\rangle$. 

Finite ancilla dissipation and dephasing induce coupling between these two states and thus they acquire finite widths. In the limit where the decoherence-induced widths are much smaller than the energy splitting between these two states, there occurs incoherent hopping between these two states as well as pure loss from these states which can be described by the following rate equation,
\begin{align}
\label{eq:two_state_rate_equation}
\dot \rho_{1} &= - (R_{11}+R_{12})\rho_{1}  + R_{21} \rho_{2} \nonumber \\
\dot \rho_{2} &= - (R_{22}+R_{21})\rho_{2}  + R_{12} \rho_{1},
\end{align} 
where $\rho_{1,2}$ refer to the population in the states $\psi_{1,2}$, respectively. $R_{12},R_{21}$ refer to the hopping rates between the two states; $R_{11}, R_{22}$ refer to the individual loss rates of the two states due to hopping to states outside the subspace. To simplify the analysis, we neglected the finite heating rate of the ancilla from lower to higher Floquet states which is small for not too strong drive. Therefore, we can restrict ourselves to the two state subspace spanned by the states $\Psi_1$ and $\Psi_2$ without considering the finite rates of hopping to higher Floquet states and then quickly hopping back. 

The rates $R_{ij}$ can be calculated using Fermi' s golden rule. The incoherent hopping between the two states is caused by the pure dephasing of the ancilla Floquet states and the hopping rates are found to be
\begin{align}
\label{eq:two_state_incoherent_hopping}
&R_{12} =  \sum_K  \frac {1}{2}\gamma |\alpha_1\alpha_2^*c_{00,-K}+\beta_1\beta_2^*c_{mm,-K}|^2 \nonumber \\
&+2 |\alpha_1\alpha_2^* (c^\dagger c)_{00,K}+\beta_1\beta_2^*(c^\dagger c)_{mm,K}|^2 \gamma_{\rm ph}(K\omega_{21}+\tilde \Omega_R),
\end{align} 
and $R_{21}$ has the same form as $R_{12}$ except that the argument of $\gamma_{\rm ph}$ changes sign. Expression~(\ref{eq:two_state_incoherent_hopping}) reduces to the pure dephasing rate of the Floquet states in Eq.~(\ref{eq:decoherence_rate}) in the weak coupling limit where  $\alpha_1\alpha_2^* = -\beta_1\beta_2^* \approx -g_a/(\omega_a - \nu_{0nK})$; an important difference is that the rate here depends on the spectrum of the dephasing noise at the Rabi splitting frequency between states $\psi_1$ and $\psi_2$: \[\tilde \Omega_R = \sqrt{(\omega_a-\nu_{0mK})^2 + (2g_a  (c^\dagger )_{m0,-K})^2}.\] 
Here we assumed that $\omega_a < \nu_{0mK}$; in the opposite case, one needs to change the sign in front of $\tilde \Omega_R$ in Eq.~(\ref{eq:two_state_incoherent_hopping}). For simplicity, we have neglected the usually weak frequency-dependence in $\gamma$ and the incoherent hopping due to much smaller intrinsic dephasing and dissipation rate of the cavity.

The individual loss rates $R_{11}$ and $R_{22}$ are found to be,
\begin{align}
R_{11} = |\alpha_1|^2 \kappa_a^{(0)}  + |\beta_1|^2 \sum_{n<m} W_{mn} \nonumber \\
R_{22} = |\alpha_2|^2 \kappa_a^{(0)} + |\beta_2|^2 \sum_{n<m} W_{mn}
\end{align}
where the rates $W_{mn}$ are given by Eq.~(\ref{eq:hopping_rate}) with $\epsilon_m$ replaced by $\epsilon_m+(\omega_a - \nu_{0mK}-\tilde\Omega_R)/2$ in $\gamma(\omega)$ and $\gamma_{\rm ph}(\omega)$ for $R_{11}$ and by $\epsilon_m +(\omega_a - \nu_{0mK}+\tilde\Omega_R)/2$ for $R_{22}$. We note the rate $R_{11}+R_{12}$ reduces to the inverse Purcell decay rate in Eq.~(\ref{eq:delta_kappa}) plus the intrinsic cavity decay rate in the weak-coupling limit. 

The solution to Eq.~(\ref{eq:two_state_rate_equation}) for the initial condition $\rho_1(0)=1,\rho_2(0) = 0$ reads,
\begin{align}
\label{eq:sol_two_state_rate_equation}
&\rho_1(t) = e^{- R_+ t/2} [\cosh(Bt/2) + R_-\sinh(Bt/2)/B] \nonumber \\
&\rho_2(t) = 2R_{12} e^{-R_+t/2}\sinh(Bt/2)/B, \nonumber \\
&R_\pm = R_{22}+R_{21}\pm(R_{11}+R_{12}),\nonumber \\
&B = \sqrt{ R_-^2+4R_{12}R_{21}}
\end{align}

In the weak coupling limit, the incoherent hopping rate between the two states is much smaller than the rate of loss from the ancilla-like state $\psi_2$: $R_{21},R_{12}\ll R_{22}$. One can neglect the hopping back from the state $\psi_2$ to $\psi_1$ and it follows from Eq.~(\ref{eq:sol_two_state_rate_equation}) that $\rho_1(t)$ decays exponentially with a rate given by the inverse Purcell decay rate in Sec.~\ref{sec:inverse_Purcell}: $\rho_1(t)\approx \exp[-(R_{11}+R_{12})t]$. 

In the strong coupling limit, $\psi_1$ and $\psi_2$ are fully hybridized ancilla-cavity states, i.e. $|\alpha_{1,2}|^2\approx |\beta_{1,2}|^2 \approx 1/2$. If we assume that the noise that leads to ancilla dephasing has a symmetric spectrum about zero frequency, then $R_{11} = R_{22}, R_{12} =  R_{21}$. It follows from Eq.~(\ref{eq:sol_two_state_rate_equation}) that $\rho_1(t)\approx e^{-R_{11}t}(1+e^{-2R_{12}t})/2$, $\rho_2(t)\approx e^{-R_{11}t}(1-e^{-2R_{12}t})/2$. After a relatively fast decay (rise) in $\rho_1(\rho_2)$ with a rate $R_{11}+2R_{12}$, the populations decay with a slower rate $R_{11}$.

\section{Scaling of the SWAP infidelity with respect to the transmon anharmonicity}
\label{sec:infidelity_scaling}

In this Appendix, we study the scaling of the SWAP infidelity in Eq.~(\ref{eq:infidelity_smallalpha}) with respect to the transmon anharmonicity $\alpha/\alpha_0$ in the limit $\alpha/\alpha_0 \ll 1$ and argue that decreasing $\alpha/\alpha_0$ reduces the infidelity while the beam-splitter rate is kept fixed. Here, $\alpha_0$ is a fixed scaling factor set by the drive detunings or the cavity detunings from the transmon ancilla; for concreteness, we choose it to be $\alpha_0 \equiv |\delta_1|$.

To maintain the same beam-splitter rate while reducing $\alpha/|\delta_1|$, we need to keep the quantity $\alpha \xi_1\xi_2/|\delta_1|$ constant. This requires increasing both $\xi_1$ and $\xi_2$ by the same factor that scales as $1/\sqrt{\alpha/|\delta_1|}$.  Taking the limit $\alpha/|\delta_1|\rightarrow 0$ then effectively corresponds to taking the classical limit $\hbar \rightarrow 0$ since $\alpha/|\delta_1| \propto \hbar$, and $\alpha\xi_1\xi_2/\delta_1$ is independent of $\hbar$. We refer the reader to the systematic semiclassical analysis in Appendix~\ref{sec:semiclassical} where we introduced the scaled Planck constant $\lambda = \alpha/2|\delta_1|$ and the dimensionless drive amplitudes $\overline \Omega_{1,2} = \sqrt{\alpha |\xi_{1,2}|^2/|\delta_1|}.$


For a fixed beam-splitter rate $g_{\rm{BS},0}$, the quantity in the square bracket of Eq.~(\ref{eq:infidelity_smallalpha}) scales as $(\alpha/|\delta_1|)^2$ for small $\alpha/|\delta_1|$. This can be seen by substituting Eqs.~(\ref{eq:dw_smallalpha},\ref{eq:gBS_dispersion_smallalpha}) into Eq.~(\ref{eq:infidelity_smallalpha}) which gives
\begin{align}
\label{eq:dispersion_small_alpha}
\delta_{\rm{BS},m}/|\delta_1| =& -2m(\alpha/|\delta_1|) [|g_a/\delta_a|^2(1+\Delta_a)\nonumber \\
&-|g_b/\delta_b|^2(1+\Delta_b)] \nonumber \\
(g_{\rm{BS},m} - g_{\rm{BS},0})/g_{\rm{BS},0} =& -2m\alpha [\delta_a^{-1}+\delta_b^{-1}+\delta_1^{-1}+\delta_2^{-1} \nonumber \\
&+(\delta_a+\delta_2)^{-1}] 
\end{align}
In addition to reducing $\alpha/|\delta_1|$, $|\delta_{\rm{BS},m}|$ can be suppressed by engineering the two cavities so that $|g_a/\delta_a|\approx |g_b/\delta_b|$. Further suppression of $\delta_{\rm BS}$ can be realized by choosing the drive parameters which modify $\Delta_a$ and $\Delta_b$ to cancel any residual difference between $|g_a/\delta_a|$ and $|g_b/\delta_b|$. This is similar to the ``$\chi$ matching" scheme presented in Ref.~\cite{Rosenblum2018a}. The ratio $(g_{\rm {BS},m}-g_{\rm {BS},0})/g_{\rm {BS},0}$ can also be suppressed by engineering the frequencies of the cavities and choosing the frequencies of the drives so that the term in the square bracket of the second line in Eq.~(\ref{eq:dispersion_small_alpha}) is small.

Now we discuss the dependence on $\alpha/|\delta_1|$ of the transition rates $W_{0m}$ in Eq.~(\ref{eq:infidelity_smallalpha}). As we discussed in Sec.~\ref{sec:heating}, the rate $W_{0m}$ has two contributions: $W_{0m} =  W_{0m}^\gamma +W_{0m}^{\gamma_{\rm ph}}$ where $W_{0m}^\gamma$ is the rate of the dissipation-induced transition and $W_{0m}^{\gamma_{\rm ph}}$ is the rate of the dephasing-induced transition. One can show that for small $\alpha/|\delta_1|$, the transition from Floquet state $\Psi_0$ to the neighboring state $\Psi_1$ is dominant over transitions to other states. For instance, $W_{02}^\gamma/W_{01}^\gamma \propto \alpha/|\delta_1|, W_{03}^\gamma/W_{01}^\gamma \propto (\alpha/|\delta_1|)^2,$ and similarly for $W_{0m}^{\gamma_{\rm ph}}$. This can be seen from the semiclassical analysis in Sec.~\ref{sec:semiclassics_heating} by doing perturbations in the parameter $\lambda.$ 

The transition rates $W_{01}^\gamma$  and $W_{01}^{\gamma_{\rm ph}}$ have very different dependence on $\alpha/|\delta_1|$. As can be seen from Eq.~(\ref{eq:semiclassical_gamma_up}), the transition rate $W_{01}^\gamma$ is independent of $\alpha/|\delta_1|$ and only depends on the ratio $\alpha|\xi_1|^2/\delta_1$ (and also $\alpha|\xi_2^2|/\delta_2$) through the squeezing parameter. In contrast, the rate $W_{01}^{\gamma_{\rm ph}}$ is inversely proportional to $\alpha/|\delta_1|$. We show the scaling of these rates with respect to $\alpha/\delta_1$ for fixed $\alpha|\xi_1|^2/\delta_1$ and $\alpha|\xi_2|^2/\delta_2$ in Fig.~\ref{fig:heating_rates_scaling}.

Combining the scaling of the rate $W_{0m}$ and dispersion in $\delta_{\rm{BS},m}$ and $g_{\rm{BS},m}$ with respect to $\alpha/|\delta_1|$, we conclude that, overall, the infidelity in Eq.~(\ref{eq:infidelity_smallalpha}) will scale as $\alpha/|\delta_1|$.


%

\end{document}